\documentclass[showpacs,twocolumn,aps,prx,floatfix,superscriptaddress,noshowpacs]{revtex4-2}
\usepackage[table]{xcolor}
\usepackage{nicematrix}

\usepackage[mathlines]{lineno}%
\usepackage{amsthm,amsmath,amssymb,graphicx,bm,color,mathrsfs,verbatim,epstopdf,dcolumn,dsfont,cancel}
\usepackage[export]{adjustbox} %

\usepackage{algorithm}
\usepackage[noend]{algpseudocode}

\usepackage{etoolbox}

\makeatletter
\AfterEndEnvironment{algorithm}{\let\@algcomment\relax}
\AtEndEnvironment{algorithm}{\kern2pt\hrule\relax\vskip3pt\@algcomment}
\let\@algcomment\relax
\newcommand\algcomment[1]{\def\@algcomment{\footnotesize#1}}

\renewcommand\fs@ruled{\def\@fs@cfont{\bfseries}\let\@fs@capt\floatc@ruled
    \def\@fs@pre{\hrule height.8pt depth0pt \kern2pt}%
    \def\@fs@post{}%
    \def\@fs@mid{\kern2pt\hrule\kern2pt}%
    \let\@fs@iftopcapt\iftrue}
\makeatother



\usepackage{graphicx}
\usepackage{soul}
\usepackage{mathtools}
\usepackage{booktabs}
\usepackage{multirow}
\usepackage{makecell}

\usepackage{balance}

\graphicspath{{figs/},{./}}

\usepackage{color,xcolor}

\definecolor{mydarkblue}{rgb}{0,0.08,0.45}
\definecolor{mydarkorange}{HTML}{804600}
\usepackage{flushend}

\usepackage{hyperref}
\hypersetup{
    colorlinks = true,
    citecolor=mydarkblue,
    urlcolor=mydarkblue
}

\makeatletter
\algnewcommand{\LineComment}[1]{\Statex \hskip\ALG@thistlm \(\triangleright\) #1}
\makeatother

\usepackage{mathtools}

\DeclarePairedDelimiter\floor{\lfloor}{\rfloor}

\usepackage{rotating} 
\definecolor{Gray}{gray}{0.7}
\definecolor{GrayBG}{gray}{0.95}
\usepackage{caption, subcaption}

\DeclareCaptionLabelFormat{opening}{#2}


\makeatletter
\renewcommand\@makecaption[2]{%
  \par
  \vskip\abovecaptionskip
  \begingroup
   \small\rmfamily
    \begingroup
     \samepage
     \flushing
     \let\footnote\@footnotemark@gobble
     \@make@capt@title{#1}{#2}\par
    \endgroup
  \endgroup
  \vskip\belowcaptionskip
}
\makeatother

\usepackage{physics}

\usepackage[normalem]{ulem}

\usepackage{cancel}

\newcommand{\eq}{\!=\!}
\newcommand{\plus}{\!+\!}

\definecolor{tblue}{HTML}{1f77b4}
\definecolor{torange}{HTML}{ff7f0e}
\definecolor{tgreen}{HTML}{2ca02c}
\definecolor{tred}{HTML}{d62728}

\newcommand{\params}{{\bm\theta}}

\begin{document}

\title{Reinforcement Learning for Many-Body Ground-State Preparation \\
    Inspired by Counterdiabatic Driving}

\author{Jiahao Yao}
\email{jiahaoyao@berkeley.edu}
\affiliation{Department of Mathematics, University of California, Berkeley, California 94720, USA}

\author{Lin Lin}
\affiliation{Department of Mathematics, University of California, Berkeley, California 94720, USA}
\affiliation{Computational Research Division, Lawrence Berkeley National Laboratory, Berkeley, California 94720, USA}
\affiliation{Challenge Institute for Quantum Computation, University of California, Berkeley, California 94720, USA}

\author{Marin Bukov}
\email{mgbukov@phys.uni-sofia.bg}
\affiliation{Department of Physics, University of California, Berkeley, California 94720, USA}
\affiliation{Department of Physics, St.~Kliment Ohridski University of Sofia, 5 James Bourchier Boulevard, 1164 Sofia, Bulgaria}

\date{\today}

\begin{abstract}
The quantum alternating operator ansatz  (QAOA) is a prominent example of variational quantum algorithms. 
We propose a generalized QAOA called CD-QAOA, which is inspired by the counterdiabatic (CD) driving procedure, designed for quantum many-body systems and optimized using a reinforcement learning (RL) approach.
The resulting hybrid control algorithm proves versatile in preparing the ground state of quantum-chaotic many-body spin chains by minimizing the energy. We show that using terms occurring in the adiabatic gauge potential as generators of additional control unitaries, it is possible to achieve fast high-fidelity many-body control away from the adiabatic regime. While each unitary retains the conventional QAOA-intrinsic continuous control degree of freedom such as the time duration, we consider the order of the multiple available unitaries appearing in the control sequence as an additional discrete optimization problem. Endowing the policy gradient algorithm with an autoregressive deep learning architecture to capture causality, we train the RL agent to construct optimal sequences of unitaries. The algorithm has no access to the quantum state, and we find that the protocol learned on small systems may generalize to larger systems. 
By scanning a range of protocol durations, we present numerical evidence for a finite quantum speed limit in the nonintegrable mixed-field spin-$1/2$ Ising and Lipkin-Meshkov-Glick models, and for the suitability to prepare ground states of the spin-$1$ Heisenberg chain in the long-range and topologically ordered parameter regimes.  This work paves the way to incorporate recent success from deep learning for the purpose of quantum many-body control.
\end{abstract}
 
\maketitle

\section{\label{sec:intro}Introduction}

The ability to prepare a quantum many-body system in its ground state is an important milestone in the quest for understanding and identifying novel collective quantum phenomena. The degree to which ground states can be confidently prepared in present-day quantum simulators, delineates the limits of our capabilities to investigate the properties of new materials or molecules, and to propose innovative technological applications based on quantum effects, such as high-temperature superconductors and superfluids, magnetic field sensors, topological quantum computers, or synthetic molecules. 

Quantum simulators, such as ultracold and Rydberg atoms~\cite{lewenstein2007ultracold,bloch2008manybody}, trapped ions~\cite{haffner2008quantum,blatt2012quantum,monroe2013scaling, devoret2013superconducting}, nitrogen vacancy centers~\cite{doherty2013nitrogen,schirhagl2014nitrogen,casola2018probing}, and superconducting qubits~\cite{devoret2013superconducting,xiang2013hybrid}, all require the development of state preparation schemes via \emph{real-time} dynamical processes.
Despite their high level of controllability, finding short protocols to prepare strongly-correlated ground states under platform-specific constraints, is a challenging problem in AMO-based quantum simulation platforms, due to the exponentially large Hilbert space dimensions of quantum many-body systems. 
On this background, speed-efficient protocols also become progressively more important for near-term quantum computing devices~\cite{arute2019quantum}, where simulation errors grow with the protocol duration due to imperfections in the implementation of the basic gate operations. %

Developing versatile methods for ground state preparation will enable quantum simulators to investigate hitherto unexplored quantum phases of matter, and determine the behavior of order parameters, correlation lengths and critical exponents.
Theoretically, although an exact mathematical expression for the ground state might be known in some models, it remains still largely unclear how to prepare it in a unitary dynamical process.
In generic models, the lack of closed-form analytical solutions motivates the use of numerical algorithms. Prominent examples for quantum state preparation include established quantum control algorithms, such as GRAPE~\cite{khaneja2005optimal} and CRAB~\cite{caneva2011chopped}, 
and variational quantum eigensolvers (VQE)~\cite{peruzzo2014variational}, such as the quantum approximate optimization ansatz (QAOA)~\cite{farhi2014quantum}.

In this study, we present a novel hybrid reinforcement learning (RL)/optimal control algorithm based on an autoregressive deep learning architecture. We improve the current state-of-the-art for digital quantum control techniques by enhancing the capabilities to find optimal protocols that prepare the ground state of quantum many-body systems. The emerging versatile algorithm combines discrete and continuous control parameters to achieve maximum flexibility in its applicability to a number of different models. 

To cope with the complexity of preparing ordered states in quantum many-body systems, we introduce a novel ansatz inspired by variational gauge potentials and counter-diabatic (CD) driving~\cite{demirplak_03,berry_09,kolodrubetz2017geometry,bukov2019geometric}. 
This allows us to excite the system away from equilibrium in a controllable manner to find short high-fidelity protocols away from the adiabatic regime.
We demonstrate that combining features of CD driving with the digital simulation character of conventional QAOA yields superior performance over a wide range of protocol durations and physical models. 
Compared to the standard counter-diabatic driving algorithms, CD-QAOA represents a more flexible ansatz which allows us to take into account
(i) experimental constraints, such as drift terms that cannot be switched off, and
(ii) control degrees of freedom not present in CD driving; 
(iii) CD-QAOA is not tied to a drive protocol which obeys specific boundary conditions (such as vanishing protocol speed).
Unlike continuous CD driving, CD-QAOA offers a simple and easy-to-apply variational ansatz without reference to the exact ground state of the system, paving the way for versatile digital quantum control. 

In particular, our RL agent constructs unitary protocols that transfer the population into the ground state of three nonintegrable spin models (spin-$1/2$ and spin-$1$ mixed-field Ising chains, and the anisotropic spin-$1$ Heisenberg chain) which feature long-range and topological order, 
and the integrable Lipkin-Meshkov-Glick (LMG) model which allows us to present simulations for a large number of particles. %
We show numerical evidence for the existence of a finite quantum speed limit in the nonintegrable mixed-field spin-$1/2$ Ising model: an almost perfect system-size scaling indicates that this behavior persists in the thermodynamic limit. 
Our RL agent has no access to unmeasurable quantum states which grow exponentially with the number of degrees of freedom in the system: this allows the protocols we find to generalize across a number of system sizes [for the spin-$1/2$ mixed-field Ising model], opening up the door to apply ideas of transfer learning to quantum many-body control. Finally, we demonstrate that the CD-QAOA ansatz has direct practical implications in digital quantum control: it leads to much shorter circuit depths while simultaneously improves the fidelity of the prepared state, which can be utilized to reduce  detrimental errors in modern quantum computers.

\section{\label{sec:qaoa}Generalized Continuous-Discrete Quantum Approximate Optimization Ansatz}

To prepare many-body quantum states, we seek a unitary process $U$ which brings the system from a given initial state $|\psi_i\rangle$ to 
the ground state $|\psi_\mathrm{GS}\rangle$ of the Hamiltonian $H$ (which we call the target state $|\psi_\ast\rangle$). Typically, Hamiltonians can be decomposed as a sum of two non-commuting parts $H\!=\!H_1\!+\!H_2$, e.g.~the kinetic and interaction energy.
We want to construct

\begin{equation}
    \label{eq:U_ansatz}
    U(\{\alpha_j\}_{j=1}^q ,\tau)\!=\!\prod_{j=1}^q U(\alpha_j, \tau_j)
\end{equation}
from a sequence $\tau$ of $q$ consecutive unitaries (or their generators) $\tau_j$ chosen from a set $\mathcal{A}$, with $\tau_j\!\neq\!\tau_{j+1}$. 
Each $U(\alpha_j, \tau_j)$ is parametrized by a continuous degree of freedom $\alpha_j$ (e.g.~time or rotation angle), i.e.~$U(\alpha_j, \tau_j)\!=\!\exp\qty(-i\alpha_j\tau_j)$.
We formulate state preparation as an optimization problem which consists of determining (i) the sequence $\tau$, and (ii) the values of the variational parameters $\alpha_j$, such that $U|\psi_i\rangle\!\approx\!|\psi_\mathrm{GS}\rangle$.

Our goal is to prepare the ground state of a Hamiltonian $H$, without having access to the ground state itself. Therefore, we use energy as a cost function
\begin{equation}
    \label{eq:energy}
    E(\{\alpha_j\}_{j=1}^q, \tau ) \!=\! \langle\psi_i|U^\dagger(\{\alpha_j\}_{j=1}^q ,\tau) H U(\{\alpha_j\}_{j=1}^q ,\tau) |\psi_i\rangle,
\end{equation}
or energy-density $E/N$ which has a well-behaved limit when increasing the number of particles $N$~\footnote{We focus on pure states, although the cost function can trivially be generalized to mixed states.}. We denote the ground state energy by $E_\mathrm{GS} = \langle\psi_\mathrm{GS}|H|\psi_\mathrm{GS}\rangle$. 

Note that conventional QAOA is recovered as a special case where one only considers two unitaries  $U_{j}\!=\!U(\alpha_j, H_{j})\!=\!\exp(-i\alpha_jH_j)$, $j\!=\! 1,2$, and $\tau$ is one of the two alternating sequences.
Whenever nested commutators of $H_j$ span the entire Lie algebra which generates transport on the complex projective space associated with the Hilbert space $\mathcal{H}$ of the system, applying QAOA is already enough to prepare any state, provided that the underlying circuit depth $q$ is sufficiently large, and the optimal $\alpha_j$ can be found~\cite{jurdjevic_72}. While true in theory, this is often impractical, since 
(i) it requires access to in principle unbounded durations, 
(ii) it increases the number of optimization parameters $\alpha_j$, and -- with it -- the probability to get stuck in a local minimum of the control landscape, and 
(iii) the condition that nested commutators of $H_j$ span the entire Lie algebra is generally not satisfied for the $H_j$'s of interest in quantum many-body physics due to, e.g., symmetry constraints.

The generalized QAOA ansatz [Eq.~\eqref{eq:U_ansatz}] allows us to utilize a larger set of unitaries $\mathcal{A}$ to construct the optimal sequence and to reduce the circuit depth $q$. Inspired by counter-diabatic (CD) driving, we find that a particularly suitable choice in the context of quantum many-body state manipulation, is given by the operators in the adiabatic gauge potential series [Sec.~\ref{sec:gauge_potentials}]. Therefore, we call the resulting algorithm CD-QAOA.
A different ansatz using more than two unitaries was considered in Ref.~\cite{zhu2020adaptive}.

Compared to conventional QAOA, CD-QAOA introduces a discrete high-level optimization to find the optimal protocol sequence $\tau$. The combined optimization landscape can be particularly difficult to navigate,
due to the existence of so-called barren plateaus where exponentially many directions have vanishing gradients~\cite{ mcclean2018barren, cerezo2020cost, grant2019initialization, pat2020loss}. Additionally, the total number of all allowed protocol sequences, $|\mathcal{A}|(|\mathcal{A}|-1)^{q-1}$~\footnote{Considering $\tau_j$ as choice of unitaries, we impose the extra constraint that, even though unitaries can be repeated in the sequence $\tau$, the same unitary cannot appear consecutively (or else one can combine the two corresponding choices $\tau_j$ into a single variable).}, scales exponentially with the number of unitaries $q$, and presents a challenging discrete combinatorial optimization problem per se; indeed, state preparation, formulated as optimization, can feature a glassy landscape~\cite{day2019glassy,bukov2018broken} [App.~\ref{app:ctrl_landscape}].
However, overcoming these potential difficulties is associated with a potential gain: CD-QAOA allows retaining the flexibility offered by continuous optimization, while increasing the number of independent discrete control degrees of freedom to $|\mathcal{A}|$; this enables us to reach larger parts of the Hilbert space in shorter durations, and with a smaller circuit depth, as compared to conventional QAOA.

Thus, we formulate ground state preparation as a two-level optimization scheme~\footnote{A similar procedure appeared recently in Ref.~\cite{li2020quantum}, although they considered a different problem setup with greedy or beam search algorithm.}. 
(1) Low-level optimization: given a fixed sequence $\tau$, we find the optimal values of $\alpha_j$ using a continuous optimization solver, e.g.~SLSQP~\footnote{In principle, one can use any optimizer which allows constraining the sum $\sum_j\alpha_j\!=\!T$.} [App.~\ref{app:slsqp}]. To cope with the associated rugged optimization landscape [App.~\ref{app:ctrl_landscape}], we run multiple realizations of random initial conditions and post-select the values which yield minimum energy. This continuous optimization problem is also present in conventional QAOA.
(2) High-level optimization: in addition to the low-level optimization, we also perform a discrete optimization for the sequence $\tau$ itself, to determine the optimal order in which unitaries from the set $\mathcal{A}$ should occur. To tackle this combinatorial problem, we formulate the high-level optimization as a reinforcement learning  (RL) problem. We learn the optimal protocol using Proximal Policy Optimization, a variant of policy gradient. The policy is parameterized by a deep autoregressive network, which allows choosing the control unitaries $U(\alpha_j, {\tau_j})$ sequentially. In practice, we sample a batch of sequences from the policy, evaluate the energy of each sequence in the low-level optimization, and apply policy gradient to update the parameters of the policy. This two-level optimization procedure is repeated in a number of training episodes until convergence [App.~\ref{app:algo}].

\section{\label{sec:gauge_potentials}Variational State Preparation inspired by Counter-Diabatic Driving}

A natural question arises as to how to choose the set $\mathcal{A}$ of unitaries for the generalized discrete-continuous QAOA ansatz. One possibility is to consider a set of universal elementary quantum gates, e.g., in the context of a quantum computer~\cite{lacroix2020improving, ding2020breaking}, and in this case $\alpha_j$ are angles of rotation.  We leave this exciting possibility for a future study, and focus here on many-body ground state preparation instead.

The complexity of many-body systems motivates the use of a physics-informed approach to defining the control unitaries in $\mathcal{A}$. Suppose we initialize the system in the ground state of the parent Hamiltonian $H(\lambda\!=\!0)$; we target the ground state of $H(\lambda\!=\!1)$, seeking the functional form of a time-dependent protocol $\lambda(t)$.
If the instantaneous ground state of $H(\lambda)$ remains gapped during the evolution, the adiabatic theorem guarantees the existence of a solution $\lambda(t)$, $t\in[0,T]$, provided $T$ is large compared to the smallest inverse gap along the adiabatic trajectory.
However, when the gap is known to close (e.g.~across a phase transition), or when the state population transfer has to be done fast, adiabatic state preparation fails.

Compared to the adiabatic paradigm, gauge potentials provide additional control directions in Hilbert space which enable paths that non-adiabatically lead to the target state in a short time.
In many-body systems, it is not known in general how to determine the exact gauge potential required for CD driving. However, it is possible to define variational approximations~\cite{sels2017minimizing,hartmann2019rapid} using an operator-valued series expansion [App.~\ref{app:varl_gauge}] similar to a Schrieffer-Wolff transformation~\cite{wurtz2020variational}, or Shortcuts to Adiabaticity methods~\cite{hegade2020shortcuts,ding2020breaking}.
Nonetheless, recent numerical simulations suggest that the exact gauge potential in generic many-body systems is a non-local operator~\cite{sels2017minimizing,pandey2020adiabatic} which renders the series expansion asymptotic.

For these reasons, here we consider the constituent terms to every order of the variational gauge potential series, $H_j$, independently, and use them to generate the set of unitaries $\mathcal{A}\!=\!\{\mathrm{e}^{-i\alpha_j H_j} \}$ for CD-QAOA \footnote{Below, we sometimes abuse notation and set $\mathcal{A}\!=\!\{ H_j \}$, denoting the set of unitaries by their generators.}.
We emphasize that our CD-QAOA ansatz is not designed to approximate the gauge potential itself, as opposed to Ref.~\cite{wurtz2021counterdiabaticity}, yet it yields similar benefits w.r.t.~preparing the target state. In Sec.~\ref{sec:comparison} we compare directly our approach with the variational gauge potential ansatz from Ref.~\cite{sels2017minimizing}.

Since CD-QAOA is a generalization of QAOA aimed to be useful in practice, we need to ensure the accessibility of the control terms $H_j$. Because they appear in the first few orders of the gauge potential series, $H_j$ are (sums of) \emph{local} many-body operators [cf.~App.~\ref{app:varl_gauge}]. Thus, in principle, there is no physical obstruction to emulate them in the lab, although this depends on the details of the experimental platform (especially for the interaction terms).
Additionally, in the context of many-body systems where energy is extensive, in order to guarantee that we do not tap into a source of infinite energy, we constrain the norm of the generators $\alpha_jH_j$: we view $\alpha_j\!\geq\!0$ as time durations, and fix $\sum_{j=1}^q\alpha_j\!=\!T$, with $T$ the total protocol duration. This keeps $\alpha_j$ on the same order of magnitude as the coupling constants in the parent Hamiltonian whose ground state we want to prepare.

\begin{table}[t!]
    \centering
    \begin{NiceTabular}{c|c}
        \hline
        {\bf short-hand notation}                       & { \bf spin operator $H_j$ }                                           \\
        \hline\hline
        $X$                                             & $\sum_i S^x_{i}$                                                      \\
        $Z$                                             & $\sum_i S^z_{i}$                                                      \\
        $Z|Z$                                           & $\sum_i S^z_{i}S^z_{i+1}$                                             \\ 
        
        $Z|Z \!+\! Z$                                       & $\sum_i J S^z_{i}S^z_{i+1} \!+\! h_z S^z_{i}$                             \\ \hdottedline
        $Y$                                             & $\sum_i S^y_{i}$                                                      \\

        $XY$                                            & $\sum_i S^x_{i}S^y_{i} \!+\! S^y_{i}S^x_{i}$                                               \\
        $YZ$                                            & $\sum_i S^y_{i}S^z_{i} \!+\! S^z_{i}S^y_{i}$                                               \\
        $X|Y$                                           & $\sum_i S^x_{i}S^y_{i+1} \!+\! S^y_{i}S^x_{i+1}$                          \\
        $Y|Z$                                           & $\sum_i S^y_{i}S^z_{i+1} \!+\! S^z_{i}S^y_{i+1}$                          \\
        $X|Y \!-\! XY$                                      & $\sum_i [S^x_{i+1} \!-\!a S^x_{i}]S^y_i \!+\! [S^y_{i+1}\!-\!aS^y_i]S^x_i$   \\
        $ Y|Z \!-\! YZ$                                     & $\sum_i  [S^z_{i+1}\!-\!bS^z_{i}] S^y_i \!+\! [S^y_{i+1}\!-\!bS^y_{i}] S^z_i$ \\
        $\hat{XY}$                                        & $\frac 1 N \sum_{i,j}S^x_{i}S^y_{j} \!+\! S^y_{i}S^x_{j}$\\

        $\hat{ZY}$                                        & $\frac 1 N \sum_{i,j} \left(S^z_{i}\!+\!\frac{I}{2}\right)S^y_{j} \!+\! S^y_{i}\left(S^z_{j}\!+\!\frac{I}{2}\right)$\\
        \hline
    \end{NiceTabular}
    \caption{\label{table:gauge_pot}
        Short-hand notation for the generators $H_j$ used to construct the set of unitaries $\mathcal{A}\!=\!\{\mathrm{e}^{-i\alpha_j H_j} \}_{j=1}^{|\mathcal{A}|}$ in CD-QAOA. 
        The $|$ indicates operators acting on neighboring sites. 
        Terms from the variational gauge potential series are shown in the lower group [cf.~App.~\ref{app:varl_gauge} for the derivation].
    }
\end{table}

\section{\label{sec:models}Many-Body Ground State Preparation}
\vskip -1em

We consider four non-integrable many-body systems of increasing complexity: the spin-$1/2$ and spin-$1$ mixed-field Ising models, the spin-$1$ Heisenberg model, 
and the integrable Lipkin-Meshkov-Glick (LMG) model where a large number of degrees of freedom is accessible in a classical simulation.
The goal of the RL agent is to prepare their ordered ground states, starting from a product state. To generate training data, we compute numerically the exact time evolution of the system.
We apply CD-QAOA using a set of unitaries built from the terms in the series expansion for the variational gauge potential.
To determine the allowed terms in the gauge potential series, cf.~Table~\ref{table:gauge_pot} (lower group), we consider the minimal set of symmetries shared by the Hamiltonian and the initial and target states [App.~\ref{app:varl_gauge}].

\subsection{\label{subsec:Ising-1/2}Mixed-Field Spin-1/2 Ising Chain}

\begin{figure}[t!]
    \includegraphics[width=1.0\columnwidth]{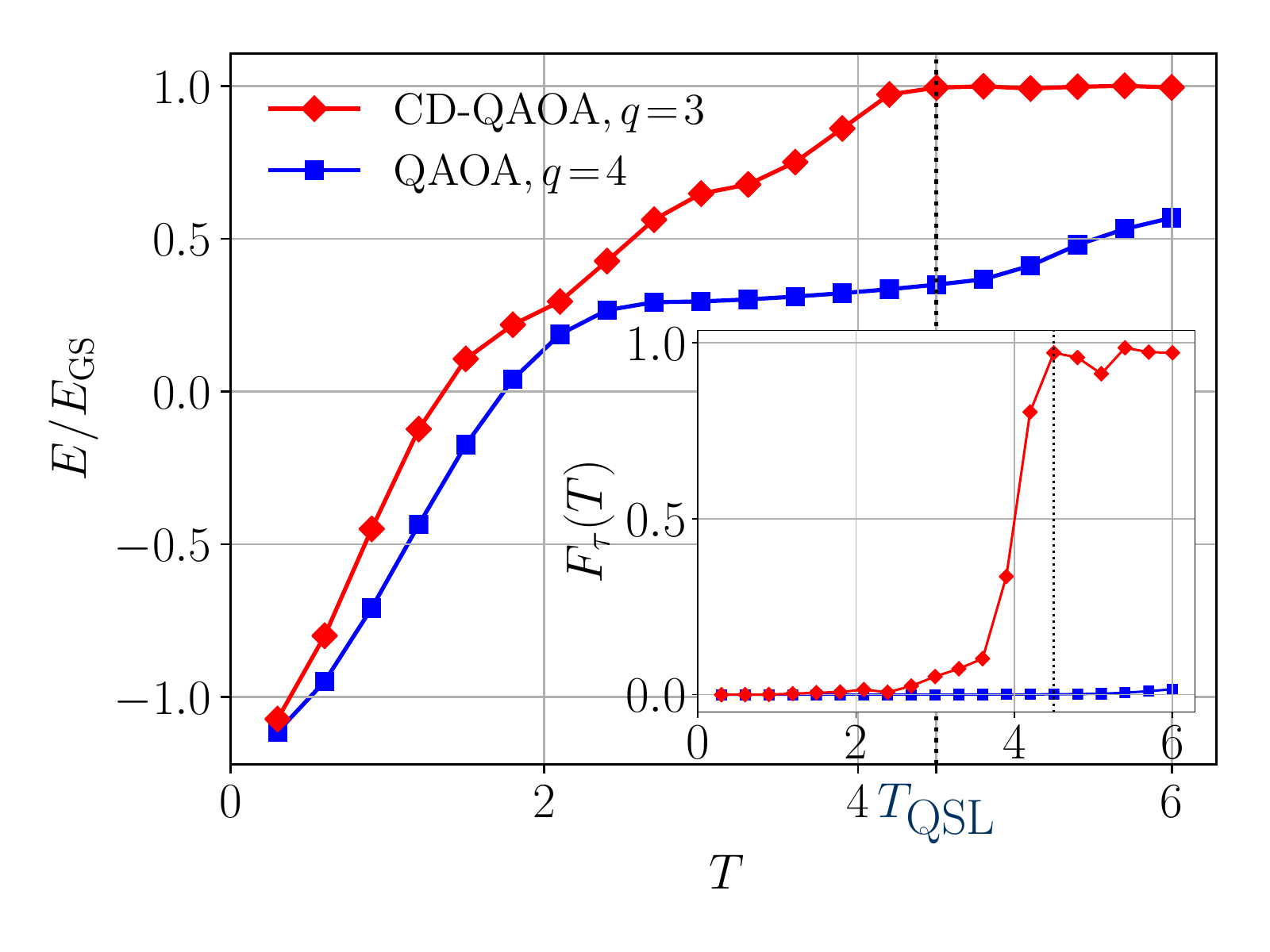}
    \caption{\label{fig:Ising_energy} Spin-$1/2$ Ising model: energy minimization and the corresponding many-body fidelity [inset] against protocol duration $T$ obtained using conventional QAOA (blue squares) and CD-QAOA (red diamonds) with circuit depths $p\!=\!q/2\!=\!2$ and $q\!=\!3$, respectively. The dotted vertical line marks the quantum speed limit $T_\mathrm{QSL}$.
    CD-QAOA outperforms conventional QAOA.
    The initial and target states are $|\psi_i\rangle\!=\!|\!\!\uparrow\cdots\uparrow\rangle$ and $|\psi_\ast\rangle\!=\!|\psi_\mathrm{GS}(H)\rangle$ for $h_z/J=0.809$ and $h_x/J=0.9045$. The alternating unitaries for conventional QAOA are generated by $\mathcal{A}_\mathrm{QAOA}=\{ Z|Z\!+\!Z, X \}$[cf.~Eq.~\eqref{eq:IM}]; for CD-QAOA, we extend this set using adiabatic gauge potential terms to $\mathcal{A}_\mathrm{CD-QAOA}=\{Z|Z\!+\!Z,X;Y,X|Y,Y|Z\}$.
    The cardinality of the CD-QAOA sequence space is $\vert \mathcal A \vert (\vert \mathcal A \vert \!-\! 1)^{q\!-\!1} = 80$. The number of spins is $N\!=\!18$ with a Hilbert space size of $\mathrm{dim}(\mathcal{H})\!=\! 7685$.
    }
\end{figure}

Consider first the antiferromagnetic mixed-field spin-$1/2$ Ising chain of $N$ lattice sites

\begin{eqnarray}
    \label{eq:IM}
    H &\!=\!&H_1\!+\!H_2,\\
    H_1\!&\!=\!&\! \sum_{j=1}^N J S^z_{j+1}S^z_j \!+\! h_z S^z_j,\quad
    H_2= \sum_{j=1}^N h_xS^x_j, \nonumber
\end{eqnarray}
where $[S^\alpha_i,S^\beta_j]=\delta_{ij}\varepsilon^{\alpha\beta\gamma}S^\gamma_j$ are the spin-$1/2$ operators. We use periodic boundary conditions and work in the zero momentum sector of positive parity.  In the following, $J\!=\!1$ sets the energy unit, and $h_z/J\!=\!0.809$ and $h_x/J\!=\!0.9045$.  We initialize the system in the $z$-polarized product state $|\psi_i\rangle\!=\!|\!\!\uparrow\cdots\uparrow\rangle$, and we want to prepare the ground state of $H$ in a short time $T$, i.e., away from the adiabatic regime. We verified that similar results can be obtained starting from $|\!\downarrow\cdots\downarrow\rangle$.

To acquire an intuitive understanding of the advantages brought by the gauge potential ansatz, consider first the non-interacting system at $J=0$, for which the control problem reduces to a single spin. Both the initial and target states lie in the $xz$-plane of the  Bloch sphere, and hence the shortest unit-fidelity protocol generates a rotation about the $y$-axis. In conventional QAOA, one would construct a $y$-rotation out of the $X$ and $Z$ terms [cf.~Table~\ref{table:gauge_pot}] present in the Hamiltonian. For a single spin, this construction is always possible due to the Euler angle representation of $\mathrm{SU}(2)$, but for the interacting spin chain this is no longer the case. The role of the gauge potential $Y$ is to `unlock' precisely this geodesic in parameter space, and make it accessible as a dynamical process. This allows preparing the target state faster, compared to the original $X,Z$ control setup. In the language of variational optimization, an accessible $Y$ term includes the shortest-distance protocol into the variational manifold, and the RL agent easily finds the exact solution [App.~\ref{app:Ising}].

For the interacting system, $J\!>\!0$, applying conventional QAOA using the two gates $U_j\!=\!\mathrm{e}^{-i\alpha_j H_j}$ with $H_1\!=\!Z|Z\!+\!Z$ and $H_2\!=\!X$ is straightforward, but it does not yield a high-fidelity protocol [Fig.~\ref{fig:Ising_energy} (blue squares)]. It was recently reported that much better energies can be obtained, using a three-step QAOA which consists of the three terms in the Hamiltonian~\eqref{eq:IM}, $Z|Z$, $X$, and $Z$, applied in a fixed order~\cite{matos2020quantifying}; invoking again an Euler angle argument provides an explanation: the $X$ and $Z$ terms effectively generate the $Y$ gauge potential term.

In stark contrast to conventional QAOA, adding just the zero-order term $H_3\!=\!Y$ from the gauge potential series [App.~\ref{app:gauge_pot_algo}], we find that CD-QAOA already gives a significantly improved protocol; this is achieved by the high-level discrete optimization which selects the order of the operators in the sequence. However, we can do better: since $|\psi_i\rangle$ is a product state while $|\psi_\ast\rangle$ is not, and because $H_3$ is a sum of single-particle terms, in order to create the target many-body correlations using a fast dynamical process, we also include the two-body first-order gauge potential terms $H_4\!=\!X|Y\!$ and $H_5\!=\!Y|Z$: this results in a nonadiabatic evolution that prepares the interacting ground state to an excellent precision [Fig.~\ref{fig:Ising_energy} (red diamonds)].

\begin{figure}[t!]
    \includegraphics[width=1.0\columnwidth]{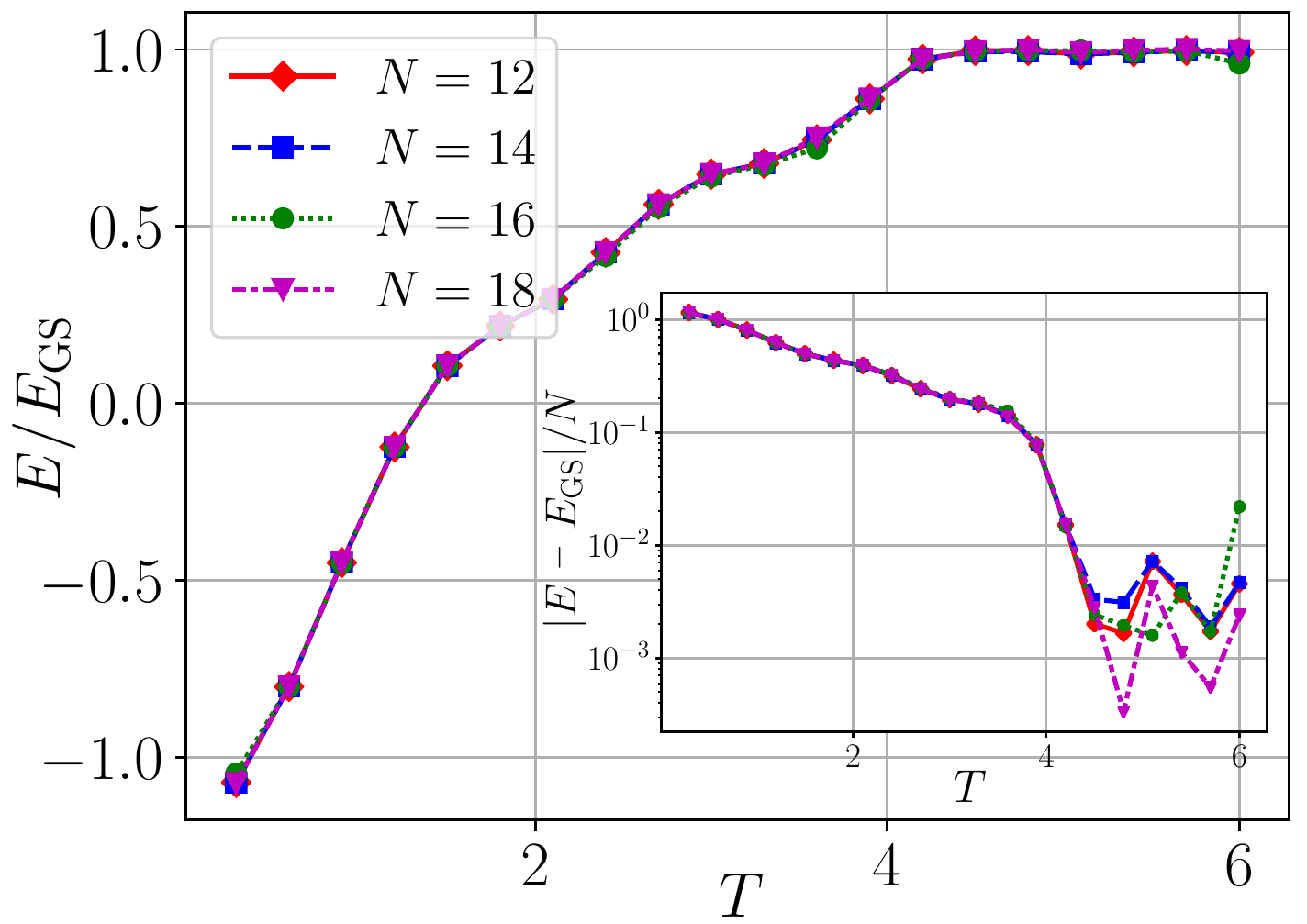}
    \caption{\label{fig:Ising_energy_scaling}Spin-$1/2$ Ising model: energy minimization and the corresponding mean absolute error [inset, log scale] against protocol duration $T$ for different system sizes using CD-QAOA with circuit depths $q\!=\!3$. system-size scaling of the variational energy density suggests the results hold for larger systems. 
    For the number of spins of $N\!=\!12, 14, 16, 18$, the Hilbert space sizes are $\mathrm{dim}(\mathcal{H})\!=\!224, 687, 2250, 7685$ respectively.
    The model parameters are the same as in Fig.~\ref{fig:Ising_energy}.
    }
\end{figure}

In Ref.~\cite{ho2019efficient}, it was shown that, in the integrable limit $h_z\!=\!0$, one can prepare the ground state of the system at the critical point using a circuit of depth $q\!=\!2N$ with conventional QAOA. Albeit for the specific initial and target states chosen, we find that it only takes CD-QAOA a depth of $q\!=\!3$ to reach the target ground state, independent of the system size $N$~\footnote{The role of the RL algorithm is to decide which three out of the five unitaries $U_j$ to apply and in which order.}. This result, though model-dependent, may come as a surprise at first sight, given that the mixed field Ising chain is a quantum chaotic system without a closed-form solution which makes it susceptible to heating away from the adiabatic limit.

Our data also reveals a finite many-body QSL at $T_\mathrm{QSL}\!\approx\! 4.5$. Importantly, this QSL appears insensitive to the system size to a very good approximation [Fig.~\ref{fig:Ising_energy_scaling}], and we expect it to persist in the thermodynamic limit. The absence of a finite QSL in conventional QAOA in the mixed-field Ising chain suggests that the observation of a QSL using CD-QAOA depends on the specific set of unitaries related to the variational gauge potential, showcasing the utility of our ansatz for many-body control.
Remarkably, we find an almost perfect system-size collapse of the target state energy density curves as a function of the total protocol duration $T$. In Sec.~\ref{sec:generalization}, we explore this feature and demonstrate the ability of the RL agent to learn on small system sizes and subsequently generalize its knowledge to control bigger systems with exponentially larger Hilbert spaces.

CD-QAOA performs successfully on the nonintegrable spin-$1/2$ mixed-field Ising chain, for a circuit depth as short as $q=3$. This shows an advantage of our ansatz, when compared to conventional QAOA. However, the small size of the sequence space, $|\mathcal{A}|(|\mathcal{A}|-1)^{q-1}\!=\!80$ at $|\mathcal{A}|=5$, poses a natural question regarding the necessity of using sophisticated search algorithms, such as RL, to find the control sequence. We now show that this is a peculiarity of the physical system, as we turn our attention to a larger sequence space.

\subsection{\label{subsec:Heisenberg}Heisenberg Spin-\texorpdfstring{$1$}{1}  Chain}

The eight-dimensional spin-$1$ group $\mathrm{SU}(3)$ provides a significantly larger space of gauge potential terms to build the optimal protocol from.
We consider a total of $|\mathcal{A}|\!=\!9$ unitaries: five are generated by the imaginary-valued terms in the gauge potential series: $Y, XY, YZ, X|Y, Y|Z$ [cf.~Table.~\ref{table:gauge_pot}], plus the two real-valued QAOA operators $H_1$ and $H_2$, which build the Hamiltonian $H\!=\!H_1\!+\!H_2$ whose ground state we target [Eq.~\eqref{eq:HM}], and the two real-valued Hamiltonian terms $X|X$ and $Z$. At $q\!=\!18$, this amounts to $|\mathcal{A}|(|\mathcal{A}|-1)^{q-1}\approx 10^{16}$ possible sequences.
The exponential scaling of the sequence space size with $q$ renders applying exhaustive search algorithms infeasible, and justifies the use of sophisticated algorithms, such as RL.

The (anisotropic) spin-$1$ Heisenberg model reads as:
\begin{eqnarray}
    \label{eq:HM}
    H &\!=\!&H_1\!+\!H_2,\\
    H_1\!&\!=\!&\! J\sum_{j=1}^N (S_{j+1}^xS_j^x \!+\! S_{j+1}^yS_j^y),\quad
    H_2= \Delta \sum_{j=1}^N S_{j+1}^zS_j^z, \nonumber
\end{eqnarray}
with the spin exchange coupling $J\!=\!1$ set as energy unit, and $\Delta$ -- the anisotropy parameter; we use periodic boundary conditions and work in the ground state sector of zero momentum and positive parity, defined by the projector $\mathcal{P}$. In the thermodynamic limit, this model features a rich ground state phase diagram including ferromagnetic (FM, $\Delta/J\!\ll\!-1$), XY ($-1\!\lesssim\!\Delta/J\!\lesssim\!0$), topological/Haldane ($0\!\lesssim\!\Delta/J\!\lesssim\!1$), and antiferromagnetic (AFM, $\Delta/J\!\gg\! 1$) order~\footnote{We define `order' in the context of phase transitions in condensed matter physics.}, with phase transitions belonging to different universality classes~\cite{chen2003ground,pollmann2010entanglement,langari2013ground}.
While the FM, XY, and AFM states are characterized by a local order parameter, the gapped Haldane state has topological order not captured by Landau-Ginzburg theory.
We consider the AFM initial state $|\psi_i\rangle\!=\!\mathcal{P}\;|\!\uparrow\downarrow\uparrow\downarrow\cdots\rangle$, and target the ground states of Eq.~\eqref{eq:HM} deep in the FM, XY, and Haldane phases, where system-size effects are the smallest.
Because CD-QAOA is not restricted to adiabatic evolution, the conventional paradigm of a closing spectral gap when transferring the population between two states displaying different order, does not apply in our non-equilibrium setup, even in the thermodynamic limit.

\begin{figure}[t!]
    \includegraphics[width=1.0\columnwidth]{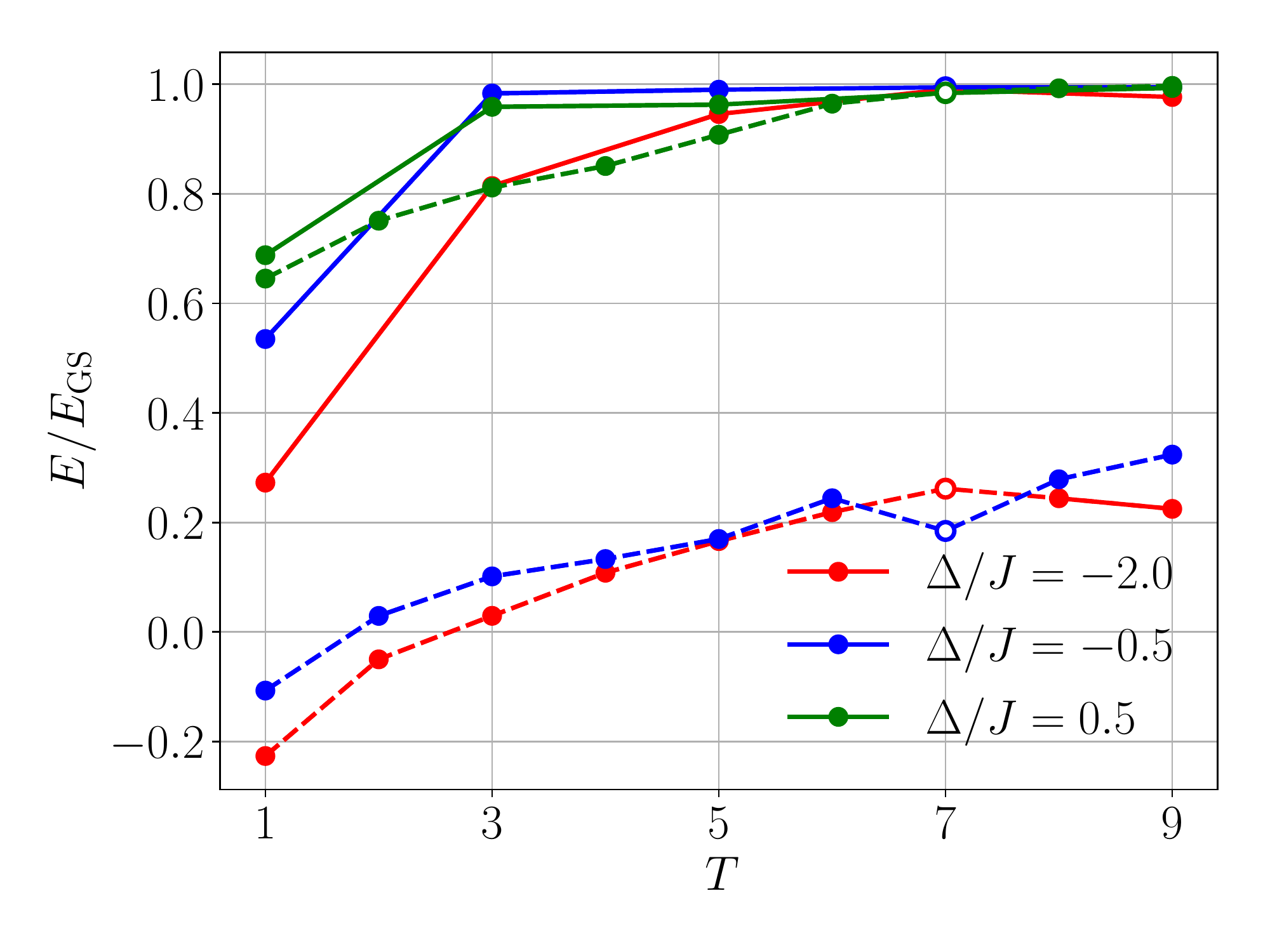}
    \caption{Heisenberg spin-$1$ chain:
    energy minimization against protocol duration $T$ using conventional QAOA (dashed lines) and CD-QAOA (solid lines) for three different states. We start from the AFM state $|\psi_i\rangle\!=\!\mathcal{P}\,|\!\!\!\uparrow\downarrow\uparrow\downarrow\cdots\rangle$ and target three different parameter regimes, corresponding to the FM ($\Delta/J\!=\!-2.0$) state, XY ($\Delta/J\!=\!-0.5$), and Haldane ($\Delta/J\!=\!0.5$) states, respectively.
    CD-QAOA outperforms conventional QAOA ($p\!=\!q/2$), more notably in the FM and XY targets where it allows us to reach close to the target state using a short protocol duration.
    The empty symbols mark the duration at which we show the evolution of the system in Fig.~\ref{fig:comparison-heisenberg-stat}.
    The alternating unitaries for conventional QAOA are generated by $\mathcal{A}_\mathrm{QAOA}=\{ H_1, H_2 \}$ [cf.~Eq.~\eqref{eq:HM}]; for CD-QAOA, we extend this set using adiabatic gauge potential terms to $\mathcal{A}_\mathrm{CD-QAOA}=\{ H_1, H_2, Z, X|X; Y, XY, YZ, X|Y \!-\! XY, Y|Z \!-\! YZ\}$.
    The circuit depths are $q\!=\!28$ ($\Delta/J\!=\!-2.0$), $q\!=\!18$ ($\Delta/J\!=\!-0.5$) and $q\!=\!18$ ($\Delta/J\!=\!0.5$).
    The cardinality of the CD-QAOA sequence space is $\vert \mathcal A \vert (\vert \mathcal A \vert \!-\! 1)^{q\!-\!1} \approx 10^{16}$ at $q=18$.
    The system size is $N=8$, where $\mathrm{dim}(\mathcal{H})=498$.
    }\label{fig:comparison-eng-heisenberg}
\end{figure}

Figure~\ref{fig:comparison-eng-heisenberg} shows a comparison between conventional QAOA with alternating sequence between the Hamiltonians $H_1$ and $H_2$, and CD-QAOA. We find that CD-QAOA shows superior performance for all three ordered ground states: while the gain over conventional QAOA for the Haldane state is already a faster protocol, we clearly see how the gauge potential terms can prove essential for reaching the ground state in the FM and XY phases within the available durations. Note that the FM target state is doubly degenerate, and minimizing the energy, it ends up in an arbitrary superposition within the ground state manifold.
Interestingly, we do not identify any distinction from preparing states with long-range and topological order, presumably due to the small system sizes that we reach in our classical simulation.

The CD-QAOA protocol sequences found by the RL agent have peculiar structures [App.~\ref{app:Heisenberg}]: some of them resemble closely the alternating sequence of conventional QAOA, with the notable difference of applying additional unitaries to rotate the state to a suitable basis, either at the beginning or at the end of the sequence. While this is formally equivalent to starting from or targeting a rotated state, the rotations use two-body operators; hence, the resulting basis does not coincide with any of the distinguished $S^x$, $S^y$ and $S^z$ directions. Variationally determining such effective bases demonstrates yet another advantage offered by the CD-QAOA ansatz.
Another kind of encountered sequence contains two different sets of alternating unitaries, similar to two independent QAOA ansatzes concatenated one after the other.
Finally, for those values of $T$, where CD-QAOA and QAOA have the same performance, we have also observed that CD-QAOA finds precisely the QAOA sequence. In this case, conventional QAOA already generates the shortest path, and the extra gauge potential terms to second-order do not give any advantage; a better performance might be expected when the three- and four-body higher-order terms from the gauge potential series are included.

Similar to other optimal control algorithms, RL agents typically find local minima of the optimization landscape; thus, there is no guarantee that the CD-QAOA protocols provide global optimal solutions; however, these sequences can serve as an inspiration to build future variational ansatzes tailored for many-body systems.

\subsection{\label{subsec:LMG}Lipkin-Meshkov-Glick Model}

The non-integrable character of the previously discussed models precludes us from applying CD-QAOA with a large number of degrees of freedom, since reliably simulating their dynamics on a classical computer is prohibitively expensive. In order to study the behavior of CD-QAOA in a large enough system which also features a quantum phase transition, we now turn our attention to an exactly solvable many-body system. 

The Lipkin-Meshkov-Glick (LMG) Hamiltonian~\cite{lipkin1965validity} describes spin-1/2 particles on a fully-connected graph of $N$ sites:
\begin{eqnarray}
	\label{eq:LMG_ham}
	H &=& H_1 + hH_2, \nonumber\\
	H_1 &=& -\frac{J}{N}\sum_{i,j=1}^N S^x_i S^x_j ,\quad H_2 = \sum_{j=1}^N \left(S^z_j+\frac{1}{2}\right),
\end{eqnarray}
where $J$ is the uniform interaction strength and $h$ the external magnetic field. In the thermodynamic limit, $N\to\infty$, the system undergoes a quantum phase transition at $h_c/J=1$ between a ferromagnetic (FM) ground state in the $x$-direction for $h/J\ll 1$, and a paramagnetic ground state for $h/J\gg 1$. The spectral gap $\Delta_\mathrm{LMG}$ between the ground state and excited states closes as $\Delta_\mathrm{LMG}(h_c)\sim N^{-1/3}$ at the critical point~\cite{botet1983large}. Realizing the LMG model is within the scope of present-day experiments with ultracold atoms~\cite{strobel2014fisher,davis2020protecting}; therefore, developing fast ground state preparation techniques can prove useful in practice.

Defining the total spin operators as $S^\alpha = \sum_{j=1}^N S^\alpha_j$, the Hamiltonian takes the form $H = -J/N \left(S^x\right)^2 + h\left(S^z + N/2\right)$. Hence, the total spin is conserved, $[H,{\bf S}\cdot{\bf S}]=0$, and the ground state symmetry sector contains a total of $N+1$ states, i.e.~$\mathrm{dim}(\mathcal{H})\!=\!N\!+\!1$, which allows us to simulate large system sizes. 

\begin{figure}[t!]
	\centering
	\includegraphics[width=1.0\columnwidth]{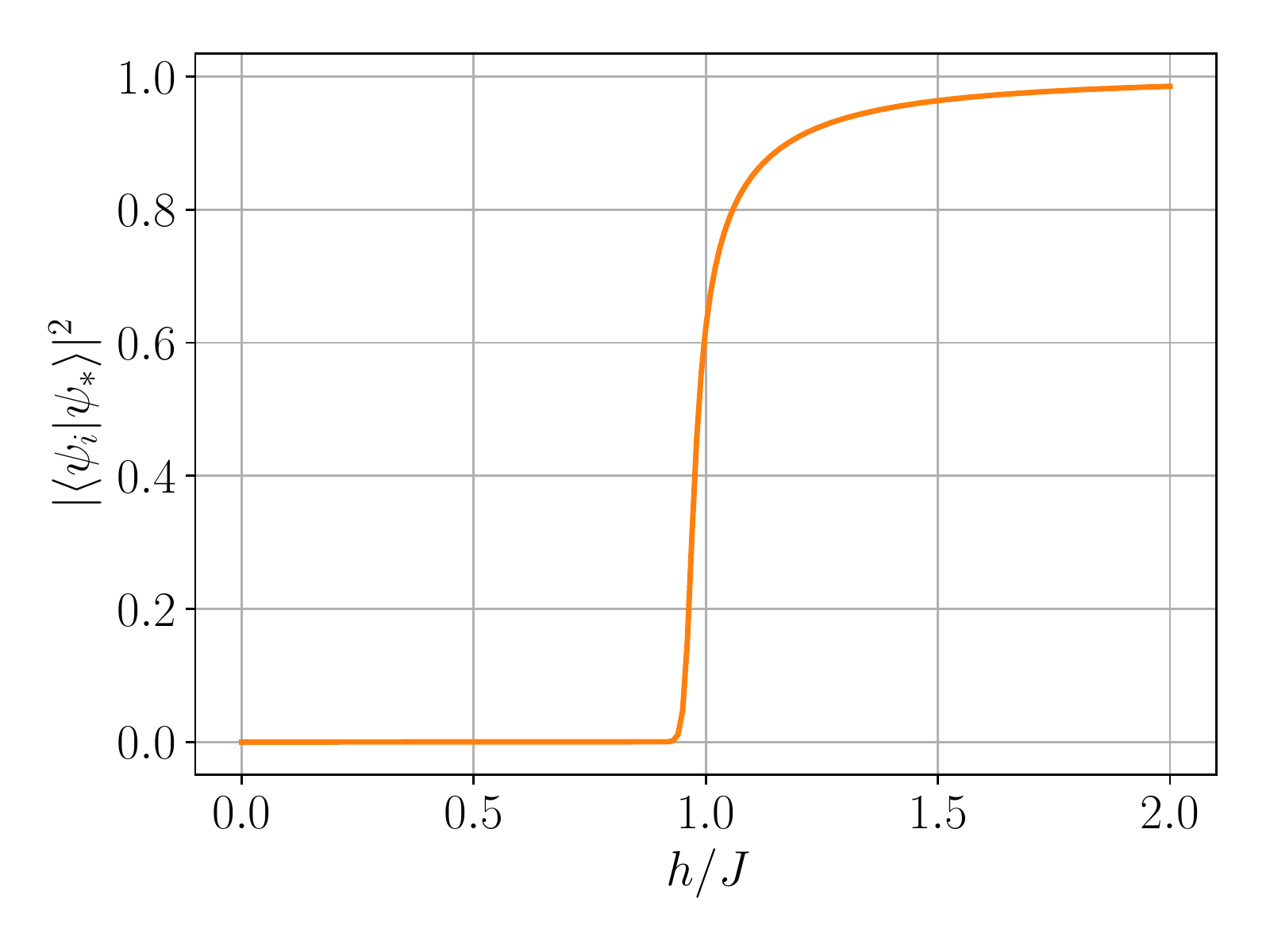}
	\caption{
		LMG model: the overlap between the initial  state $\ket{\psi_i}$ and the target state $\ket{\psi_\ast}$ is vanishingly small in the ferromagnetic phase $h/J\ll 1$, which motivates the parameter choice for the target state. In the vicinity of the critical point, the overlap increases and approaches unity in the limit $h/J\to\infty$. 
		Note that, in the FM phase, the ground state is doubly degenerate, in which case the overlap is computed w.r.t.~the ground state manifold: $|\langle\psi_i |\psi_\ast^{(1)}\rangle|^2 + |\langle\psi_i |\psi_\ast^{(2)}\rangle|^2$. In the paramagnetic phase, the ground state is unique. 
		We used $N=501$ spins.
	}
	\label{fig:LMG_states}
\end{figure}

Our goal is, starting from the $z$-polarized paramagnetic initial state, $|\psi_i\rangle=|\!\!\downarrow\downarrow\cdots\rangle$, to target an arbitrary superposition in the doubly-degenerate FM ground state manifold, at fixed values of the external field $h/J$ which controls the magnitude of the transversal fluctuations on top of the ferromagnetic order. 
Figure~\ref{fig:LMG_states} shows that the overlap of the initial and target states is vanishingly small in the FM phase, and approaches quickly unity across the critical point into the paramagnetic phase. Therefore, we choose to prepare ground states in the FM phase where the problem naturally appears more difficult. 

\begin{figure}[t!]
	\centering
	\includegraphics[width=1.0\columnwidth]{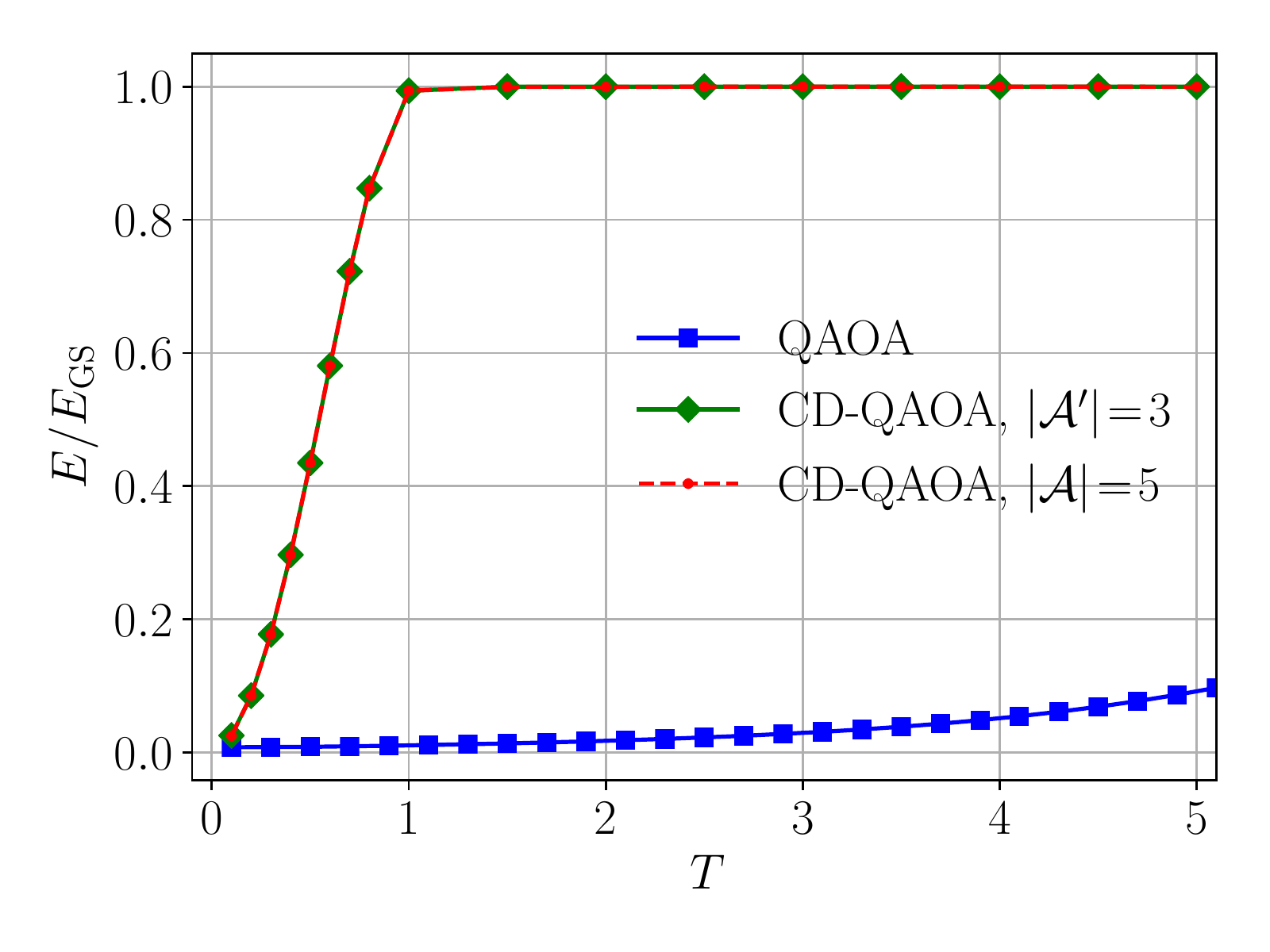}
	\caption{
		LMG model: energy minimization against protocol duration $T$ using conventional QAOA (blue square) and CD-QAOA (red dashed line, green solid line). We start from the $z$-polarized state $|\psi_i\rangle=|\!\!\downarrow\downarrow\cdots\rangle$ and target ground state of LMG Hamiltonian~\eqref{eq:LMG_ham}. CD-QAOA significantly outperforms conventional QAOA for short durations. The alternating unitaries for conventional QAOA are generated by $\mathcal{A}_\mathrm{QAOA}=\{ H_1, H_2 \}$ [cf.~Eq.~\eqref{eq:LMG_ham}]; for CD-QAOA, we extend this set using adiabatic gauge potential terms to $\mathcal{A}_\mathrm{CD-QAOA}=\{ H_1, H_2 ; Y, \hat{XY}, \hat{YZ}\}$ and $\mathcal{A'}_\mathrm{CD-QAOA}=\{ H_1, H_2 ; Y\}$. The external field is  $h/J\eq 0.5$; the circuit depth is $q\!=\!8$, and the system size is $N=501$, where effective Hilbert dimension $\mathrm{dim}(\mathcal{H})\!=\!502$.
	}
	\label{fig:LMG_energy}
\end{figure}

\begin{figure}[t!]
	\centering
	\includegraphics[width=1.0\columnwidth]{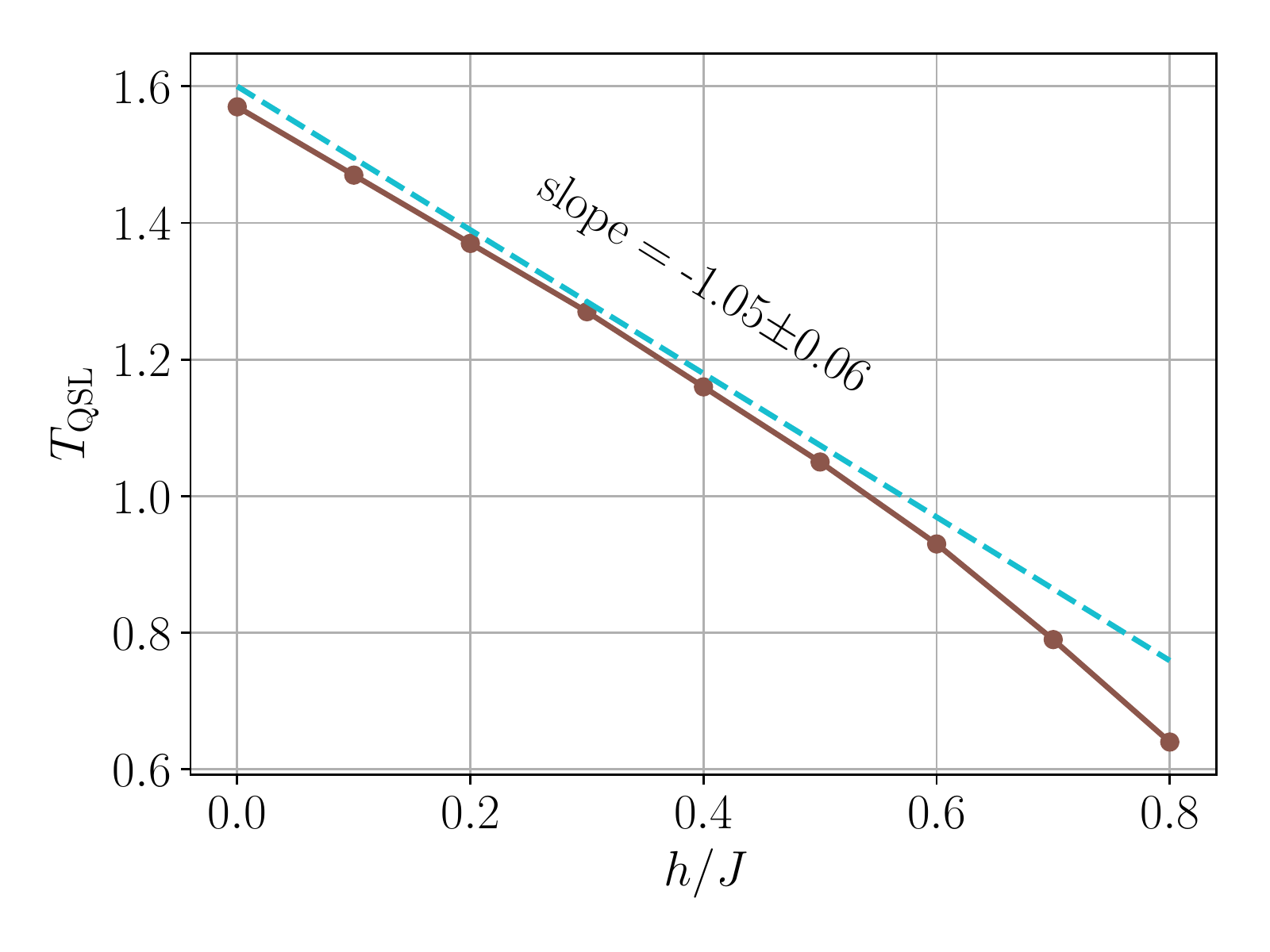}
	\caption{
		LMG model: Quantum speed limit, $T_\mathrm{QSL}$, as a function of the transverse field $h$, for a target state in the ferromagnetic phase. At $h/J=0$, we have $T_\mathrm{QSL}=\pi/2$, which is the angle required to turn the $z$-polarized initial state into the $x$-ferromagnet. For finite $h/J$ quantum fluctuations in the target ferromagnetic ground state decrease the angle required to transfer the population from the initial state, which results in a smaller value of $T_\mathrm{QSL}$.
		The dashed cyan line is a least squares fit for small values of $h/J$, suggesting the behavior $T_\mathrm{QSL}(h) = -h/J + \pi/2 + \mathcal{O}(h^2)$.
		We used $N=501$ spins.  
	}
	\label{fig:LMG_QSL}
\end{figure}

Figure~\ref{fig:LMG_energy} shows a comparison between CD-QAOA and QAOA on the LMG model at $h/J = 0.5$ for $N\eq 501$ spins [more $h/J$ values are shown in App.~\ref{app:LMG_physics}]. First, note the superior performance of CD-QAOA, as compared to conventional QAOA in a range or short durations $T$ in the nonadiabatic driving regime. We applied CD-QAOA with two different sets of generators: $\mathcal{A}=\{H_1,H_2; Y\}$, and $\mathcal{A}'=\{H_1,H_2; Y, \hat{XY}, \hat{ZY}\}$ [cf.~Table~\ref{table:gauge_pot}] and found that, for the LMG model, the higher-order two-body terms $\hat{XY}, \hat{ZY}$ do not offer any advantage deep in the FM phase. 
This observation can be understood as follows: to turn the $z$-polarized initial state into the $x$-ferromagnet, it is sufficient to perform a rotation about the $y$-axis, which coincides precisely with the single-body term in the gauge potential series expansion [cf.~App.~\ref{app:LMG}]. Indeed, for all protocol durations smaller than the quantum speed limit, $T<T_\mathrm{QSL}$, the RL agent finds that the optimal protocol consists of a single ${Y}$-rotation, while for $T\geq T_\mathrm{QSL}$ the optimal protocol is degenerate, and typically involves the various terms from $\mathcal{A}$. This finding allows us to extract the QSL as a function of the external field $h$, cf.~Fig.~\ref{fig:LMG_QSL}.

Close to the critical point $h_c$, we observe strong sensitivity in the best found protocols to system-size effects, and a single ${Y}$-rotation is no longer optimal below the QSL. Interestingly, at the critical point (and in the paramagnetic phase), the optimal protocol is given by QAOA: in this regime, despite the larger set of terms $\mathcal{A}$ we use in CD-QAOA, the RL agent correctly identifies the sequence of alternating $H_1$ and $H_2$ terms as optimal, which shows the versatility of CD-QAOA: the algorithm can always select a smaller effective subspace of actions when this is advantageous in the parameter regime of interest.

\section{\label{sec:comparison}Comparison with Counter-Diabatic Driving}

To compare and contrast the CD-QAOA ansatz with CD and adiabatic driving~\cite{sels2017minimizing}, consider the driven spin-$1$ Ising model~\footnote{We deliberately use a different form in Eq.~\eqref{eq:H_var_vs_qaoa} as compared to Eq.~\eqref{eq:IM}; the former may appear more natural in quantum many-body physics, where the transverse-field Ising model $H_1$ can be mapped to free fermions.}:
\begin{eqnarray}
    \label{eq:H_var_vs_qaoa}
    H(\lambda) \!&=&\! \lambda(t)H_1 \! + \! H_2, \\
    H_1 \!&=&\! \sum_{j=1}^N J S^z_{j+1}S^z_j + h_x S^x_j,\qquad H_2\!=\! \sum_{j=1}^N h_zS^z_j, \nonumber
\end{eqnarray}
where $\lambda(t)\!=\! \sin^2\left(\frac{\pi t}{2T}\right)$, $t\in[0,T]$, is a smooth protocol satisfying the boundary conditions for CD driving: $\lambda(0)\!=\! 0$, $\lambda(T)\!=\! 1$, $\dot\lambda(0)\!=\! 0 \!=\! \dot\lambda(T)$.
The initial state is the ground state at $t\!=\!0$, i.e.~$|\psi_i\rangle\!=\!|\!\!\downarrow\cdots\downarrow\rangle$, while the target state is the ground state of the Ising model at $t=T$ for $h_z/J=0.809$ and $h_x/J=0.9045$.
Unlike the setup in Sec.~\ref{subsec:Ising-1/2}, adiabatic state preparation following the protocol $\lambda(t)$, suggests using the QAOA generators $\mathcal{A}_\mathrm{QAOA}=\{H_1,H_2\}$.

\begin{figure}[t!]
    \includegraphics[width=1.0\columnwidth]{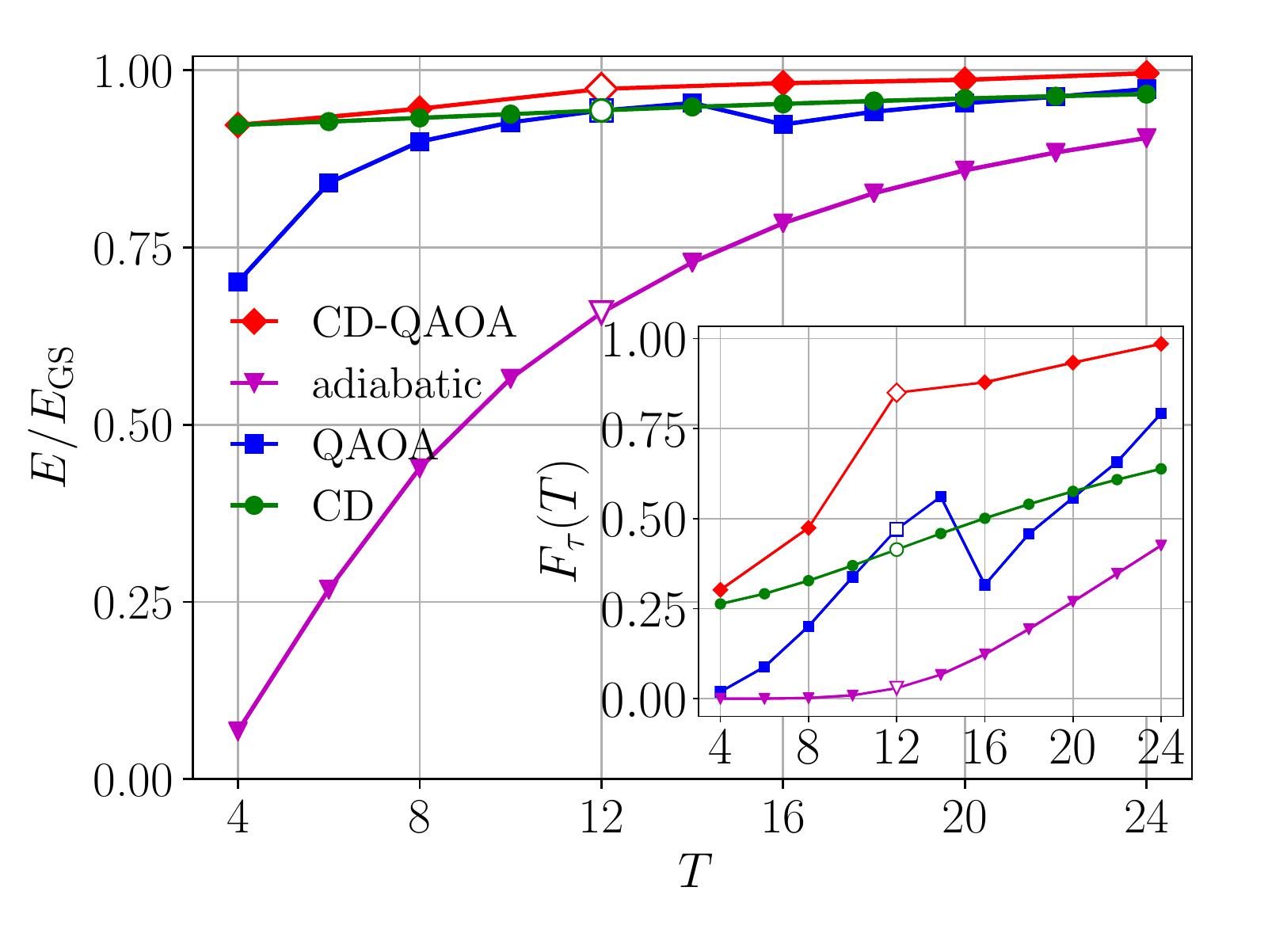}
    \caption{Spin-$1$ Ising model:
        energy minimization and the corresponding many-body fidelity [inset] against different protocol duration $T$ for four different optimization methods: CD-QAOA (red line), conventional QAOA (blue line), variational gauge potential (green) and adiabatic evolution (magenta).
        The empty symbols mark the duration for which the evolution of physical quantities is shown in Fig.~\ref{fig:comparison-ising1-stat}. 
        The initial and target states are $|\psi_i\rangle\!=\!|\!\!\downarrow\cdots\downarrow\rangle$ and $|\psi_\ast\rangle\!=\!|\psi_\mathrm{GS}(H)\rangle$ for $h_z/J=0.809$ and $h_x/J=0.9045$.
        The alternating unitaries for conventional QAOA are generated by $\mathcal{A}_\mathrm{QAOA}=\{ H_1, H_2\}$ [cf.~Eq.~\eqref{eq:H_var_vs_qaoa}];
        for CD-QAOA, we extend this set using adiabatic gauge potential terms to $\mathcal{A}_\mathrm{CD-QAOA}=\{ H_1, H_2; Y, XY, YZ, X|Y , Y|Z \}$.
        The variational gauge potential in CD driving uses all five imaginary-valued gauge potentials $\{Y,XY, YZ,X|Y,Y|Z \}$.
        The CD- and adiabatic driving simulations are both based on the smooth protocol function $\lambda(t)\!=\! \sin^2\left(\frac{\pi t}{2T}\right)$, with a time-discretization step $\Delta t\!=\! 0.2$.
        The value of $q\!=\! 20$ and the size of sequence space is $\vert \mathcal A \vert (\vert \mathcal A \vert  \!-\! 1)^{q\!-\!1} \!\approx\! 10 ^ {15}$.
        The system size is $N\!=\!8$, where $\mathrm{dim}(\mathcal{H})\!=\!498$.
    }
    \label{fig:comparison-eng}
\end{figure}

Figure~\ref{fig:comparison-eng} shows a comparison between different methods using the best found energy density (main figure), and the corresponding many-body fidelity (inset).
Let us focus on CD-QAOA and QAOA first. As expected, CD-QAOA (red) performs better for short durations $T$, since it contains conventional QAOA (red) as an ansatz, i.e.~$\mathcal{A}_\mathrm{QAOA}\!\subsetneq\!\mathcal{A}_{\mathrm{CD\text{-}QAOA}}$.
We emphasize that such a performance is not guaranteed in practice, since it is conceivable that the RL agent gets stuck in a local minimum associated with lower energy than the QAOA solution [App.~\ref{app:ctrl_landscape}], e.g., if the deep autoregressive network architecture is not expressive enough, or if the learning rate schedules are not well-tuned to the problem.
Unlike the spin-$1/2$ Ising model, here we cannot clearly identify a finite QSL, as the CD-QAOA energy keeps improving with increasing circuit depth $q$ [App.~\ref{app:algo}].

To construct the counter-diabatic Hamiltonian $H_\mathrm{CD}\!\approx\! H(\lambda)\!+\!\dot\lambda \mathcal{X}(\{\beta_j\})$ for Eq.~\eqref{eq:H_var_vs_qaoa}, we make a variational ansatz~\cite{sels2017minimizing} for the gauge potential $\mathcal{X}$, and solve for the optimal parameters $\beta_j$ numerically [App.~\ref{app:varl_gauge}].
We note the following differences between this approach and CD-QAOA:
(i) the variational gauge potential depends on time $t$ continuously, which requires further discretization when performing a gate-based implementation.
(ii) the number of variational parameters in the standard variational gauge potential method is $N_T|\mathcal{A}|$ with $N_T$ the number of steps used to discretize the time interval $[0,T]$; instead, in CD-QAOA, we have $q$ variational parameters.
(iii) the variational gauge potential method does not constrain the magnitude of the variational coefficients $\beta_j$, and hence the time-averaged norm of $H_\mathrm{CD}$ over the protocol can grow indefinitely; especially for short durations $T$ this typically gives a higher fidelity. By contrast, in CD-QAOA the time-averaged norm of the unitary generators $\alpha_jH_j$ summed along the sequence, is kept bounded via the constraint $\sum_j\alpha_j\!=\!T$. Nonetheless, in practice, we find that these norms are on the same order of magnitude for all methods considered [App.~\ref{app:spin-1}].

As anticipated, Fig.~\ref{fig:comparison-eng} shows that CD driving performs better than adiabatic driving, and the two agree in the limit of large $T$.
Moreover, we see explicitly that the CD and QAOA solutions are far from the adiabatic regime.
Not surprisingly, CD driving outperforms conventional QAOA for small $T$, as it can increase the values of the variational parameters (and with it the norm) indefinitely.
However, CD-QAOA consistently outperforms CD driving in the entire $T$-range; the contrast is especially pronounced in the many-body fidelity [Fig.~\ref{fig:comparison-eng}, inset].
CD-QAOA makes use of the variational power of QAOA, combining it with physics-motivated input from CD driving.

Table~\ref{tab:comparison-eng-N10} shows a comparison with the best obtained energies for $N=10$ spin-$1$ particles (qutrits): the superior performance of CD-QAOA remains despite the exponentially growing Hilbert space size. Reaching significantly larger system sizes is infeasible with the present-day computational power: we note that this a feature of the quantum system rather than a drawback of CD-QAOA, cf.~discussion on LMG model in Sec.~\ref{subsec:LMG}.

\begin{table}
    \centering
    {
    {\renewcommand{\arraystretch}{1.35}%
        \begin{tabular}{@{\;}c@{\quad}||>{\columncolor[gray]{0.9}}c@{\qquad}c@{\qquad}c@{\qquad}c@{\;}}
        
        \Xhline{2.5\arrayrulewidth}
         \multirow{2}{*}{$T$} & \multicolumn{4}{c}{ $E / E_{\mathrm{GS}} \ [N\eq10]$} \\ 
        
         & CD-QAOA & CD & QAOA & adiabatic\\\hline\hline
        4 & \textbf{0.943837} & 0.923199 & 0.79534 & 0.067807\\ \hline
        8 & \textbf{0.961383} & 0.933067 & 0.93386 & 0.438856\\ \hline
        12 & \textbf{0.990415} & 0.942857 & 0.96275 & 0.658182\\\Xhline{2.5\arrayrulewidth}
        \end{tabular}
    }}
    \caption{Spin-$1$ Ising model:
        comparison of the best obtained energy ratio $E/E_\mathrm{GS}$ after optimization, for four different optimization methods: CD-QAOA, variational CD driving, conventional QAOA, and adiabatic evolution, at $T=4,8,12$ for $N=10$ qutrits, where $\mathrm{dim}(\mathcal{H})\!=\!3219$.
        The remaining setup and parameters are the same as in Fig.~\ref{fig:comparison-eng}.
}
    \label{tab:comparison-eng-N10}
\end{table}

We emphasize that CD-QAOA features some important advantages as compared to CD driving:
(1) Due to the nested commutators in the definition of time-ordered exponentials, the QAOA dynamics can effectively implement total unitaries $U(\{\alpha_j\}_{j=1}^q ,\tau)$ generated by effective non-local operators; therefore, CD-QAOA can, in principle, realize a nonlocal effective Hamiltonian as an approximation to the true CD Hamiltonian, thereby overcoming convergence issues related to operator-valued series expansions.
(2) CD-QAOA lifts the boundary constraint present in adiabatic and CD driving where the initial and target Hamiltonians are eigenstates of $H(0)$ and $H(1)$, respectively; an interesting open question is whether a local effective Hamiltonian exists, which captures the evolution of the system in this case. Examining the evolution of the entanglement entropy and other local observables induced by the optimal protocol, suggests that this is indeed the case [App.~\ref{app:spin-1}].
(3) One can add any control unitary to the set $\mathcal{A}$, not just terms related to gauge potentials: CD-QAOA has high flexibility to accommodate experimental constraints.
(4) determining the variational gauge potential in CD-driving requires using the exact ground state in order to minimize the action, which can be a significant drawback when the ground state is not known or cannot be computed.

\section{\label{sec:generalization}Transfer Learning and Generalization of the RL Algorithm to different System Sizes}

The scale collapse in the energy density of the spin-$1/2$ Ising model presents a testbed for the transfer learning capabilities of RL. In transfer learning, the RL agent learns to control one physical system, and is then used to manipulate another. In our case, the two systems are given by the same Ising model at two different system sizes. Note that transfer learning would have not been possible, had we defined the learning problem using the full quantum states, because the latter are vectors in Hilbert space whose size grows exponentially with $N$.

\begin{figure}[t!]
  \vspace{-1em}
    \includegraphics[width=1.0\columnwidth]{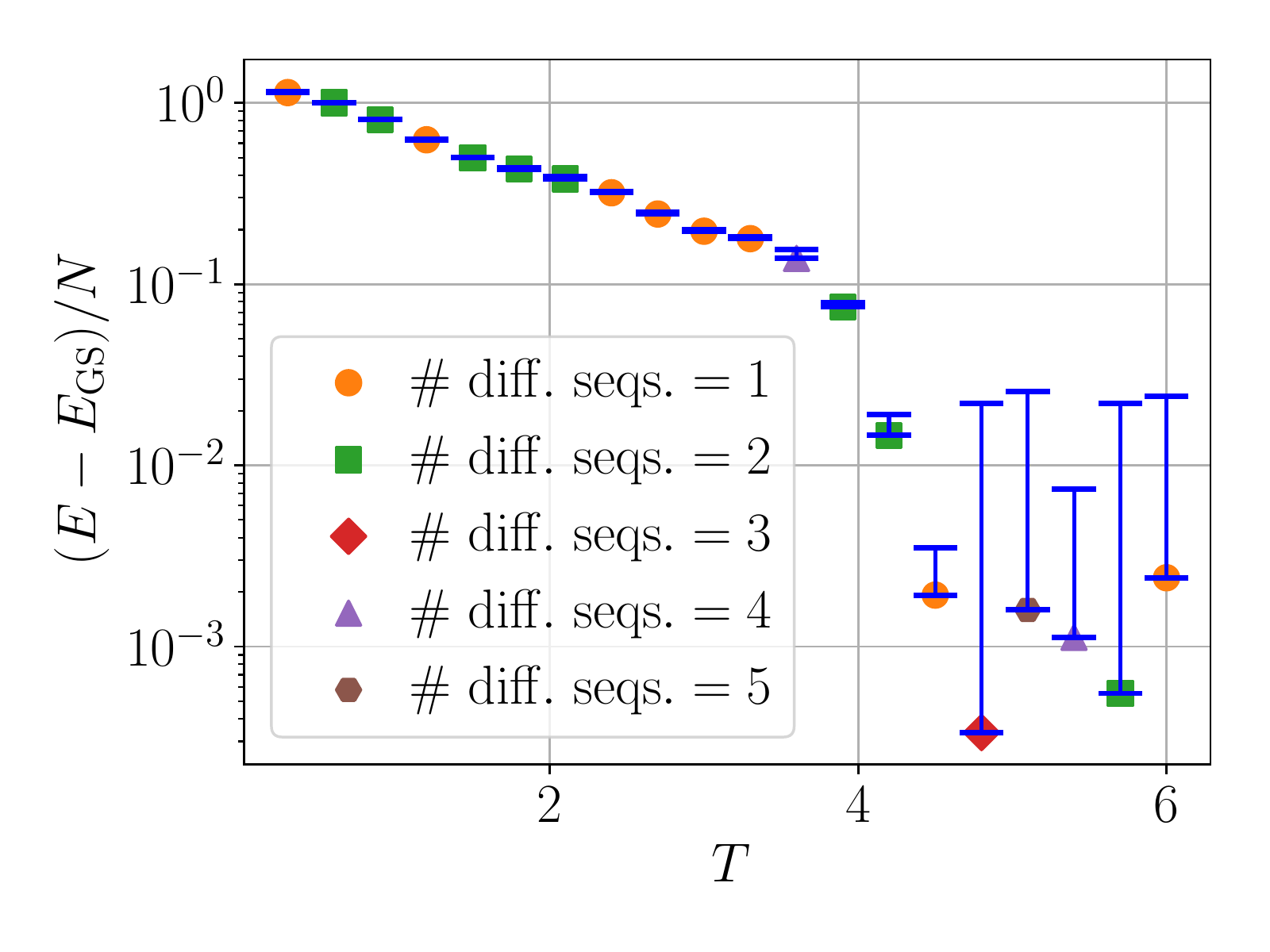}
       \vspace{-3em}
    \caption{\label{fig:Ising_gen}Spin-$1/2$ Ising model: Protocol generalization across various system sizes. The marker types show the number of different protocols found by the RL agent at a fixed $T$ across all system sizes $N\!=\! 6,10,12,14,16$ and $18$. Each protocol is applied to every system size $N$ at a fixed $T$ which results in a set of cost function values; the error bars designate the range between the largest and smallest cost function value. The parameters are the same as in Fig.~\ref{fig:Ising_energy}.
    }
    \vspace{-1em}
\end{figure}

To apply transfer learning, consider first a fixed protocol duration $T$. For every fixed system size $N$, we first train a different RL agent. Next, we build the set of protocols across all system sizes, found by these agents, and determine the number of unique protocols [cf.~legend in Fig.~\ref{fig:Ising_gen}]. Finally, we apply all unique protocols to all system sizes available, and store the energy densities they result in. This leaves us with a set of energy density values for every fixed $T$. The error bars in Fig.~\ref{fig:Ising_gen} show the best and the worst protocols over this set. Observe that, below the QSL, there are only a few points $T$ where the best control protocol is the same across all system sizes. Transfer learning works well, as can be seen by the small error bars. In this regime, the RL agent generalizes its knowledge and learns universal features of the protocol, required to control the Ising model. In contrast, for $T>T_\mathrm{QSL}$, there are many more protocols giving approximately similar ground state energies. While the corresponding energies are similar in value, the agent does not generalize. Nevertheless, we checked that, in this regime, training on smaller system sizes still provides a useful pre-training procedure for learning on larger systems.

\section{\label{sec:outro}Discussion/Outlook}

We analyzed many-body ground state preparation using unitary evolution in the spin-$1/2$ Ising model, the spin-$1$ anisotropic Heisenberg and Ising models, and the fully connected LMG spin-$1/2$ model. We introduced the CD-QAOA ansatz: an RL agent optimizes the order of unitaries in the protocol sequence, generated from terms in the adiabatic gauge potential series, and obtains short high-fidelity protocols away from the adiabatic regime. The resulting algorithm combines the strength of continuous and discrete optimization into a unified and versatile control framework. 
We find that our CD-QAOA ansatz outperforms consistently both conventional QAOA, and variational CD driving across different models and protocol durations. An interesting open question is whether one can use CD-QAOA to find a nonlocal approximation to the variational gauge potential itself, which is beyond the scope of asymptotic series expansions. Another straightforward application of CD-QAOA would be imaginary time evolution~\cite{beach2019making}.

For the nonintegrable spin-$1/2$ Ising chain, we reveal the existence of a finite quantum speed limit. Moreover, we find a remarkable system-size collapse of the energy curves suggesting that the sequences found by the agent hold in the thermodynamic limit; this is corroborated by numerical experiments on transfer learning which demonstrate that one can train the agent on one system size while it generalizes to larger systems. 
In the Heisenberg spin-$1$ system, CD-QAOA allows preparing long-range and topologically ordered ground states, even when the initial state does not belong to the phase of the target state. The optimal protocols found by the RL agent contain nontrivial basis rotations, intertwined with alternating QAOA-like subsequences, suggesting new ans\"atze for more efficient variants of CD-QAOA. Numerical studies of nonequilibrium quantum many-body systems, in turn, suffer from limitations related to the exponentially large dimension of the underlying Hilbert space: future work can investigate dynamics beyond exact evolution.

Compared to conventional QAOA, using terms from the variational gauge potential series has higher expressivity, which results in much shorter, yet better performing, circuits. This method can be used, e.g., to reduce the cumulative error in quantum computing devices. However, gauge potential terms are not always easy to realize in experiments since they implement imaginary-valued terms which break time-reversal symmetry; that said, it is often possible to generate such terms using auxiliary real-valued operators via a generalization of the Euler angles, or by means of change-of-frame transformations~\cite{sels2017minimizing}. Moreover, as we have demonstrated, CD-QAOA admits non-gauge potential terms as building blocks for control sequences, e.g., universal gate sets. Other experimental constraints, such as the presence of drift terms, which cannot be switched off, can also be conveniently incorporated by redefining the set of unitaries $\mathcal{A}$. 

Finally, let us remark that RL provides only one possible set of algorithms to explore the exponentially large space of protocol sequences; in practice, one can apply other discrete optimization techniques, e.g.~genetic algorithms and search algorithms like Monte-Carlo Tree Search~(MCTS).

\emph{Acknowledgments.---}We wish to thank A.~Polkovnikov, Dong An and Yulong Dong for valuable discussions. This work was partially supported by the Department of Energy under Grant No. DE-AC02-05CH11231, No. DE-SC0017867 and by a Google Quantum Research Award (L.L., J.Y.), and by the NSF Quantum Leap Challenge Institute (QLCI) program through grant number OMA-2016245 (L.L.). 
M.B.~was supported by the U.S. Department of Energy, Office of Science, Office of Advanced Scientific Computing Research, under the Accelerated Research in Quantum Computing (ARQC) program, the Quantum Algorithm Teams Program, the U.S. Department of Energy under cooperative research agreement DE-SC0009919, the Emergent Phenomena in Quantum Systems initiative of the Gordon and Betty Moore Foundation, and the Bulgarian National Science Fund within National Science Program VIHREN, contract number KP-06-DV-5.
We used SLSQP implemented in SciPy for the QAOA solver, NumPy, and TensorFlow and TensorFlow Probability for the deep learning simulations;
we used \href{https://github.com/weinbe58/QuSpin#quspin}{Quspin} for simulating the dynamics of the quantum systems~\cite{weinberg2017quspin, weinberg2019quspin}.
The authors are pleased to acknowledge that the computational work reported on in this paper was performed on Savio3 Condo of Berkeley Research Computing (BRC).

\clearpage
\appendix

\section{\label{app:algo}High-level optimization: Policy Gradient using Deep Autoregressive Networks}

Recently, progress made in machine learning (ML)~\cite{dunjko2018machine,mehta2019high,carleo2019machine,carrasquilla2020machine} has raised the question as to how we can harness such modern advances to improve techniques to manipulate quantum systems.
Examples of ML applications include model-based optimization~\cite{sung2020exploration}, differentiable programming~\cite{schafer2020differentiable} and Bayesian inference~\cite{sauvage2019optimal} quantum control, cavity control~\cite{fosel2020efficient}, designing quantum end-to-end learning schemes~\cite{wu2020end} and measurement-based adaptation protocols~\cite{albarran2018measurement}, as well as applications to quantum error-correction~\cite{fosel2018reinforcement,nautrup2019optimizing}.

Reinforcement learning (RL) algorithms~\cite{sutton2018reinforcement, rose2020reinforcement}, such as policy gradient \cite{niu2019universal, august2018taking,  porotti2019coherent}, Q-learning \cite{bukov2018reinforcement, bukov2018day} and AlphaZero \cite{dalgaard2020global}, have recently attracted the attention of physicists, and in particular how they can be combined with physically motivated VQEs for improved performance.
In RL, policy gradient  has been proposed as an alternative optimizer for QAOA showcasing the robustness of RL-based optimization to both classical and quantum sources of noise~\cite{yao2020policy}; a related study applied Proximal Policy Optimization (PPO) to prepare the ground state of the transverse-field Ising model~\cite{wauters2020reinforcement}. The QAOA ansatz with policy gradient has been applied to efficiently find optimal variational parameters for unseen combinatorial problem instances on a quantum computer~\cite{khairy2019reinforcement}; Q-learning was used to formulate QAOA into an RL framework to solve difficult combinatorial problems~\cite{garcia2019quantum}, and in the context of digital quantum simulation~\cite{bolens2020reinforcement}. 

In the following, we introduce the details of the Reinforcement Learning algorithm used for the high-level optimization in this work. 

\subsection{\label{app:RL}Reinforcement Learning Basics}

Reinforcement learning (RL) comprises a class of machine learning algorithms where an agent learns how to solve a given task through interactions with its environment using a trial-and-error approach~\cite{sutton2018reinforcement}. It is based on a Markov Decision Process (MDP) defined by the tuple $(\mathcal S, \mathcal A,p,R)$ where $\mathcal S$ and $\mathcal A$ represent the state and action spaces, $p: \mathcal S\times \mathcal S \times \mathcal A \rightarrow [0,1]$ defines the transition dynamics, and $R: S\times A \rightarrow \mathbb{R}$ is the reward function that describe the environment. Let $\pi(a_j|s_j): \mathcal A \times  \mathcal S \rightarrow [0, 1]$ denote a stochastic policy that defines the probability distribution of choosing an action $a_j\in\mathcal{A}$ given the state $s_j \in \mathcal S$. Rolling out the policy $\pi(a_j|s_j)$ in the environment can also be viewed as sampling a trajectory $ \tau \sim \mathbb P^{\pi}( \cdot )$ from the MDP,
where $\mathbb{P}^{\pi}( \tau ) = p_0(s_1)\pi(a_1|s_1)p(s_2|s_1, a_1)\cdots \pi(a_{q}| s_q)p(s_{q+1}|s_{q}, a_{q})$ is the probability for the trajectory $\tau$ to occur, $q$ sets the episode or trajectory length, and $p_0$ is the distribution of the initial state; an example for a trajectory is $\tau=(s_1, a_1, ...., a_{q}, s_{q+1})$.
The goal in RL is to find a policy that maximizes the expected return:
\begin{equation}
    J(\params) = \mathbb{E}_{\tau \sim \mathbb P^{\pi}}\left[\sum_{j=1}^{q}  R(s_j,a_j)\right].
    \label{eq:RL}
\end{equation}

To maximize the expected return $J(\params)$, we use policy gradient -- an RL algorithm, which is (i) on-policy, i.e.~trajectories have to be sampled from the current policy $\pi_\mathrm{\bf\params}$:~$\pi\!=\!\pi_\mathrm{\bf\params}$, and (ii) model-free, i.e.~the agent does not need to have a model for the environment dynamics: $p(s'|s,a)$ is assumed unknown for the purpose of finding the optimal policy. Highly expressive function approximators, such as deep neural networks, help parametrize the policy using variational parameters $\params$.
Policy gradient gradually improves the expected return in a number of iterations (or training episodes), by increasing the probability for actions that lead to higher rewards, and decreasing the probability for actions that lead to lower rewards, until it reaches a (nearly) optimal policy.

We mention in passing that we use interchangeably the terms return and cost function (the latter being the negative of the former): the goal of the RL agent is thus to maximize the expected return, or to minimize the cost function.

\begin{figure}[t!]
	\centerline{
	\includegraphics[width=1.1\columnwidth]{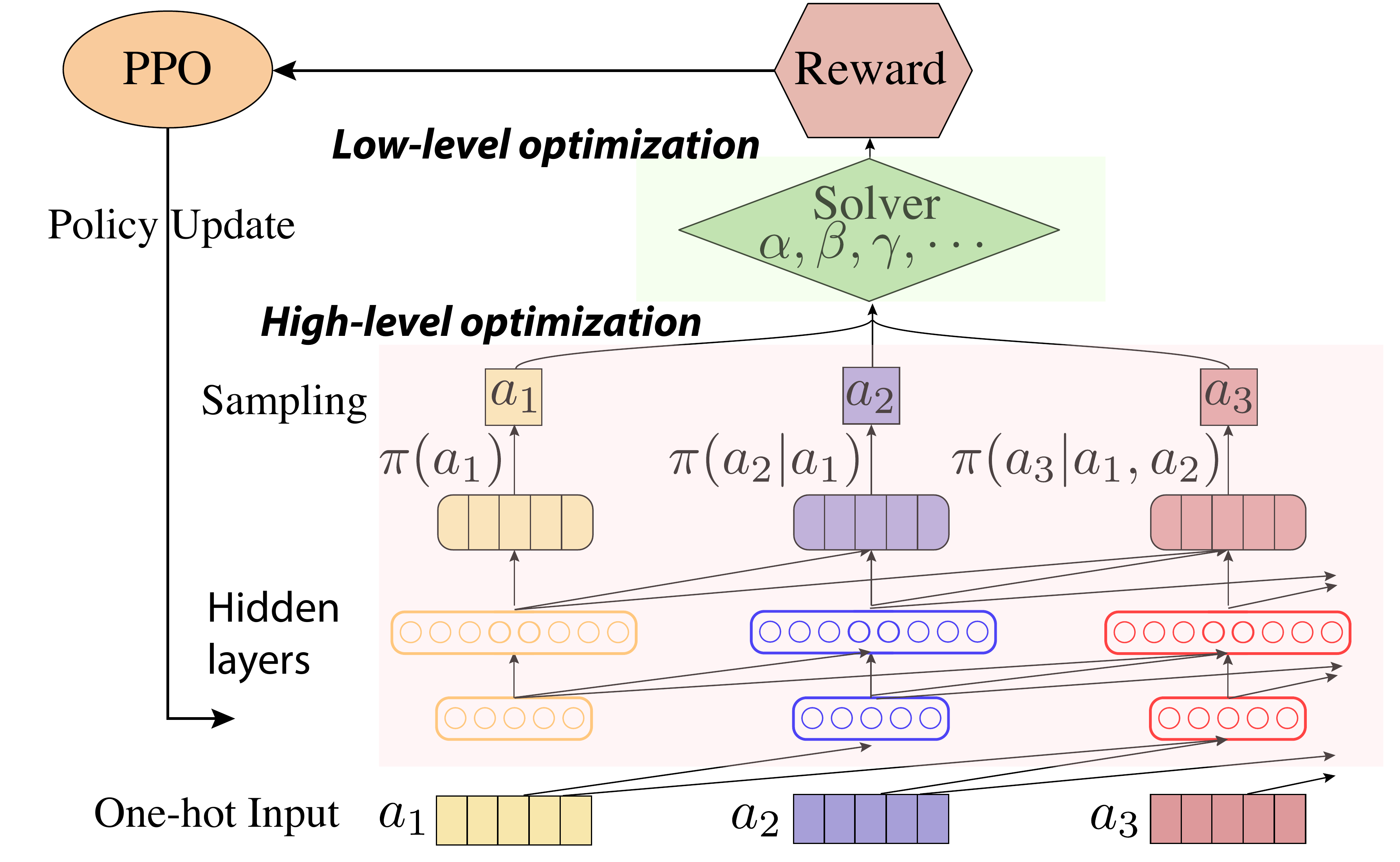}
	}
	\caption{ 
		Schematics of CD-QAOA with an autoregressive policy network.
		The ancestral sampling procedure used for training is displayed in Fig.~\ref{fig:ar-sampling}. 
		The details of the network structure and its training hyperparameters are shown in Table~\ref{tab:model_hp}. 
	}
	\label{fig:schematics}
\end{figure}

\begin{figure*}[t!]
\centering
	\includegraphics[width=\textwidth]{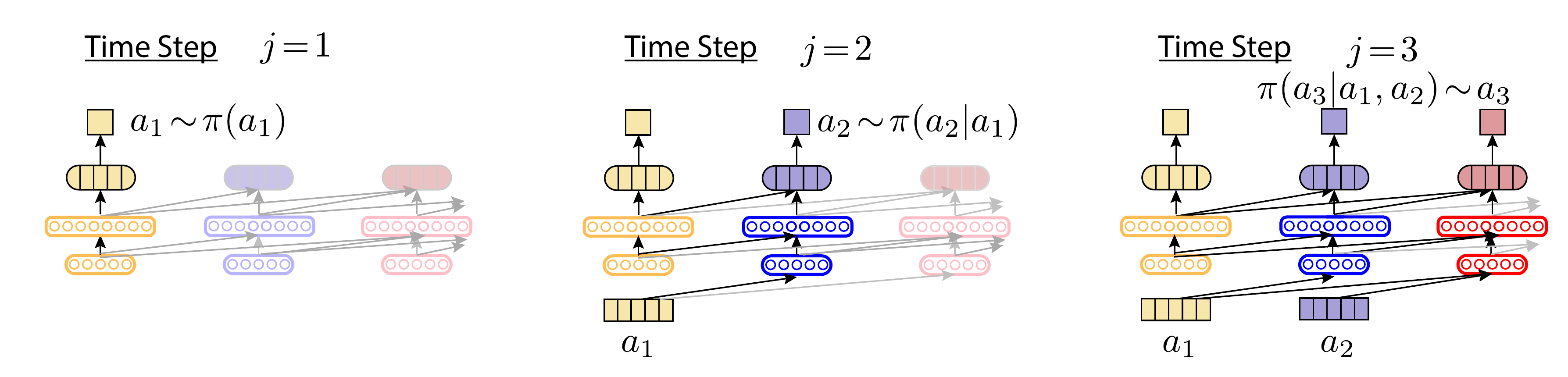}
	\caption{ The exact sampling algorithm for CD-QAOA with an autoregressive policy network, where faded nodes and connections represent unused nodes and connections. The action at each time step is generated sequentially, by computing its respective conditional categorical distribution, and sampling according to that. Notice that only a single column is processed at each time step, and in order to sample a complete sequence of actions in an episode one needs to make a forward pass through the network architecture $q$ times.}
	\label{fig:ar-sampling}
\end{figure*}

\subsection{\label{app:RL_env}Policy Gradient Reinforcement Learning for Quantum Many-Body Systems}

{\bf Actions:} To apply the reinforcement learning formalism to quantum control, we identify taking actions at each time step within a learning episode, with selecting unitaries one at a time within the circuit depth $q$. Choosing the same unitary at two consecutive time steps is prohibited because the same actions can be merged resulting in a lower effective circuit depth $q-1$. At the initial time step $j=1$, the quantum wavefunction is given by the initial state $| \psi_{i}\rangle$; for each intermediate protocol step $j$, the action $a_{j}\!=\!H_{j}$ is chosen according to the policy $\pi_\mathbb{\params}$.
Note that the RL agent only selects the generator $H_j$ out of the set of available actions $\mathcal{A}$ (or alternatively -- which unitary to apply). In other words, unlike Ref.~\cite{yao2020policy}, the RL part of CD-QAOA is \emph{not} concerned with finding the corresponding optimal duration $\alpha_j$; one can think of this low-level continuous optimization as being part of the environment [cf.~App.~\ref{app:slsqp}]\footnote{It is also possible to define an RL framework for hybrid continuous-discrete control where optimization is entirely based on RL, cf.~Ref.~\cite{yao_in_prep}.}. At the end of the episode, the quantum state is evolved by applying the entire generated circuit $U(\{\alpha_j\}_{j=1}^{q} ,\tau)$ to the initial quantum state $|\psi_i\rangle$.

{\bf States:} Since the initial state $| \psi_{i}\rangle$ is fixed and thus the quantum state at any time step $j$ is uniquely determined by the previous actions taken, here we represent the RL state by concatenating all the previous actions up to step $j$~\cite{bukov2018quantum}. One reason for this is that, in many-body quantum systems, the number of components in the quantum state scales exponentially with the system size $N$, which quickly leads to a computational bottleneck for the simulation on classical computers. A second advantage of this choice is that the first layer of the underlying deep neural network architecture, which parametrizes the policy, will not depend on the system size $N$ either, which allows the algorithm to handle a large number of degrees of freedom. Using the quantum state would not be viable on quantum computers either, because quantum states are unphysical mathematical constructs that cannot be measured. Therefore, we can simplify the form of trajectories to consist of actions only, e.g.~$\tau=(a_1,a_2,\dots,a_q)$.

{\bf Rewards:} The reward $R_{j}=R(s_j,a_j)$ is chosen as the negative energy density at the end of the episode:
$$
    R_{j}=\left\{\begin{array}{cl}0,                              & \text { if } j<q \\
        -E(\{\alpha_j\}_{j=1}^q, \tau )/ N, & \text { if } j=q.\end{array}\right. 
$$
We use energy density, since it is an intensive quantity which has a well-defined limit when increasing the number of particles $N$. In all figures, we show the relative energy $E/E_\mathrm{GS}$ for clarity (the ground state energy $E_\mathrm{GS}$ is typically negative in our models), but the RL agent is always trained with the (negative) energy density $-E/N$.
Rewards can also be other observables or nonobservable quantities, such as the overlap squared between two quantum states (fidelity), or the entanglement entropy.

Notice that the reward is sparse: only at the end of the episode is the negative energy density given as a reward; there is no instantaneous reward during the sequence [and thus we can use interchangeably the terms reward and total return]. This is motivated by the quantum nature of the control problem, where a projective measurement results in a wavefunction collapse.

\subsection{\label{app:RL_policy}Policy Parametrization using an Autoregressive Neural Network}

An essential part of the policy gradient algorithm is the definition of the policy $\pi_{\params}$. It is common to parametrize the policy with a highly expressive function approximator, such as a neural network. In our setup, we use a deep autoregressive network, which has recently been used in physics applications of learning to generate samples from free energies in statistical mechanics models~\cite{wu2019solving}, and variational approximators for quantum many-body states~\cite{sharir2020deep}. This architecture is selected to incorporate causality by factorizing the total probability into a product of conditional probabilities:
\begin{equation}
    \pi_{\params}( a_{1}, a_{2}, \cdots, a_{q}) = \pi_{\params}(a_{1}) \prod_{j=2}^{q} \pi_{\params}(a_{j} | a_{1}, \cdots, a_{j-1}),
    \label{eq:policy}
\end{equation}
where the marginal distribution $ \pi_{\params}(a_{1})$ and the conditional distribution $\pi_{\params}(a_{j} | a_{1}, \cdots, a_{j-1})$ are discrete categorical distributions over $\mathcal A$.
This kind of parametrization explicitly tells how the actions taken in the earlier steps of an episode affect the actions selected later on during the same episode.
Such a causal requirement would not be necessary, had we used the full quantum state, which would make the dynamics of the environment Markovian.
Each of the conditional probabilities in Eq.~\eqref{eq:policy} can be modeled explicitly using the autoregressive neural network architecture, which naturally allows the policy to depend on all the previous actions only. The structure of the policy network is shown in Fig.~\ref{fig:schematics}, the sampling of the autoregresssive policy is depicted in Fig.~\ref{fig:ar-sampling} and the hyperparameters of the algorithm (including the number of parameters) are given in Table~\ref{tab:model_hp}.

\subsection{\label{app:RL_training}Training Procedure: Proximal Policy Optimization (PPO)}

In each iteration of the policy gradient algorithm, a batch of sampled trajectories $\{\tau^k\}=\{(a_{1}^{k}, \cdots, a_{q}^{k})\}_{k=1}^{M}$ are rolled out (i.e.~sampled) from the current policy, where $M$ is the batch/sample size. Then, the return $R(\tau^{k})$ corresponding to trajectory $\tau^k$ is computed as
$$R(\tau^{k}) = \sum_{j=1}^q R^{k}_j = -E(\{\alpha_j^{k}\}_{j=1}^q, \tau^k )/ N.$$
To compute the energies, we use the low-level optimization to determine the best-estimated values of $\alpha_j$, given a sequence $\tau$, see App.~\ref{app:slsqp}. To minimize the chance of getting stuck in a suboptimal local minimum, each sequence is evaluated multiple times, starting from a different initial realization for the $\alpha_j$-optimizer, and the best result is selected [App.~\ref{app:ctrl_landscape}]. 

\begin{figure}[t!]
    \includegraphics[width=0.5\textwidth]{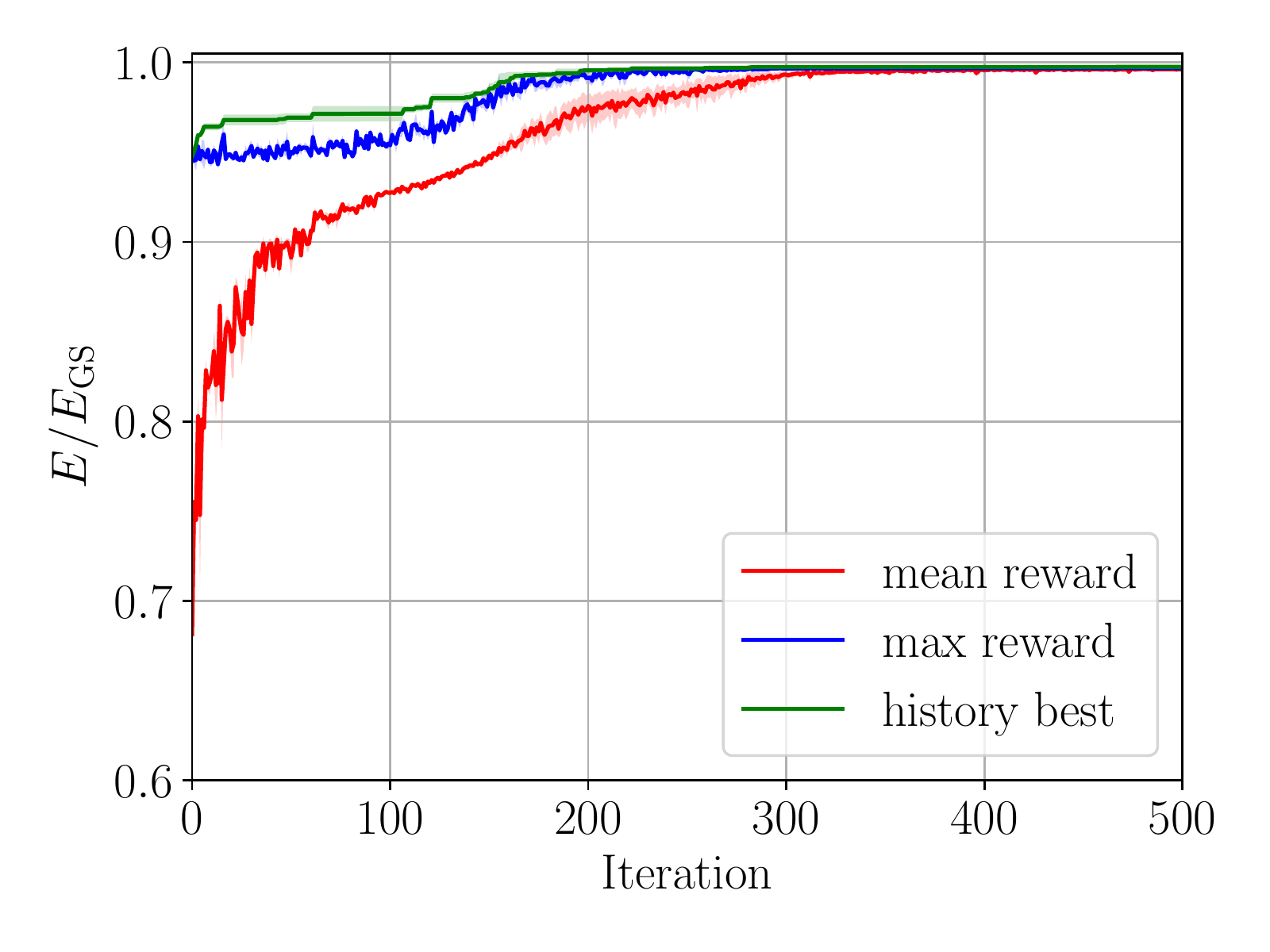}
    \caption{\label{fig:nn-training} Spin-$1$ Ising model:
    training curves for CD-QAOA with energy minimization as a cost function.
    The mean negative energy density (red) is computed for a sample generated using the policy at the current iteration; max (blue) is the maximum within the sample; the history best (green) is the best-encountered policy during the entire training process (i.e., considering all iterations).
    Each curve shows the average out of three simulations corresponding to three different seed values for the high level RL optimization; the fluctuations around the seed-averages are shown as a narrow shaded area.
    The total duration is $T\!=\!28$ and the number of spin-$1$ particles is $N\!=\!8$. The initial and target states are $|\psi_i\rangle\!=\!|\!\downarrow \cdots\downarrow\rangle$ and $|\psi_\ast\rangle\!=\!|\psi_\mathrm{GS}(H)\rangle$ for $h_z/J=0.809$ and $h_x/J=0.9045$. The CD-QAOA action space is $\mathcal{A}_{\text{CD-QAOA}}=\{Z|Z\!+\!X,Z; Y,XY,YZ,X|Y,Y|Z \}$, and we use $q\!=\!20$. 
    }
\end{figure}

\begin{figure}[t!]
    \includegraphics[width=1.0\columnwidth]{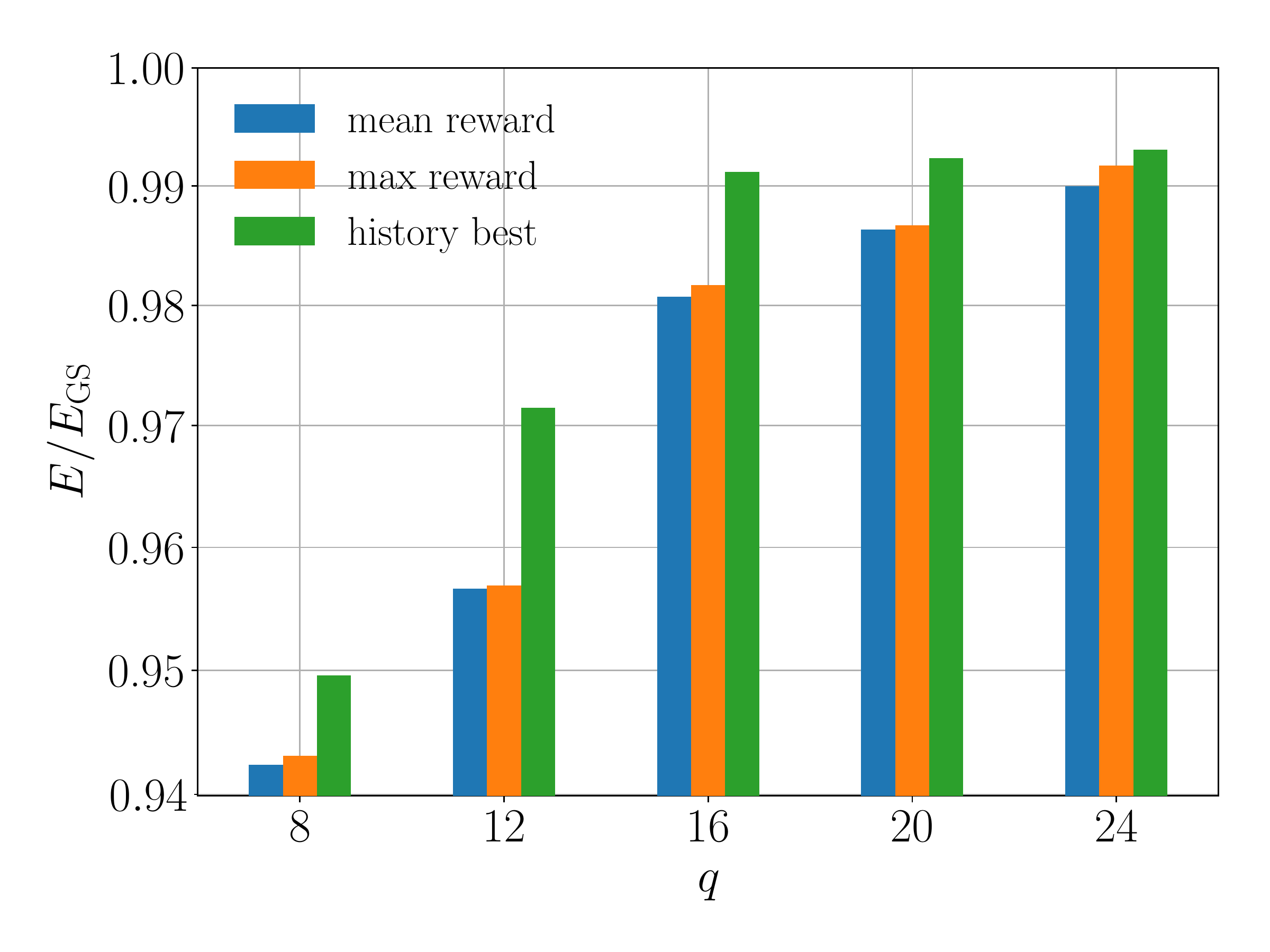}
    \caption{\label{fig:Ising_q_scaling}Spin-$1$ Ising model:
    energy minimization against different circuit depths $q$ using CD-QAOA. 
    The mean negative energy density (blue) is computed for a sample generated using the final, learned policy; max (orange) is the maximum within the sample; the history best (green) is the best encountered policy during the entire training process (i.e., considering all iterations). The total duration $T\!=\!20$ and the values of $q$ ranges from 8 to 24. 
    The other model parameters are the same as in Fig.~\ref{fig:nn-training}.  
    }
\end{figure}

For every iteration, we can define three quantities for a fixed batch of samples: 
(i) mean reward (over the current batch), 
(ii) max reward (over the current batch), and
(iii) history best (best-encountered reward over all the previous iterations). 
These quantities measure the performance of the learned policy, and are shown in Fig.~\ref{fig:nn-training}. Figure~\ref{fig:Ising_q_scaling} shows the scaling of these quantities for the spin-$1$ Ising chain, as a function of the episode length $q$. The performance of CD-QAOA increases because the action space for a larger value of $q$ always contains as a subset the action space for a smaller $q$.

In order to improve the policy represented by the autoregressive network, the RL algorithm interacts with the quantum environment by querying the reward for samples from the current policy. Each trajectory is assigned a reward, once the simulation of the quantum dynamics is complete [note that, as of present date, the simulation may be more expensive if evaluated on a quantum computer]. Thus, it is advantageous to reduce the sample size needed to learn the policy, i.e., to improve the sample efficiency. 

The vanilla policy gradient method is known for its poor data efficiency. Thus, we adopt Proximal Policy Optimization (PPO)~\cite{schulman2017proximal}, a more robust and sample-efficient policy gradient type algorithm. To be more specific, we use the following clipped objective function:
\begin{eqnarray}
        \mathcal G (\params)=
        \mathbb{E}_{\tau \sim \pi_{ \params_t}} \bigg[&& \min \big\{
         \rho_\params (\tau) A_{\params_t}(\tau),\label{eqn:ppo}\\
         && \operatorname{clip}\left(\rho_\params(\tau), 1-\epsilon, 1+\epsilon\right) A_{\params_t}(\tau)  \big\} \bigg]. \nonumber
\end{eqnarray}

Here, $\tau =( a_{1}, a_{2}, \cdots, a_{q})$ is the action sequence sampled from the previous policy $\pi_{\params_{t}}$ [cf.~Algorithm~\ref{alg:ARPG}]. Typically, the policy from the last iteration is chosen to be the old policy;
$\rho_\params(\cdot) = \frac{\pi_\params(\cdot)}{\pi_{\params_{t}}(\cdot) }$ is the importance sampling weight between the new policy $\pi_\params$ and the old policy $\pi_{\params_{t}}$;
$A_{\params_{t}}(\tau)= R(\tau) - b$ is the advantage function, where $b$ is called a baseline: the advantage measures the reward gain of choosing a specific action, w.r.t.~the baseline. For example, a simple baseline can be the average reward, e.g., $b=\mathbb E_{\tau \sim \pi_{\params_{t}}} [R(\tau)]$, and then the advantage measures how much better (or worse) an action is w.r.t.~the average; in the numerical experiments, we use an exponential moving average [cf.~App.~\ref{app:RL_technical} for the details].

Further, the clip function,
$$\mathrm{clip}(r, x,y)= \max \big( \min \left(r, x \right) , y\big),$$ 
clips the value of $r$ within the interval $[x, y]$, which is used to restrict the likelihood ratio in the range $[1-\epsilon, 1+\epsilon]$; this prevents the policy update from deviating too much from the old policy after one gradient update. The clipped objective function is designed to improve the policy as well as to keep it within some vicinity of the last iteration, whence the name Proximal Policy Optimization.

We update the network parameters $\params$ by ascending along the gradient of the RL objective $\mathcal G(\params)$. To provide intuition about the PPO objective, consider the following limiting case.
If we only have the first term in the objective, i.e.~$\mathcal G_1(\params)= \mathbb{E}_{\tau \sim \pi_{ \params_{t}}} [ \rho_\params (\tau) A_{\params_{t}}(\tau) ]$, we obtain the following gradient:
\begin{eqnarray*}
    \nabla_\params \mathcal G_1(\params) &=& \mathbb{E}_{\tau \sim \pi_{ \params_{t}}} [ \nabla_\params \rho_\params (\tau) A_{\params_{t}}(\tau) ]\\
    &=& \mathbb{E}_{\tau \sim \pi_{ \params_{t}}} \left[ \frac{\nabla_\params \pi_\params (\tau)}{\pi_{\params_{t} (\tau)}}  A_{\params_{t}}(\tau) \right].
\end{eqnarray*}
Since we are taking the gradient with respect to $\params$, it will pass through $\pi_{\bf\params_{t}}$ and $A_{\params_{t}}(\tau)$.
Furthermore, whenever the parameters $\params\approx\params_{t}$, the gradient above is identical to the policy gradient:
\begin{eqnarray*}
    \nabla_\params \mathcal G _1(\params) &\approx& \mathbb{E}_{\tau \sim \pi_{\bf \params}} \left[ \frac{\nabla_\params \pi_\params (\tau)}{\pi_{\params (\tau)}}  A_{\params}(\tau) \right] \nonumber\\
    &=& \mathbb{E}_{\tau \sim \pi_{\bf \params}} [\nabla_\params \log \pi_\params (\tau)  A_{\params}(\tau) ].
\end{eqnarray*}

\begin{figure}[H]
	\centering
	\includegraphics[width=\columnwidth]{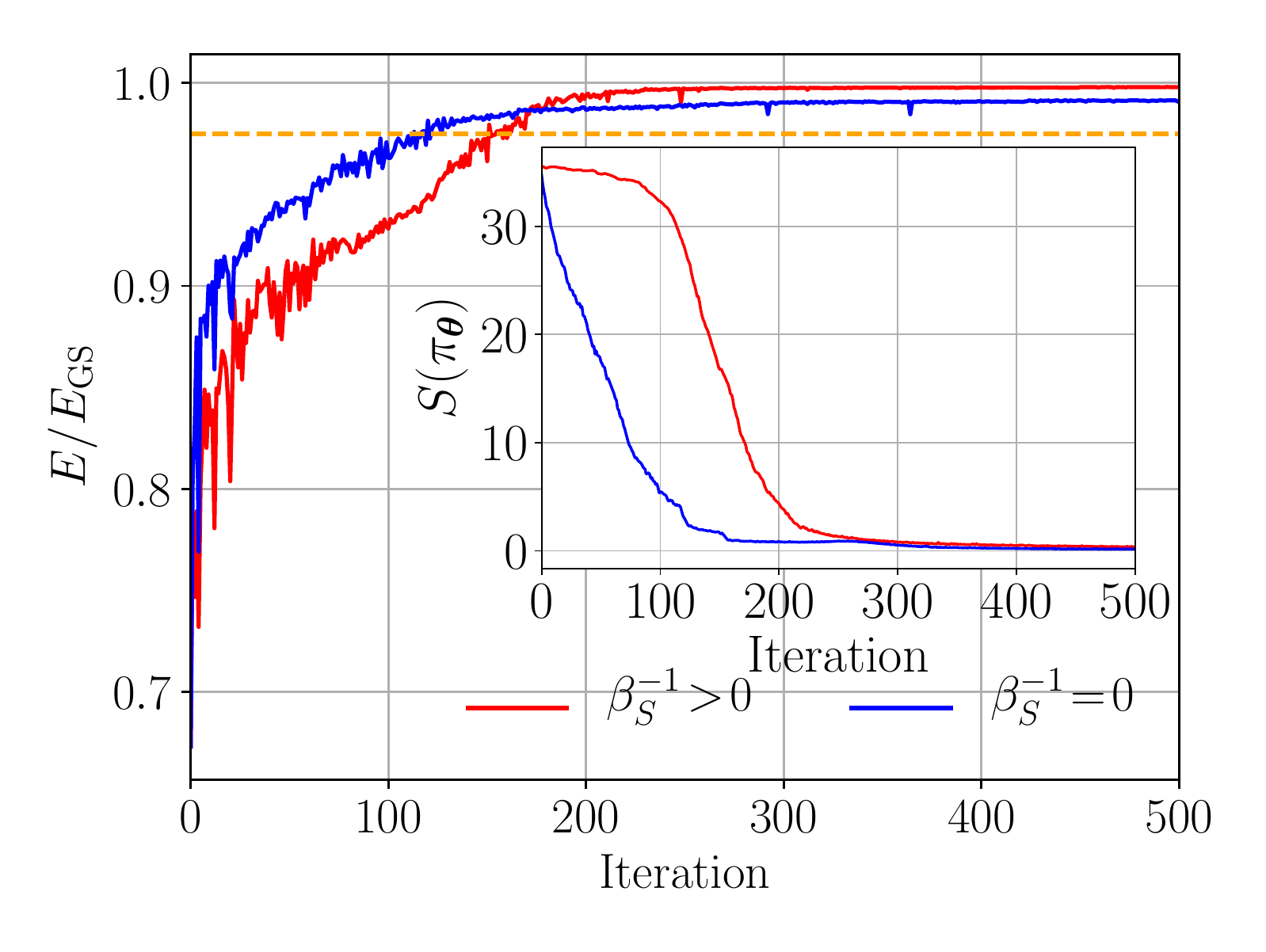}
	\caption{ Spin-$1$ Ising model:
		Comparison of the mean reward with ($\beta_{S, \{0\}}^{-1}\eq 0.1$) and without ($\beta_S^{-1}\eq 0$) the entropy bonus during training. For comparison, the dashed horizontal line marks the performance of QAOA.
		The inset shows the evolution of the policy information entropy during training. 
		Adding entropy gives more room for the RL agent to explore the space of policies instead of directly exploiting the knowledge it obtains. As becomes clear from the figure, the RL algorithm with the entropy bonus achieves a better final performance at the end of training, at the cost of suffering an intermediate lower reward at the beginning of training. 		
		The simulation parameters are the same as in Fig.~\ref{fig:nn-training}.
	}
	\label{fig:maxEnt}
\end{figure}

However, PPO performs multiple gradient updates on the sampled data, rendering policy learning more sample efficient~\cite{schulman2017proximal}.

\subsubsection{\label{app:entropy}Incentivizing Exploration using Entropy}

Maintaining a balance between exploration and exploitation is another major challenge for the reinforcement learning algorithm. Too much exploration prevents the agent from adopting the best strategy it knows so far; on the contrary, too much exploitation limits the agent from attempting new actions and achieving a potentially higher reward. Therefore, it is more appropriate for the agent to explore substantially in the initial iterations of the training procedure, and to gradually switch over to exploitation towards the end of the training procedure.

In order to incentivize the agent to explore the action space at the beginning of training, we include an entropy `bonus' ~\cite{ziebart2010modeling, haarnoja2018soft2} to the PPO objective from Eq.~\eqref{eqn:ppo}.
To do this, consider the maximal-entropy objective, where the agent aims to maximize the sum of the total reward and the policy entropy $S$ [cf.~Eq.~\eqref{eq:entropy}]:

\begin{widetext}
    \begin{align}
    \mathcal J (\params) &\eq \mathcal G(\params) \plus \beta^{-1}_{{S}} \mathcal S(\pi_\params) \label{eqn:ppo-maxent} \\
    &\eq \mathbb{E}_{\tau=(a_1, \cdots, a_q)\sim \pi_{ \params_{t}}} \bigg[\min \{
        \rho_\params (\tau) A_{\params_{t}}(\tau),\;
        \operatorname{clip}\left(\rho_\params(\tau), 1-\epsilon, 1+\epsilon\right) A_{\params_{t}}(\tau)  \}+ \beta^{-1}_{{S}} \sum_{j=1}^q \mathcal S\big(\pi_\params(\;\cdot\; | a_{1}, \cdots, a_{j-1})\big)  \bigg],
    \nonumber
    \end{align}
\end{widetext}
where $\mathcal S\big(\pi_\params(\,\cdot\, | a_{1}, \cdots, a_{j-1})\big) \equiv \mathcal S\big(\pi_\params(\,\cdot\, )\big)$, for $j\!=\!1$. 
The trade-off between exploration and exploitation is controlled by the coefficient $\beta^{-1}_{S}$, which carries a meaning analogous to temperature in statistical mechanics: for $\beta^{-1}_{S}\to 0$ (or $\beta_{S}\to \infty$), any exploration is limited to the intrinsic probabilistic nature of the policy; if training is successful, it is expected that, for deterministic environments, the policy eventually converges to a delta distribution (over the action space) at the later training iterations; this may deteriorate exploration and learning. However, in the opposite limit, $\beta^{-1}_{S}\to\infty$ (or $\beta_{S}\to 0$), every action is selected with equal probability, and the values of the policy $\pi$ become irrelevant. Therefore, in practice, we use a decay schedule for the inverse temperature $\beta^{-1}_{S}$ to gradually reduce exploration [see App.~\ref{app:RL_technical}]. 

Since the marginal distribution $ \pi_{\params}(\,\cdot \,)$ and the conditional distribution $\pi_{\params}(\,\cdot \, | a_{1}, \cdots, a_{j-1})$ are discrete categorical distributions over $\mathcal A$, we can compute a closed form expression for the entropy of the categorical distribution policy. For trajectory $\tau^i\!=\!(a_1^i,\cdots, a_q^i)$, the $j$-th term in the entropy bonus simplifies to
\begin{align}
&\mathcal S\big(\pi_{\params}(\cdot | a^i_{1}, \cdots, a^i_{j-1})\big)  \label{eq:entropy} \\
& \!=\! -\sum_{a\in \mathcal A} \pi_{\params}( a| a^i_{1}, \cdots, a^i_{j-1} )\log \pi_{\params}( a| a^i_{1}, \cdots, a^i_{j-1} ). \nonumber
\end{align}

We emphasize that the entropy considered here is the Shannon or information entropy associated with the policy as a probability distribution, and should be contrasted with the thermodynamic entropy, associated with the logarithm of the density of protocol configurations (a.k.a. density of states) in the optimization landscape. The Shannon entropy helps exploration in the space of policies, and thus the annealing of the corresponding Lagrange multiplier,  $\beta^{-1}_{{S}}$, is not related to thermal annealing in the optimization/energy landscape in a straightforward manner. Moreover, notice that the policy optimization is part of the classical postprocessing of the quantum data, i.e., it does not compromise the nature of the quantum data which is fed to the algorithm in form of rewards.

Figure~\ref{fig:maxEnt} shows a comparison of PPO with and without entropy, as controlled by the value of the temperature $\beta^{-1}_{{S}}$. Introducing the policy information entropy keeps the policy a bit broader in the initial stages of training which enhances exploration; towards the end of training the information entropy is not needed: therefore, we gradually “anneal” $\beta^{-1}_{{S}}$, cf.~App.~\ref{app:RL_technical}.

\subsection{Technical Details}
\label{app:RL_technical}

We train the CD-QAOA algorithm for $500$ epochs/iterations with a mini-batch size of $M=128$. Throughout the training, we sample trajectories according to the marginal and conditional policy distributions given by the autoregressive network.

We use Adam to perform gradient descent on the objective in Eq.~\eqref{eqn:ppo-maxent} with the default parameters $\beta_1=0.9$ and $\beta_2=0.999$, which define the exponential decay rate for the first and second moment estimates, respectively. The learning rate is initialized as $\alpha_{\{\mathrm{lr},0\}}=0.01$ and decays by a factor of $0.96$ every $50$ steps in a staircase fashion. To be more precise, the learning rate at the $k$-th iteration with the exponential decay reads as $\alpha_{\mathrm{lr},\{k\}} = 0.01 \cdot 0.96 ^{\floor*{ k/ 50} }$. The subscript $\{k\}$ denotes the iteration/episode number.

We also introduce an exponential decay schedule for the pre-factor [a.k.a.~temperature] $\beta_S^{-1}$ of the entropy bonus from Eq.~\eqref{eqn:ppo-maxent}. The temperature initializes at $\beta_{S,\{0\}}^{-1}=0.1$ and decays by a factor of 0.9 every 10 steps. At the $k$-th iteration, the temperature is $\beta^{-1}_{S, \{k\}} = 0.1 \cdot 0.9 ^{ k/ 10 }$. Eventually, the temperature is annealed to zero. 

We estimate the advantage function by $A_{\params_{\mathrm{old}}}(\mathbf \tau)= R(\mathbf \tau) - b$, where $b$ is the baseline used to reduce the variance of the estimation. Our baseline $b$ uses an exponential moving average (EMA) of the previous rewards. EMA stabilizes the training and also leverages the past reward information to form a lagged baseline. In practice, we find that the RL algorithm can achieve better rewards compared with using the average of current samples as the baseline. To be more specific, the exponential moving baseline update is $b_{\{k\}} = \eta b_{\{k-1\}} + (1\!-\!\eta) \bar R_{\{k\}} $, where $b_{\{0\}} = 0$ and $\eta = 0.95$. Here, $\bar R_{\{k\}}$ is the sample average of the reward at the $k$-th iteration, i.e.~$\bar R_{\{k\}}\!=\!\frac{1}{M} \sum_{i\!=\!1}^M R_{\{k\}}^i(\tau^i)$.

In terms of policy optimization, we perform multiple steps of ADAM on the objective [Eq.~\eqref{eqn:ppo-maxent}]. The gradient update steps are $4$ per minibatch. The clipped parameter in the objective is set to $\epsilon\!=\!0.1$.

The hyperparameters of the algorithm are listed in Table~\ref{tab:model_hp}.

\begin{widetext}

\begin{algorithm}[H]
    {
	\caption{ CD-QAOA with autoregressive network based policy}
	\label{alg:ARPG}
	\begin{algorithmic}[1]
		\Require batch size $M$, learning rate $\eta_t$, total number of iterations $T_{\mathrm{iter}}$, exponential moving average coefficient $m$, entropy coefficient $\beta^{-1}_{{S}}$, PPO gradient steps $K$.
		\State Generate and select the gauge potential sets $\mathcal A$ using Algo.~\ref{alg:basis-gen}. 
		\State Initialize the autoregressive network and initialize the moving average $\hat R\eq 0$.
		\For {$t=1,..,T_{\mathrm{iter}}$}
		\State Autoregrssively sample a batch of discrete actions of size $M$, denoted as $B$:
		\vspace{-0.6em}
		$$\tau ^k \eq (a^k_{1}, a^k_{2}, \cdots, a^k_{q}) \sim \pi_{\boldsymbol{\theta}}\left(a_{1}, a_{2}, \cdots, a_{q}\right),  \ k = 1, 2, \cdots, M.$$
		\vspace{-1.5em}
		\State Apply the SLSQP solver to the lower-level continuous optimization (cf. App~\ref{app:slsqp}):
		$$\min_{\{\alpha_j^k\}_{j=1}^q} \; \left\{ N^{-1}E(\{\alpha_j^k\}_{j=1}^q, \tau^k ) \bigg| \sum_{j=1}^q \alpha_j^k = T;\ 0 \leq \alpha_j^k \leq T \right\}.$$
		\State Use the negative energy density as the return and compute the moving average:
		\vspace{-1.5em}
		$$R_k = -N^{-1}E(\{\alpha_j^k\}_{j=1}^q, \tau^k ), \quad \hat R = m \cdot \hat R + (1-m) \cdot \frac{1}{M}\sum_{k=1}^M R_k.$$
		\vspace{-1.0em}
		\State Compute the advantage estimates 
		$ A_k = R_k - \hat R$.
		\State Initialize the parameter $\params_{t+1}^{[1]} \eq \params_{t}$.
		\For {$\kappa\eq1,..,K$}
		\State Evaluate the likelihood of samples using the parameters from last iteration and current iteration, i.e. $\pi_{\params_t}(\tau^k)$, $\pi_{\params_{t+1}^{[\kappa]}}(\tau^k)$,  and compute the importance weight $\rho_k^{[\kappa]}\eq \pi_{\params_{t+1}^{[\kappa]}}(\tau^k) / \pi_{\params_t}(\tau^k) $.
		\State Use the advantage estimate and importance weight to compute $\mathcal G_k , \mathcal S_k $, following Eq.~\eqref{eqn:ppo} and Eq.~\eqref{eq:entropy}. 
		\State Compute the CD-QAOA objective Eq.~\eqref{eqn:ppo-maxent} and backpropagate to get the gradients: 
		\[ \nabla_{\params}\mathcal J(\params_{t+1}^{[\kappa]})= \frac{1}{M}\sum_{\{a_j^{\{k\}}\}_{j=1}^q \in B } \nabla_\params \bigg[ \mathcal G^{[\kappa]}_k + \beta^{-1}_{{S}} \mathcal S^{[\kappa]}_k \bigg]. \] 
		\State Update weights $\params_{t+1}^{[\kappa+1]}\leftarrow\params_{t+1}^{[\kappa]} + \eta_t\nabla_{\params}\mathcal J(\params_{t+1}^{[\kappa]})$.
		\EndFor
		\State Update the parameter $\params_{t+1} \leftarrow \params_{t+1}^{[K+1]}$
		\EndFor
	\end{algorithmic}
	}
\end{algorithm}

\end{widetext}

\begin{table}
    \begin{center}
        \begin{tabular}{l@{\hspace{.22cm}}|l@{\hspace{.22cm}}}
            \hline
            \textbf{Parameter}                                       & \textbf{Value}              \\
            \hline\hline
            optimizer                                                & Adam \citep{kingma2014adam} \\
            \hline
            learning rate ($\eta_{\{0\}}$)             & $1 \cdot 10^{-2}$           \\
            learning rate decay steps                                & 50                          \\
            learning rate decay factor                               & 0.96                        \\
            learning rate decay style                                & Staircase                   \\
            \hline
            RL temperature ($\beta_{S, \{0\}}^{-1}$)                 & $1 \cdot 10^{-1}$           \\
            RL temperature decay steps                               & 10                          \\
            RL temperature decay factor                              & 0.9                         \\
            RL temperature decay style                               & Smooth                      \\
            \hline
            baseline exponential moving decay factor ($m$)        & 0.95                        \\
            gradient steps (PPO)                                     & 4                           \\
            clip parameter $\epsilon$                                & 0.1                         \\
            \hline
            number of hidden layers                                  & 2                           \\
            number of hidden units per layer   ($d_\mathrm{hidden}$) & 112                         \\
            nonlinearity                                             & ReLU                        \\
            \hline
            number of samples per minibatch ($M$)                    & 128                         \\
            \hline
        \end{tabular}
        \caption{Hyperparameter values for training the autoregressive deep learning model. In the case of $|\mathcal A_{\mathrm{CD-QAOA}} | \!=\!9, q\!=\!18$ [cf.~Eq.~\eqref{fig:comparison-eng-heisenberg}] the total number of parameters is 24431; for $|\mathcal A_{\mathrm{CD-QAOA}} | \!=\!7, q\!=\!20$ [cf.~Eq.~\eqref{fig:comparison-eng}] the total number of parameters is 21985.
        }
        \label{tab:model_hp}
    \end{center}
    \vskip -1em
\end{table}

\section{\label{app:slsqp} Low-level optimization: finding optimal protocol time steps \texorpdfstring{$\alpha_j$}{a}}

In order to determine the values of the time steps $\alpha_j$, we proceed as follows. 
For any given sequence of actions (or protocol sequence) $\tau\!=\!(a_1, \cdots, a_q)$ of total duration $T$, we solve the following low-level optimization problem:
\begin{equation}
\label{eq:slsqp}
    \min_{\{\alpha_j\}_{j=1}^q} \; \left\{ N^{-1}E(\{\alpha_j\}_{j=1}^q, \tau ) \bigg| \sum_{j=1}^q \alpha_j = T;\ 0 \leq \alpha_j \leq T \right\}
\end{equation}
where $q$ is the sequence length (circuit depth), $N$ is the system size, and $E(\cdot)$ is the energy of the final quantum state [cf.~Eq.~\eqref{eq:energy}] after evolving the initial quantum state $|\psi_i\rangle$ according to the fixed protocol $\tau$.

Note that the $\alpha_j$-optimization is both bounded and constrained. It fits naturally into the framework of the Sequential Least Squares Programming (SLSQP). SLSQP solves the nonlinear problem in Eq.~\eqref{eq:slsqp} iteratively,
using the Han-Powell quasi-Newton method with a Broyden–Fletcher–Goldfarb–Shanno (BFGS)\cite{nocedal2006numerical} update of the B-matrix (an approximation to the Hessian matrix), and an $L1$-test function within the step size.

During each iteration of the policy update, a batch of trajectories $\{\tau^i\}=\{(a_{1}^{i}, \cdots, a_{q}^{i})\}_{i=1}^{M}$ is sampled. Each trajectory sequence $\tau^i$ is assigned a reward, by solving the optimization problem in Eq.~\eqref{eq:slsqp}. Since performing the low-level optimization in Eq.~\eqref{eq:slsqp} is independent of the high-level optimization discussed in App.~\ref{app:algo}, we run the former concurrently to boost the efficiency of the algorithm. We distribute every sequence $\tau^i\!=\!(a_1^{i}, \cdots, a_q^{i})$ to a different worker process and aggregate the results back to the master process in the end. In practice, we use the batch size $M\!=\!128$, and we distribute the simulation on $4$ nodes with $32$ cores each, so that each core solves only one optimization at a time.

Recently, it was demonstrated that it is possible to perform the continuous optimization on par with the discrete one, which eliminates the need to use a solver and results in a fully RL optimization approach~\cite{yao_in_prep}.

\section{\label{app:scaling}Scaling with the number of particles $N$, the protocol duration $T$, and the circuit depth $q$}

\begin{table}
	\centering
	{
		{\renewcommand{\arraystretch}{1.35}%
			\begin{tabular}{@{\;}c@{\quad}|@{\quad}c@{\quad}|@{\quad}c@{\quad}||c@{\qquad}c@{\;}}
				\Xhline{2.5\arrayrulewidth}
				$N$ & $T$ & $q$ & $t_{\mathrm{solver}}$  (sec/iter) & $t_{\mathrm{RL}}$  (sec/iter)\\ 
				\hline\hline
				10 &  20 & 20 &   $ 57.254 \pm 13.829 $ &  $ 0.042 \pm 0.005 $ \\
				8 &  20 &  20 &     $ 17.24 \pm 2.554 $ &  $ 0.055 \pm 0.024 $ \\
				6 &  20 &  20 &    $ 10.559 \pm 3.963 $ &  $ 0.028 \pm 0.004 $ \\
				4 &  20 &  20 &    $ 6.021 \pm 5.149 $ &  $ 0.027 \pm 0.002 $ \\
				\hline\hline
				10 &  28 & 20 &    $ 68.55 \pm 19.044 $ &  $ 0.055 \pm 0.019 $ \\
				10 &  24 & 20 &    $ 61.425 \pm 15.171 $ &  $ 0.038 \pm 0.009 $ \\
				10 &  20 & 20 &   $ 57.254 \pm 13.829 $ &  $ 0.042 \pm 0.005 $ \\
				10 &  16 & 20 &  $ 49.043 \pm 12.447 $ &  $ 0.041 \pm 0.007 $ \\
				10 &  12 & 20 &    $ 39.33 \pm 13.976 $ &  $ 0.038 \pm 0.006 $ \\
				10 &   8 & 20 &   $ 24.689 \pm 14.348 $ &  $ 0.033 \pm 0.008 $ \\
				10 &   4 & 20 &      $ 7.023 \pm 2.651 $ &  $ 0.025 \pm 0.001 $ \\
				\hline\hline
				8 &  20 &  24 &   $ 20.723 \pm 3.903 $ &  $ 0.065 \pm 0.024 $ \\
				8 &  20 &  20 &    $ 17.24 \pm 2.554 $ &  $ 0.055 \pm 0.024 $ \\
				8 &  20 &  16 &   $ 12.626 \pm 3.129 $ &  $ 0.024 \pm 0.004 $ \\
				8 &  20 &  12 &    $ 8.641 \pm 2.654 $ &   $ 0.02 \pm 0.003 $ \\
				8 &  20 &   8 &     $ 5.511 \pm 2.18 $ &  $ 0.016 \pm 0.002 $ \\
				8 &  20 &   4 &    $ 2.092 \pm 1.312 $ &  $ 0.011 \pm 0.002 $ \\
				\Xhline{2.5\arrayrulewidth}
			\end{tabular}
			
	}}
	\caption{
		Wall clock running time of the two-level CD-QAOA optimization steps for the with different system sizes $N$, protocol durations $T$, and circuit depths $q$. The right-hand side of the table shows the time used for the lower-level solver (column $t_{\mathrm{solver}}$) and the time spent for the high-level RL algorithm (column $t_{\mathrm{RL}}$) at every successful iteration. 
		The total cost can then be obtained by multiplying the time for $t_{\mathrm{solver}}$ by the appropriate number of repetitions (e.g., continuous solver realizations, policy sample batch size, PPO training episodes, etc.), taking into account any parallelization if present. 
		Every number represents an average over $40$ independent runs with the corresponding standard deviation shown; the significant deviation in $t_{\mathrm{solver}}$ is caused by the random initial solver state used which causes the algorithm to take a different number of steps to converge within the given tolerance, cf.~App.~\ref{app:ctrl_landscape}. This test is carried out on a single processor Intel Core i7-8700K CPU 6-core 3.70GHz.
	} 
	\label{tab:timing}

\end{table}

Next, we discuss the computational scaling of CD-QAOA. While there are a number of (hyper-)parameters in the algorithm, here we focus on the system size $N$, the protocol duration $T$, and the circuit depth $q$ -- which are physically the most relevant ones. We also consider the continuous and discrete optimization steps separately (the continuous step being also an essential part of conventional QAOA). 

When it comes to the continuous optimization performed by a solver [cf.~App.~\ref{app:slsqp}], the main computational cost comes from the quantum evolution itself.
The basic operation inside the solver is a multiplication of the matrix exponential $\mathrm{exp}(-i\alpha_j H_j)$ by the state $\ket{\psi_i}$. The Hamiltonian $H_j$ is stored as a sparse matrix, and the action of the matrix exponential onto the quantum state, $\mathrm{exp}(-i\alpha_j H_j)\ket{\psi_i}$, can be evaluated without computing the matrix exponential itself with the help of a sparse matrix-vector product; this operation scales exponentially with the system size $N$, i.e.~$O(\exp(c N))$ for some constant $c$.
If we denote the sequence length (a.k.a.~circuit depth) by $q$, then the total cost for evaluating a single value of the continuous angle $\alpha$ scales as $O(q\exp(c N))$. 
We stress that this cost is also incurred by conventional QAOA.

For the discrete optimization performed using reinforcement learning [App.~\ref{app:RL_training}], notice first that the machine learning model is agnostic to the physical quantum model, because we do not use information about the quantum model to train the policy, cf.~App.~\ref{app:RL_env}. 
Because the policy input is, by construction, independent of the quantum state, the input layer of the neural network architecture is shielded from the exponential growth of the physical Hilbert space with $N$. Hence, the deep neural network is \emph{independent} of the Hilbert space dimension.
Further, we use an autoregressive network model which scales linearly with the sequence length $q$, and also linearly with the size of the available action set $|\mathcal{A}|$. 
Thus, the total computational cost for the reinforcement learning optimization scales as $O(q |\mathcal{A}|)$. The scaling of the neural network with the variational networks parameters (weights and biases) is trivially given by the matrix-vector multiplication, as is the case for typical ML deep networks, and is also independent of the physics of the controlled system. 

\begin{figure*}[t!]
    \includegraphics[width=1.0\textwidth]{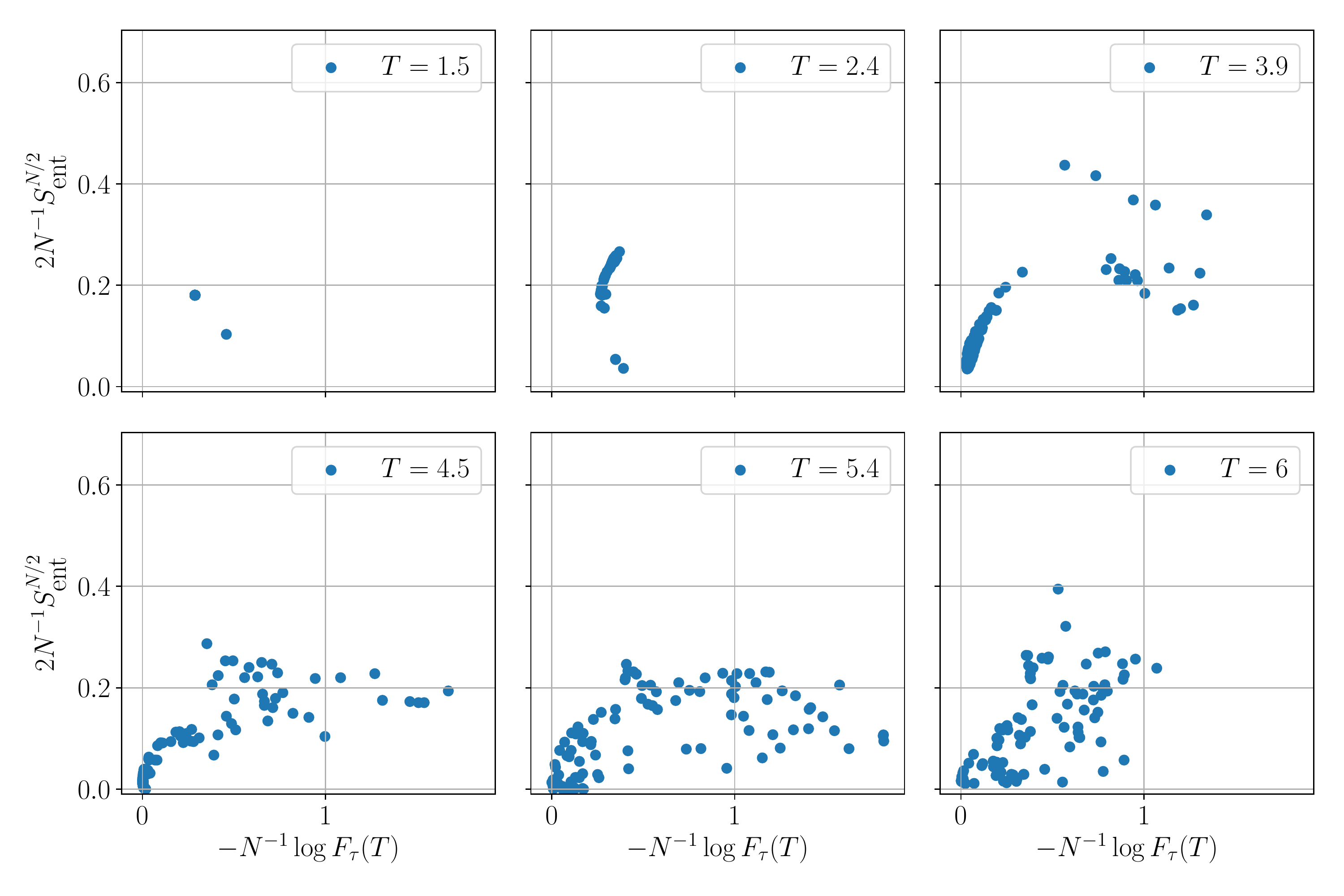}
    \caption{\label{fig:Ising_landscape}Spin-$1/2$ Ising model: Visualization of the continuous optimization landscape for the durations $\alpha_j$ in the fidelity-entanglement entropy plane, for the best sequence found by the RL agent [see App.~\ref{app:ctrl_landscape}]. Each point corresponds to a local minimum, obtained using the SLSQP optimizer, starting from a uniformly drawn random initial condition.
        The system size is $N=16$, and the rest of the parameters are the same as in Fig.~\ref{fig:Ising_energy}.
    }
\end{figure*}

A comparison of the wall clock time for the discrete and continuous optimization steps is provided in Table~\ref{tab:timing}. We distinguish between the continuous solver optimization and the discrete RL optimization, and show the average times for \emph{one} successful step of each in the two columns on the right-hand-side. The total cost can then be obtained by multiplying the time for $t_{\mathrm{solver}}$ by the appropriate number of repetitions (e.g., continuous solver initial conditions, policy sample batch size, PPO training episodes, etc.), and by multiplying the time for $t_{\mathrm{RL}}$ by the number of PPO iterations, thereby taking into account any parallelization if used; for instance, the most expensive simulation we performed ran for about 109 hours on four nodes (Intel Xeon Skylake 6130 32-core 2.1 GHz) to produce the $N=10$, $T=12$, $q=20$ data point shown in Table~\ref{tab:comparison-eng-N10}.

We emphasize that the time $t_{\mathrm{solver}}$ required for the continuous optimization is an essential part of conventional QAOA, and is the current limiting factor for reaching large system sizes, as is the case in merely all simulations of quantum dynamics on classical computing devices. In sharp contrast, the cost for training the deep autoregressive network is $N$-independent, and $t_{\mathrm{RL}}$ per iteration is negligible; however, the choice of RL algorithm can strongly impact the number of iterations.
CD-QAOA is, thus, suitably designed for potential applications on quantum simulators and quantum computers which will enable accessing large system sizes bypassing the exponential bottleneck intrinsic to simulations of quantum dynamics on classical devices.

\section{\label{app:ctrl_landscape}Many-Body Control Landscape}

Let us briefly address the question about how hard the many-body ground state preparation problems are, that we introduced in the main text. To this end, recall that CD-QAOA has a two-level optimization structure: 
(i) discrete optimization to construct the optimal sequence of unitaries [App.~\ref{app:algo}], and 
(ii) continuous optimization to find the best angles, given the sequence, to minimize the cost function [App.~\ref{app:slsqp}]. 
Here, we focus exclusively on the continuous optimization landscape, and postpone the discrete landscape to a future study.

The RL agent learns in batches/samples of $M\!=\!128$ sequences, which sample the current policy at each iteration step and provide the data set for the policy gradient algorithm. To evaluate each sequence in the batch, we use SLSQP to optimize for the durations $\alpha_j$ in a constrained and bounded fashion: $\sum_j\alpha_j=T$ and $0 \leq \alpha_j \leq T$ [cf.~App.~\ref{app:slsqp}]. This provides us with the full unitary $U(\{\alpha_j\}_{j=1}^q ,\tau)$; applying it to the initial state we obtain the reward value for this sequence. This procedure repeats iteratively as the RL agent progressively discovers improved policies.

Once the RL agent has learned an optimal sequence, i.e.~after the optimization procedure is complete, we focus on the best sequence from the sample, and examine how difficult it is to find the corresponding durations $\alpha_j$ using SLSQP. To this end, we draw $q$ values at random from a uniform distribution over the interval $[0,T/q]$, and use them as initial conditions for the $\alpha_j$, to initialize the SLSQP optimizer with. We use the same $q$ as the circuit depth so that the initial durations $\alpha_j^{(0)}$ are, on average, equal. We then repeat this procedure $P$ times, and generate a sample $\mathcal{M}=\big\{ \{\alpha_j^n\}_{j=1}^q\big\}_{n=1}^{P}$ of the local minima in the optimization landscape for $\alpha_j$'s.
The larger $P$, the better our result for the true reward assigned to $\tau$ is. 

Notice that, in the beginning of the training, the RL agent is still in the exploration stage and the reward estimation does not need to be too accurate; this reward estimation needs to be more accurate as the agent switches over exploitation during the end of the training. In order to make the algorithm computationally more efficient, we introduce a linear schedule for the number of realizations of the $\alpha_j$-optimizer, starting from 3 with an increment of 1 every 30 iteration steps, i.e.~$P_{\{k\}}^{\mathrm{tot}}\!=\!3 + \floor{k / 30}$, where subscript $k$ indicates the iteration number for the RL policy optimization. In order to further save time in the reward estimation, we also introduce some randomness here by sampling $P_{\{k\}}$ from a uniform distribution over $1, 2, \cdots, P_{\{k\}}^{\mathrm{tot}}$.

Even though they all correspond to the same sequence, every local minimum in $\mathcal{M}$ represents a potentially different protocol, since the durations $\alpha_j$ will cause the initial quantum state to evolve into a different final state. We can evaluate for every protocol in $\mathcal{M}$  the negative log-fidelity, $-\log F_\tau(T)$, and entanglement entropy of the half chain, $S_\mathrm{ent}^{N/2}$. Since the target state for the Ising model is an ordered ground state, it has area-law entanglement. Figure~\ref{fig:Ising_landscape} shows a cut through the landscape in the fidelity-entanglement entropy plane for a few different durations $T$ for the spin-$1/2$ Ising model. The better solutions are located in the lower left corner. The proliferation of local minima across the quantum speed limit has recently been studied in the context of RL~\cite{day2019glassy} and QAOA~\cite{matos2020quantifying}. This behavior indicates the importance of running many different SLSQP realizations, or else we may mis-evaluate the reward of a given sequence and the policy gradient will perform poorly.

Figure~\ref{fig:Ising_landscape} also provides a plausible explanation for the destruction of the scaling collapse for $T\gtrsim T_\mathrm{QSL}$ [Fig.~\ref{fig:Ising_energy_scaling}]. Although the precision of the SLSQP optimizer is set at $10^{-6}$, the energy curves for large durations no longer fall on top of each other with a larger relative error. Hence, the occurrence of many local minima of roughly the same reward, which correspond to different protocols, effectively removes any universal features from the obtained solution; therefore, different system size simulations end up in different local minima.

\section{\label{app:varl_gauge}Variational Gauge Potentials}

Consider the generic Hamiltonian
\begin{equation}
    H(\lambda) = H_0 + \lambda H_1,
\end{equation}
with a general smooth function $\lambda=\lambda(t)$. We define a state preparation problem where the system is prepared in the ground state of $H_0$ at time $t=0$, and we want to transfer the state population in the ground state of $H$ by time $t=T$.

Unlike adiabatic protocols, counter-diabatic driving relaxes the condition of being in the instantaneous ground state of $H(\lambda)$ during the evolution. The idea is to reach the target state in a shorter duration $T$ (compared to the adiabatic time) at the expense of creating controlled excitations [w.r.t.~the instantaneous $H(\lambda)$] during the evolution, which are removed before reaching the final time $T$. To achieve this, one can define a counter-diabatic Hamiltonian $H_\mathrm{CD}$. In general, the original $H(\lambda)$ differs from $H_\mathrm{CD}$, whose ground state the system follows adiabatically:
\begin{equation}
    H_\mathrm{CD}(\lambda)=H(\lambda)+\dot\lambda A_\lambda,
\end{equation}
where $A_\lambda$ is the gauge potential; $A_\lambda$ is defined implicitly as the solution to the equation~\cite{kolodrubetz2017geometry}
\begin{equation}
    [\partial_\lambda H +i [A_\lambda, H], H]=0.
    \label{eq:gauge_pot}
\end{equation}
The boundary conditions $H_\mathrm{CD}(\lambda(0))=H(\lambda(0))$ and $H_\mathrm{CD}(\lambda(T))=H(\lambda(T))$ impose the additional constraint $\dot\lambda(0)=0= \dot\lambda(T)$ which suppresses excitations at the beginning and at the end of the protocol.

Using Eq.~\eqref{eq:gauge_pot}, one can convince oneself that the gauge potential $A_\lambda$ of a real-valued Hamiltonian $H$ is always imaginary-valued~\cite{kolodrubetz2017geometry}.

For generic many-body systems, it has recently been argued that the gauge potential $A_\lambda$ is a nonlocal operator~\cite{sels2017minimizing}. Nevertheless, one can proceed by constructing a variational approximation $\mathcal{X}\approx A_\lambda$, which minimizes the action
\begin{equation}
    \label{eq:action}
    \mathcal{S}(\mathcal{X}) = \langle G^2(\mathcal{X})\rangle-\langle G(\mathcal{X})\rangle^2, \quad G(\mathcal{X}) = \partial_\lambda H + i[\mathcal{X},H].
\end{equation}
For ground state preparation, $\langle\cdot\rangle = \langle\psi_\mathrm{GS}(\lambda)|\cdot|\psi_\mathrm{GS}(\lambda)\rangle$ is the instantaneous ground state expectation value w.r.t.~$H(\lambda)$.
More generally, one can use $\langle\cdot\rangle = \mathrm{Tr}(\rho_\mathrm{th}\times(\cdot))$, where $\rho_\mathrm{th}\propto\exp(-\beta H)$ is a thermal density matrix at temperature $\beta^{-1}$: for $\beta \to \infty$, we recover the ground state expectation value; for $\beta\to 0$ all eigenstates are weighted equally.

We mention in passing that alternative schemes to approximation the adiabatic gauge potential have also been considered~\cite{hegade2020shortcuts}.

\subsection{Spin Hamiltonians}

\subsubsection{Real-valued Spin-\texorpdfstring{$1/2$}{1/2} Hamiltonians}

Let $H$ now be a real-valued spin-$1/2$ Hamiltonian with translation and reflection invariance. Such a system is given, e.g., by the mixed-field Ising model, discussed in the main text. We now construct an ansatz for the variational gauge potential $\mathcal{X}$ which obeys these symmetries, and is imaginary valued. %

We can organize the terms contained in $\mathcal{X}$ according to their multi-body interaction type, as follows. The only single-body imaginary valued term we can write is $\sum_j \beta_j S^y_j$. Translation and reflection symmetries, whenever present in $H$, further impose that the coupling constant $\beta_j=\beta$ be site-independent, i.e.~spatially uniform. Hence, this is the zeroth-order term in our variational gauge potential construction, cf.~Eq.~\eqref{eq:chi_Ising}.

Next, we focus on the two-body terms. Because the exact $A_\lambda$ is imaginary valued for real-valued Hamiltonians, we may only consider interaction terms where $S^y$ appears precisely once: $S^xS^y$ and $S^yS^z$. For spin-$1/2$ systems, the two operators have to act on different sites, or else one can further simplify their product to single-body operators using the algebra for Pauli matrices. Once again, translation invariance dictates that the coupling constants are uniform in space, while reflection invariance requires us to take a symmetric combination. Imposing further that the interaction be short-range (we want to construct the most local variational ansatz), we arrive at
\begin{eqnarray}
    \label{eq:chi_Ising}
    \mathcal{X}(\{\beta_l^{(k)}\}) \!=\! \sum_j && \beta^{(0)}_{0}(\lambda) S^y_j  +  \beta^{(0)}_{1}(\lambda)\left(S^x_{j+1} S^y_j \!+\! S^y_{j+1} S^x_j\right) + \nonumber\\
    &&+ \beta^{(1)}_{1}(\lambda)\left(S^z_{j+1} S^y_j \!+\! S^y_{j+1} S^z_j\right).
\end{eqnarray}
The coefficients $\beta^{(k)}_{l}$ are the variational parameters that we need to determine to find the approximate CD protocol. To find their optimal values, we minimize the action $\mathcal{S}(\mathcal{X})$~\cite{kolodrubetz2017geometry}. Note that, since we do not have a closed-form expression for the instantaneous ground state of $H(\lambda)$, we do the minimization numerically at every fixed time $t$ along the protocol $\lambda(t)$ [cf.~App.~\ref{app:CD_numerical_min}].

We can, in principle, add the next order terms to the series; however, they will either be less local, or consist of three- and higher-body interactions, which is hard to implement in experiments.

\subsubsection{Real-valued Spin-\texorpdfstring{$1$}{1} Hamiltonians}

The situation is more interesting for spin-$1$ systems: the eight-dimensional Lie algebra $\mathfrak{su}(3)$, which generates SU(3), contains three distinct imaginary-valued directions, which form a closed subalgebra $\mathfrak{su}(2)\subsetneq \mathfrak{su}(3)$, and hence there is more room to generate imaginary-valued combinations. To find all imaginary-valued terms consistent with a set of symmetries, we use QuSpin's functionality to implement an algorithm [App.~\ref{app:gauge_pot_algo}] that lists them for generic bases~\cite{weinberg2017quspin,weinberg2019quspin}.

\textbf{\textit{Translation and Reflection Symmetric spin-$1$ Hamiltonians}}, such as the spin-$1$ Ising and Heisenberg models, have a similar expansion to their spin-$1/2$ counterparts, but allow for more terms. Restricting the expansion to two-body terms, we have
\begin{widetext}
    \begin{eqnarray}
        \label{eq:chi_Heisenberg_TP}
        \mathcal{X}(\{\beta^{(k)}_l\}) \!=\!  \sum_j  &\bigg[&  \beta^{(0)}_{0}(\lambda) S^y_j   +
            \beta^{(0)}_{1}(\lambda)\left(S^x_{j} S^y_j \!+\! S^y_{j} S^x_j\right) \!+\! \beta^{(0)}_{2}(\lambda)\left(S^z_{j} S^y_j \!+\! S^y_{j} S^z_j\right)+
            \\
            &&\!+\! \beta^{(1)}_{0}(\lambda)\left([S^x_{j+1} -a S^x_{j}]S^y_j \!+\! [S^y_{j+1} -aS^y_j]S^x_j\right)
            \!+\! \beta^{(1)}_{1}(\lambda)\left( [S^z_{j+1}-bS^z_{j}] S^y_j \!+\! [S^y_{j+1}-bS^y_{j}] S^z_j\right)  \bigg].  \nonumber
    \end{eqnarray}
\end{widetext}
where the constants $a$ and $b$ are chosen so that all five terms are mutually orthogonal w.r.t.~the scalar product induced by the trace (i.e.~Hilbert-Schmidt) norm; this ensures the linear independence of the constituent terms.
Note that the three imaginary-valued on-site terms correspond precisely to the imaginary-valued $\mathfrak{su}(2)\subsetneq \mathfrak{su}(3)$.

Adding \textbf{\emph{Magnetization Conservation and Spin Inversion Symmetry}} further reduces the allowed terms in the series. Therefore, one has to restrict to three- and four-body terms:
\begin{widetext}
    \begin{eqnarray}
        \label{eq:chi_Heisenberg_MTPZ}
        \mathcal{X}(\{\zeta^{(k)}_l\}) \!=\!  \sum_j
        \zeta^{(2)}_{0}(\lambda)\left( i S^+_jS^-_{j+1}S^z_{j+2} +  i S^z_jS^-_{j+1}S^+_{j+2} + \mathrm{h.c.} \right) +
        \zeta^{(3)}_{0}(\lambda)\left( iS^-_jS^z_j S^+_{j+1}S^z_{j+1}   +  i S^+_jS^z_j S^-_{j+1}S^z_{j+1} + \mathrm{h.c.} \right),  \nonumber\\
    \end{eqnarray}
\end{widetext}
Because these terms are multi-body and less local, we refrain from using them in CD-QAOA in the present study. We merely list them here for completeness.

As explained in the main text, to apply CD-QAOA for many-body ground state preparation, we consider the constituent terms in $\mathcal{X}$ as independent generators $\{H_j\}_{j=1}^{|\mathcal{A}|}$. This comes in contrast to the variational gauge potential method where the ratios between the coefficients $\beta^{(k)}_{l}$ play an important role.

\begin{figure}[t!]
    \includegraphics[width=1.0\columnwidth]{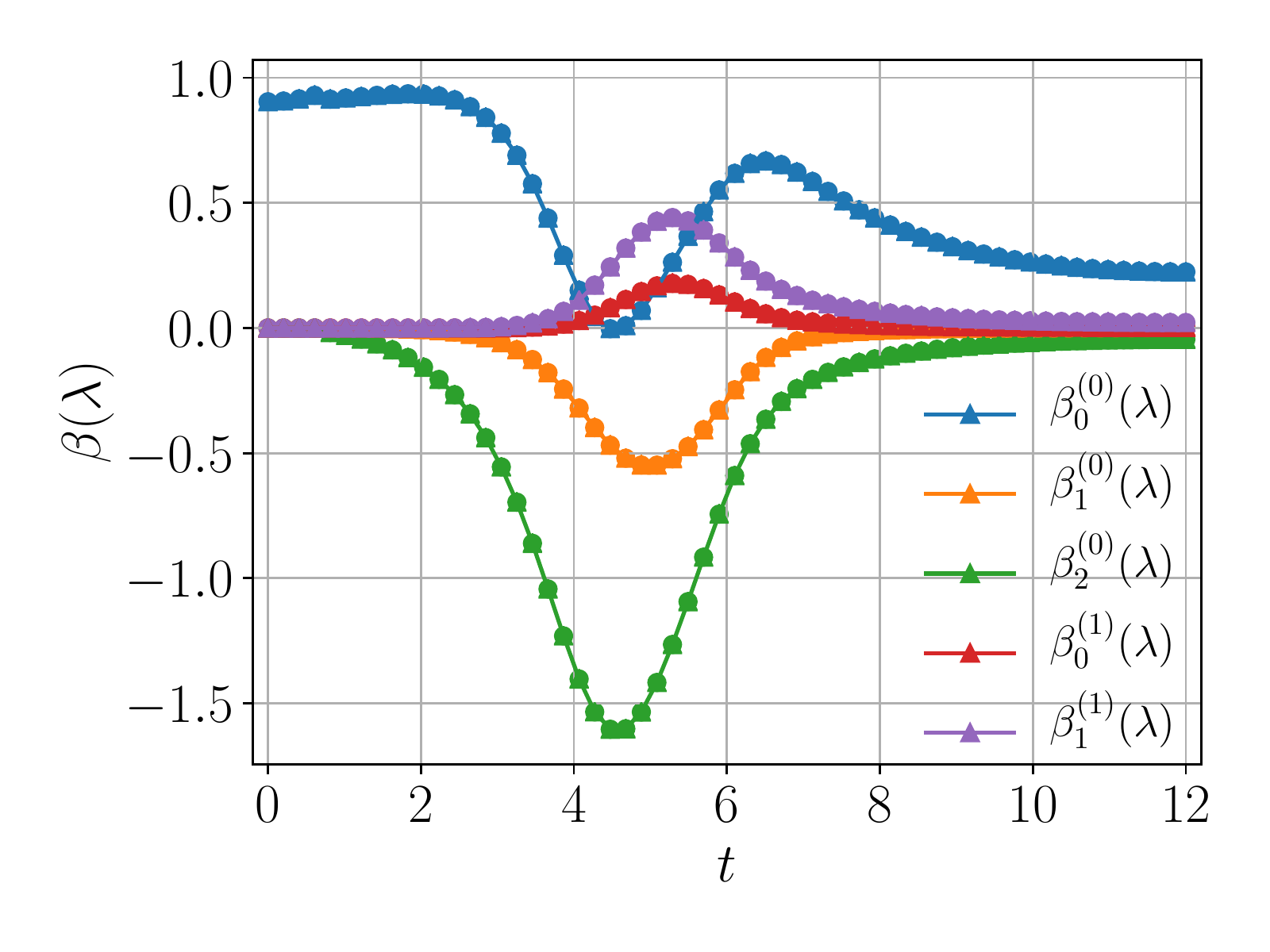}
    \caption{\label{fig:beta_vs_time} Spin-$1$ Ising chain.
        Time dependence of the optimal coefficients $\beta^{(k)}_{l}(\lambda(t))$ in the variational gauge potential (Eq.~\ref{eq:chi_Heisenberg_TP}) with translation and reflection symmetry, determined from the procedure in App.~\ref{app:CD_numerical_min}. 
        The total duration $T=12$ with the time discretization step $\Delta t\!=\!0.2$, and the system size $N=8$. 
        The protocol we used is $\lambda(t)\!=\! \sin^2\left(\frac{\pi t}{2T}\right)$. The other model parameters are the same as in Fig.~\ref{fig:comparison-eng}.  
    }
\end{figure}

\subsubsection{\label{app:LMG}Variational Gauge Potential Ansatz for the Lipkin-Meshkov-Glick Model}

As explained in the main text, the Lipkin-Meshkov-Glick (LMG) Hamiltonian, cf.~Eq.~\eqref{eq:LMG_ham}, models homogeneously interacting spin-$1/2$ particles on an all-to-all connected graph in the presence of an external field. Here, we compute the lowest-order terms appearing in the series for the variational gauge potential $\mathcal X$, going beyond Ref.~\cite{hatomura2017shortcuts}. 

The starting point is the LMG Hamiltonian
\begin{equation}
	H = -\frac{J}{N} \left(S^x\right)^2 + h\left(S^z + N/2\right).
\end{equation}
We introduce two bosonic modes, $s$ and $t$, where $S^z=t^\dagger t - N/2 = n_t-N/2$ and $S^+ = t^\dagger s$, and cast the LMG Hamiltonian in the form
\begin{equation}
	H = h t^\dagger t - \frac{J}{4N}\left(t^\dagger s + s^\dagger t\right)^2.
\end{equation}

Recalling once again that real-valued Hamiltonians have imaginary-valued gauge potentials, and that gauge potentials do not have diagonal matrix elements, we make the following ansatz:
\begin{equation}
	\label{eq:LMG_gauge_pot}
	\mathcal{X}(\{\beta^{(k)}_l\}) \!=\!  \beta^{(0)}_{0}(\lambda){Y}  +  \beta^{(1)}_{1}(\lambda)  \hat{XY}  +  \beta^{(0)}_{1}(\lambda)  \hat{ZY},
\end{equation}
where
\begin{eqnarray}
    {Y}&=& S^y = \frac{i}{2}\left(s^\dagger t - t^\dagger s\right), \nonumber\\
	\hat{XY} &=&  \frac 1 N \qty(S^xS^y + S^yS^x )= -\frac{i}{2N}\left[(t^\dagger s)^2 - (s^\dagger t)^2\right], \nonumber\\
	\hat{ZY} &=& \frac 1 N \qty(\left(S^z+\frac{N}{2}\right)S^y + S^y\left(S^z+\frac{N}{2}\right)) \nonumber\\
	&=& \frac i {2N} \qty( s ^\dagger t^\dagger t t - s  t^\dagger t t ^\dagger + s^\dagger t t ^\dagger t - st^\dagger t^\dagger t ).
\end{eqnarray}

To compute the matrix elements of the gauge potentials, we define the basis 
$$|N,n_t\rangle = \frac{\left(t^\dagger \right)^{n_t} \left(s^\dagger \right)^{N-n_t}}{\sqrt{n_t!(N-n_t)!}}|0\rangle,\quad \mathrm{with}\quad n_t=0,\dots,N. $$ 
The gauge potentials have the following non-zero matrix elements (plus their conjugates to make the operators hermitian):
\begin{widetext}
\begin{eqnarray}
	\langle N,n_t| {Y} |N,n_t+1\rangle &=& -\frac{i}{2}\sqrt{(n_t+1)(N-n_t)}, \nonumber\\
	\langle N,n_t| \hat{XY} |N,n_t+2\rangle &=& \frac{i}{2N}\sqrt{(n_t+2)(n_t+1)(N-n_t-1)(N-n_t)}, \nonumber\\
	\langle N,n_t| \hat{ZY} |N,n_t+1\rangle &=& \frac{i}{2N}(2n_t+1)\sqrt{(n_t+1)(N-n_t)}.
\end{eqnarray}
\end{widetext}

\subsection{\label{app:CD_numerical_min}Numerical Minimization to obtain the Variational CD Protocol}

Since the action $\mathcal{S}$ in Eq.~\eqref{eq:action} is quadratic in the variational parameters $\beta_j$, it is possible to derive a generic linear system, whose solutions are the optimal parameters of the variational gauge potential within CD driving~\cite{mzaouali2021work}.

Suppose that  $\mathcal X = \sum_{j=1}^{r}\beta_{j} H_{j}$ is given by a linear combination of $r$ gauge potential terms. Then, it is straightforward to see that
\begin{equation}
    G(\mathcal X) = \partial_\lambda H + \sum_{j=1}^{r} i[ H_{j},H] \beta_{j}.
\end{equation}
Defining the operator-valued quantities $B_0=\partial_\lambda H$ and $B_j=i[ H_{j},H]$  and setting $\beta_0=1$, we arrive at the following expression for the variational action
\begin{eqnarray}
    \mathcal{S}(\mathcal{X}) &=&  \left\langle \left( B_{0} + \sum_{j} B_{j} \beta_{j} \right) ^{2}\right\rangle - \left(\langle B_{0} + \sum_{j} B_{j} \beta_{j} \rangle \right) ^{2} \nonumber\\
    & =& \sum_{ i, j=0}^{r} \bigg (\langle B_{i} B_{j} \rangle - \langle B_{i} \rangle \langle B_{j} \rangle\bigg) \beta_{i}\beta_{j},
\end{eqnarray}
which is a quadratic form in the unknown coefficients $\beta_j$. To find the minimum of $\mathcal{S}(\mathcal{X})$ w.r.t.~$\beta_j$, we can take the derivative and set it to zero, to obtain the linear system of equations for the optimal $\beta_j$:
\begin{equation}
    \sum_{k}  \mathcal{M}_{jk} \beta_{k}  = -  \mathcal{M}_{0j}
\end{equation}
where $\mathcal{M}_{jk} = \langle B_{j} B_{k} \rangle +  \langle  B_{k} B_{j}\rangle - 2 \langle  B_j \rangle \langle B_k \rangle$.
Solving the system we obtain the minimum $\{\beta_j\}_{j=1}^r$ of the variational action $\mathcal{S}$.

The ground state expectation values in the above procedure, as well as the Hamiltonian $H(\lambda(t))$ depend implicitly on time $t\in[0,T]$ via the protocol $\lambda(t)$. Therefore, to find the time dependence of $\beta_j(t)$, we discretize the time interval $[0,T]$ into $N_T$ time steps, and repeat the procedure at every time step. This yields $\beta_j(t_i)$ at the time steps $t_i$. To recover the full functional dependence, we use a fine discretization mesh, and apply a linear interpolation to $\beta_j(t_i)$. Alternatively, notice that the coefficients $\beta_j=\beta_j(\lambda(t))$ depend on time $t$ only implicitly via the protocol $\lambda$. Therefore, it is also possible to discretize the range of $\lambda(t)$ instead.

For the spin-$1$ Ising model, the time-dependence of $\beta_j$ is shown in Fig.~\ref{fig:beta_vs_time}. This defines $H_\mathrm{CD}$ which generates the CD evolution. In Sec.~\ref{sec:comparison} and App.~\ref{app:spin-1}, we compare variational CD driving to CD-QAOA and conventional QAOA.

\subsection{\label{app:gauge_pot_algo}Algorithm for Generating Gauge Potential Terms in the Presence of Lattice Symmetries}

Finally, we also show the algorithm we used to determine the terms appearing in the gauge potential expansions in Eqn.~\eqref{eq:chi_Ising},~\eqref{eq:chi_Heisenberg_TP}, and~\eqref{eq:chi_Heisenberg_MTPZ}, which obey a fixed set of symmetries.

In general, one can represent any local operator of the kind $J_{i_1,\cdots,i_l}O_{i_1}^{\gamma_1}\cdots O_{i_l}^{\gamma_l}$ as a triple ($\mathcal Y$, $\mathcal I$, $J$), where $J=J_{i_1,\cdots,i_l}$ is the coupling coefficient constant, $\mathcal{I}=(i_1,\cdots\!,i_l)$ is the set of sites the operators act on, and $ \mathcal Y=(\gamma_1,\cdots\!,\gamma_l)$ defines the types of operators that act on the corresponding sites; the triple ($\mathcal Y$, $\mathcal I$, $J$) can then be used to construct the operator.

In the following, we refer to the separate terms appearing in the gauge potential series as `Hamiltonians' $H_j$, i.e.~$\mathcal{X}=\sum_j \beta_j H_j$; a Hamiltonian is defined as $H = \sum_{(i_1,\cdots,i_l)\in \Lambda}J_{i_1,\cdots,i_l}O_{i_1}^{\gamma_1}\cdots O_{i_l}^{\gamma_l}$, where $\Lambda$ is the lattice graph. As we argued above, real-valued Hamiltonians have purely imaginary-valued gauge potentials; thus, the coefficient $J$ is chosen to be purely imaginary.

We build the series for the variational gauge potential $\mathcal{X}$ recursively: we first consider a set $\mathcal L_\mathrm{elem}$ of elementary operators $O$ --- the building blocks for the expansion: e.g., for the spin-$1$ chains, these can be the spin-$1$ operators $\mathcal L_\mathrm{elem}=\{S^+, S^-, S^z\}$.
We want to construct the terms in the expansion for $\mathcal{X}$ iteratively at a fixed order $l$, e.g.~$l=1$ comprises single-body terms, $l=2$ -- two-body terms, etc. We also assume that we have access to a routine which checks if a trial list of operators obeys a given lattice symmetry; if not, the same routine returns the missing operators to be added to the original list, so that the symmetry is now satisfied [e.g.,~such a routine is used in QuSpin~\cite{weinberg2017quspin,weinberg2019quspin}].

The pseudocode we developed is shown in Algorithm~\ref{alg:basis-gen}.
To construct multi-body terms at a fixed order $l$, we define combinations of the elementary operators, and store them in the list $\mathcal L_\mathrm{op}$; the way these combinations are built can be used to implement constraints, such as particle/magnetization conservation, etc. This is implemented via the product operator (Line~\ref{alg:prod} of Algorithm~\ref{alg:basis-gen}). It generates all possible combinations of selecting $l$ elementary operators with replacement.
The sets of lattice sites that the operators from $\mathcal L_\mathrm{op}$ act on, are stored in the list $\mathcal L_\mathrm{sites}$ (Line~\ref{alg:site} of Algorithm~\ref{alg:basis-gen}).
Then, for each trial triple  ($\mathcal Y$, $\mathcal I$, $J$), we make use of the routine to check the symmetry and record any operators which do not respect it. We append these, so-called \emph{missing operators}, to the original list, and we keep checking the symmetry condition until we obtain all operators that satisfy the symmetry (Line~\ref{alg:sym_beg}\,-\ref{alg:sym_end} of Algorithm~\ref{alg:basis-gen}). The finite number of combinations guarantees a termination in a finite number of steps.

\begin{algorithm}[H]
    \caption{Generation of variational gauge potential}\label{alg:basis-gen}
    \algcomment{\centering \fontsize{7.2pt}{2em}\selectfont \texttt{product}: \href{https://docs.python.org/2/library/itertools.html\#itertools.product}{Cartesian product}, \texttt{equivalents}: equivalent mod scalar, \\ \vspace{-0.32cm} \texttt{missing operator}: the operator missed for the symmetry requirement}
    \begin{algorithmic}[1]
        \Require a list of required symmetries $\mathcal L_{\text{sym}}$, order $l$, a list of elementary operator types $\mathcal L_{\text{elem}}$.
        \State Initialize empty list for gauge potential terms $\mathcal L_{\text{gauge}}$.
        \State Generate all possible combinations of local operators at order $l$
        \vspace{-0.5em}
        $$\mathcal L_{\text{op}} = \text{product}(\mathcal L_{\text{elem}}, \text{repeat}=l).$$
        \vspace{-1.5em}
        \label{alg:prod}
        \State Enumerate all possible combinations of lattice sites $\mathcal L_\mathrm{sites}$ the $l$-th order operators act on.\label{alg:site}
        \For{$\mathcal Y$ in $\mathcal L_{\text{op}}$}
        \For{$\mathcal I$ in $\mathcal L_{\text{sites}}$}
        \State Initialize an empty list $\mathcal L_{H}$
        \State Set $J = i\; $($i=\sqrt{-1})$.
        \State Append ($\mathcal Y$, $\mathcal I$, $J$) to  $\mathcal L_{H}$.
        \State Set the flag $IsSym = \textit{False}.$
        \While{$IsSym$ is $False$} \label{alg:sym_beg}
        \State Set $IsSym = \textit{True}.$
        \For{sym in $\mathcal L_{\text{sym}}$}
        \If{exists missing operator
            ($\mathcal Y'$, $\mathcal I'$, $J'$) }
        \State Set $IsSym = \textit{False}.$
        \State Append ($\mathcal Y'$, $\mathcal I'$, $J'$) to $\mathcal L_{H}$. \label{alg:sym_end}
        \EndIf
        \EndFor
        \EndWhile
        \State Build Hamiltonian $H$ using the triplets in $\mathcal L_{H}$.
        \If{ $H$ or equivalents  not included in $\mathcal L_{\text{gauge}}$} \label{alg:equiv}
        \State Append  $H$ to $\mathcal L_{\text{gauge}}$ .
        \EndIf
        \EndFor
        \EndFor
        \State \textbf{Return} the list of gauge potential basis  $\mathcal L_{\text{gauge}}$.
    \end{algorithmic}
\end{algorithm}

In order to avoid repeating previously identified Hamiltonians, we discard \emph{equivalent} Hamiltonians (Line~\ref{alg:equiv} of Algorithm~\ref{alg:basis-gen}): two Hamiltonians are called equivalent when one is a scalar times the other. Since here we consider imaginary-valued gauge potentials, the multiple constant should be real. To test whether the Hamiltonians $H_1$ and $H_2$ are equivalent in practice, it suffices to test whether $H_1$ is equal to $\pm \frac{\Vert H_1 \Vert }{\Vert H_2 \Vert} H_2$, where we use the Hilbert-Schmidt norm.

\section{\label{app:misc}CD-QAOA for Many-Body State Preparation}

Here, we provide a supplementary discussion on the performance of CD-QAOA for many-body pure state preparation using the quantum spin chains introduced in the main text. We refer the reader to the main text for the definition of various model parameters; the short-hand spin operator notation used is defined in Table~\ref{table:gauge_pot}.

\subsection{\label{app:Ising}Spin-\texorpdfstring{$1/2$}{1/2} Ising Chain}

\begin{figure}[t!]
    \includegraphics[width=1.0\columnwidth]{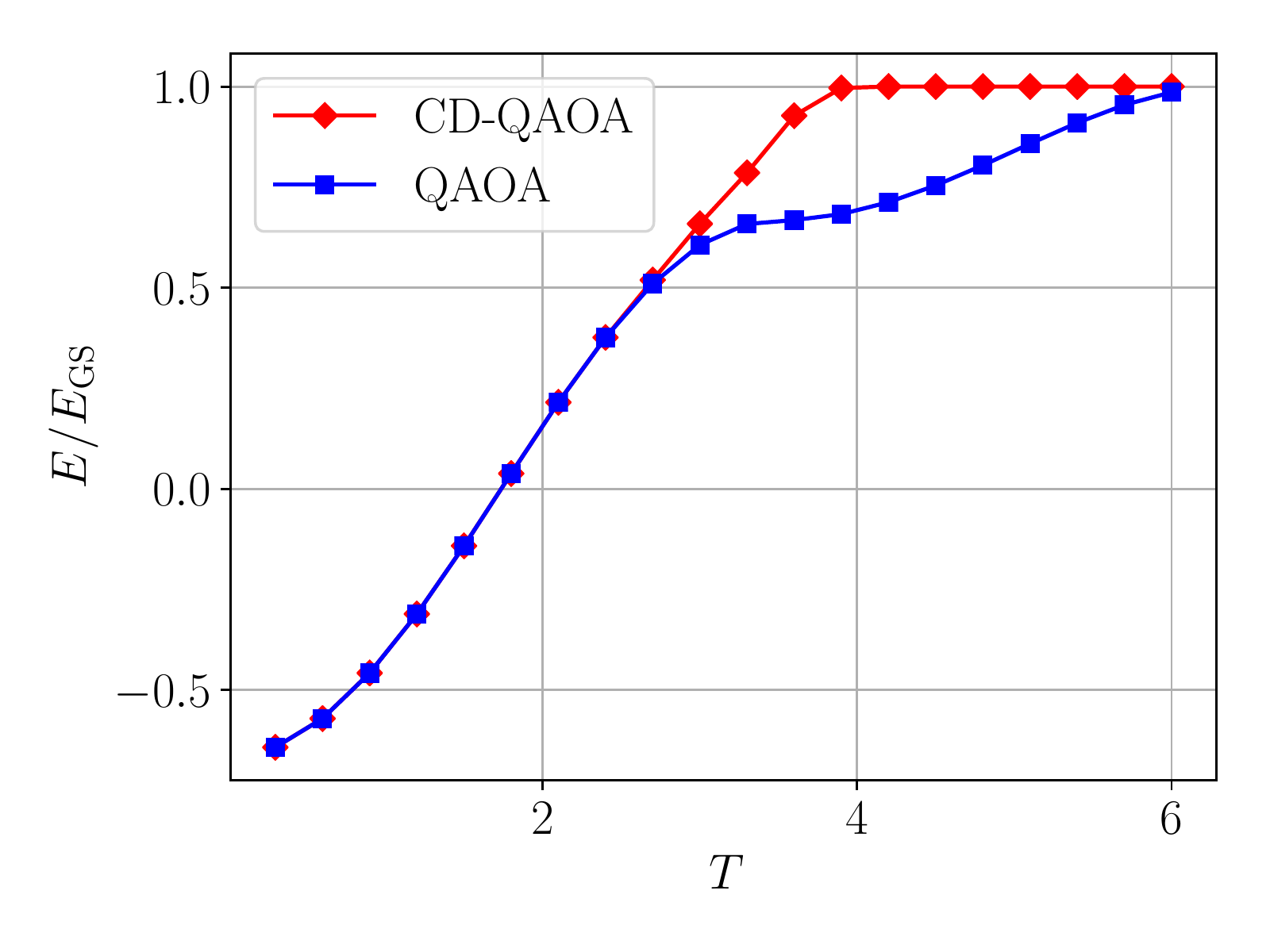}
    \caption{\label{fig:Ising_single_spin}
        Single spin-$1/2$ state preparation: energy density against protocol duration for CD-QAOA with $\mathcal{A}_\mathrm{CD-QAOA}=\{Z, X, Y\}$ (red) and conventional QAOA with $\mathcal{A}_\mathrm{QAOA}=\{Z, X\}$ (blue). The values of $q$ is 3 for both methods. For conventional QAOA, we trained two possible alternating patterns \big(i.e.~$(Z\to X\to Z)$ and $(X\to Z\to X)$\big) and pick the best one for the comparison. The model parameters are the same as in Fig.~\ref{fig:Ising_energy} with $J=0$.
    }
\end{figure}

First, we show results for the single-spin problem ($J=0$):
\begin{equation}
    H \!=\!H_1\!+\!H_2,\qquad
    H_1\!=\! h_z S^z,\quad
    H_2=h_xS^x.
\end{equation}
In Fig.~\ref{fig:Ising_single_spin}, we clearly see that CD-QAOA [red curve] has a smaller quantum speed limit $T_\mathrm{QSL}\approx 4.0$ than conventional QAOA [blue]; this is anticipated, since CD-QAOA has a larger control space at its disposal. Moreover, we find that, for $T<T_\mathrm{QSL}$, CD-QAOA only makes use of a single $Y$ rotation by setting the durations $\alpha_j$ associated with any other unitaries from the set $\mathcal{A}$, to zero. As mentioned in the main text, conventional QAOA tries to represent this $Y$-rotation by means of Euler angles, i.e.~composed of $X$ and $Z$ rotations; in general, this results in a higher duration cost to complete the population transfer (leading to a larger $T_\mathrm{QSL}$). However, for short durations $T$, a $Y$-rotation can be exactly obtained using a proper sequence of the $X$ and $Z$ terms. For these reasons, we find an exact agreement between the two curves for small values of $T\lesssim 3$.

\begin{figure}[t!]
    \includegraphics[width=1.0\columnwidth]{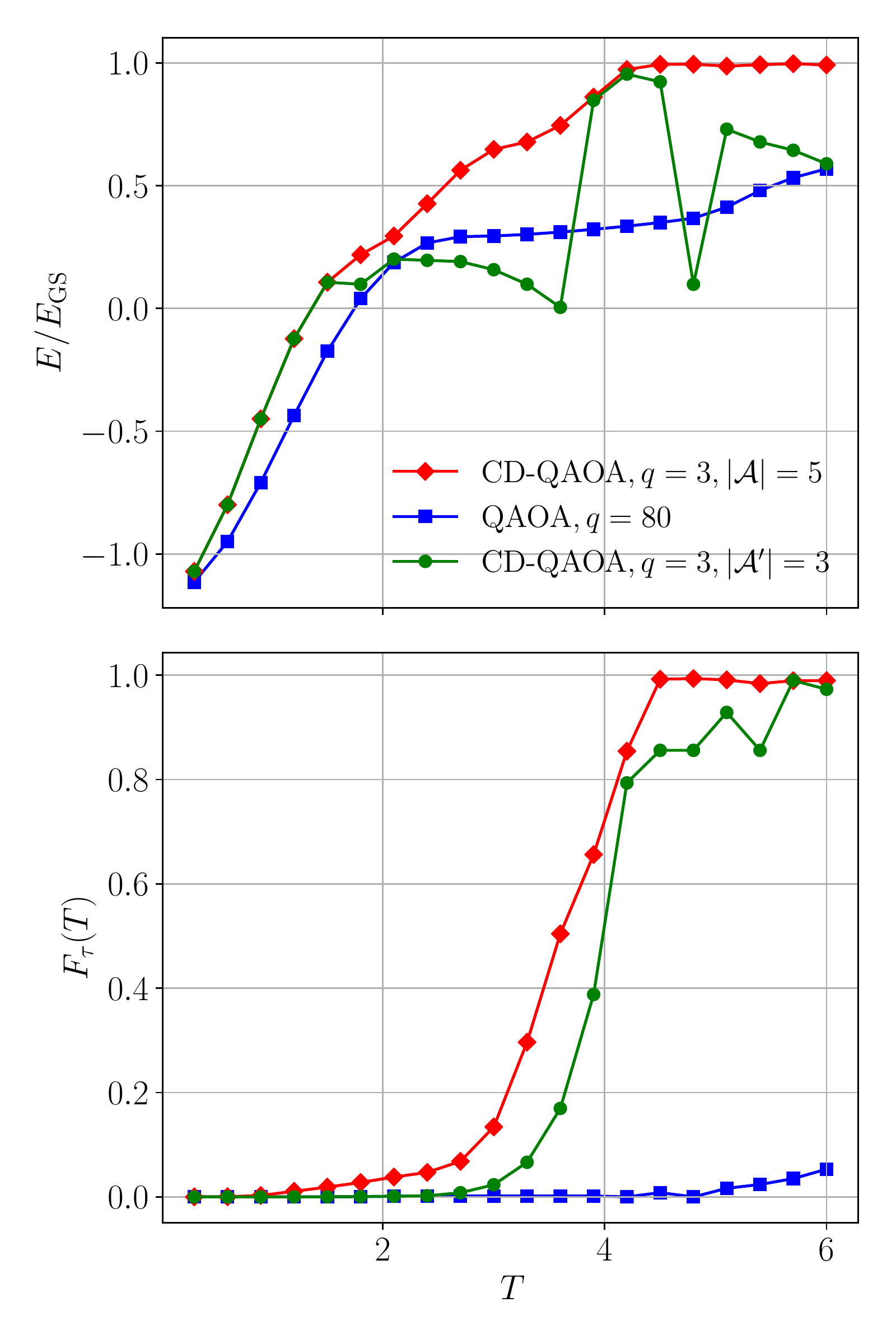}
    \caption{\label{fig:Ising_Edensity_all}
        Spin-$1/2$ Ising model:
        energy minimization (top) and many-body fidelity maximization (bottom) against protocol duration $T$.
        We compare CD-QAOA with $\mathcal{A}_\mathrm{CD-QAOA}=\{Z|Z+Z,X; Y, X|Y, Y|Z\}$ (red), CD-QAOA with $\mathcal{A}'_\mathrm{CD-QAOA}=\{Z|Z+Z,X;Y\}$ (green), and conventional QAOA with $\mathcal{A}_\mathrm{QAOA}=\{Z|Z+Z,X\}$ (blue). The model parameters are the same as in Fig.~\ref{fig:Ising_energy} with the number of spins $N\!=\!14$.
    }
\end{figure}

Let us now switch on the spin-spin interaction strength $J>0$; consider the spin-$1/2$ Ising chain
\begin{eqnarray}
    H &\!=\!&H_1\!+\!H_2,\\
    H_1\!&\!=\!&\! \sum_{j=1}^N J S^z_{j+1}S^z_j \!+\! h_z S^z_j,\quad
    H_2= \sum_{j=1}^N h_xS^x_j. \nonumber
\end{eqnarray}
Figure~\ref{fig:Ising_Edensity_all} [top] shows a comparison of the best learned energies, between conventional QAOA, and CD-QAOA for two sets  ($\mathcal{A}, \mathcal{A}'$) with different number of unitaries: $|\mathcal{A}|=5, |\mathcal{A}'|=3$ [see caption]. We find that additionally using only the single-particle gauge potential term $Y$ [green line], typically accessible in experiments, one can already obtain a higher-fidelity protocol than QAOA to prepare the ground state.
Interestingly, for short protocol durations $T$, the two-body gauge potential terms, present in $\mathcal{A}$ but not in $\mathcal{A}'$, do not contribute to improving the energy of the final state, as can be seen from the agreement of the red and green lines for $T\lesssim 1.5$. This suggests that single-particle processes dominate over many-body processes when it comes to lowering the energy of the $z$-polarized initial state, and implies that the target ground state is single-particle-like (i.e.~close to a product state). The non-smooth behavior of the green curve at larger durations, is attributed to the ruggedness of the control landscape, as different runs of the SLSQP optimizer may get stuck in one of the many suboptimal local minima [App.~\ref{app:ctrl_landscape}].

One may wonder if it is possible to prepare the ground state by straightforward fidelity maximization. We define the many-body fidelity to transfer the population to the target state using the unitary process $U(\{\alpha_j\}_{j=1}^q ,\tau)$, with $\sum_{j=1}^q \alpha_j=T$, as
\begin{equation}
    \label{eq:fidelity}
    F_\tau(T) = F(\{\alpha_j\}_{j=1}^q, \tau ) \!=\! |\langle\psi_*|U(\{\alpha_j\}_{j=1}^q ,\tau) |\psi_i\rangle|^2.
\end{equation}
The fidelity can be less relevant from the perspective of many-body physics because (i) the many-body fidelity is typically exponentially suppressed, and (ii) it requires a reference to the ground state itself (which we seek) in order to be computed. However, the fidelity of a quantum process is a widely used benchmark in quantum computing; it also provides a better measure (than energy density) for the distance between two states in the Hilbert space $\mathcal{H}$.

Figure~\ref{fig:Ising_Edensity_all} [bottom] shows the many-body fidelity for $N\!=\!14$ spins. Unlike the inset of Fig.~\ref{fig:Ising_energy} from the main text (where we show the fidelity associated with the protocol obtained using energy density minimization), here we use the fidelity as a reward function for QAOA. We observe that optimizing the fidelity behaves quantitatively similar to optimizing the energy density. We would like to emphasize here once again the advantage of the gauge potential ansatz: the conventional QAOA simulation is done using $q=80$ variational parameters $\alpha_j$ [yet no significant improvement is observed for $q\geq 4$, cf.~Fig.~\ref{fig:Ising_energy}], while CD-QAOA requires only $q=3$ variational parameters.

\begin{figure}[t!]
    \includegraphics[width=1.0\columnwidth]{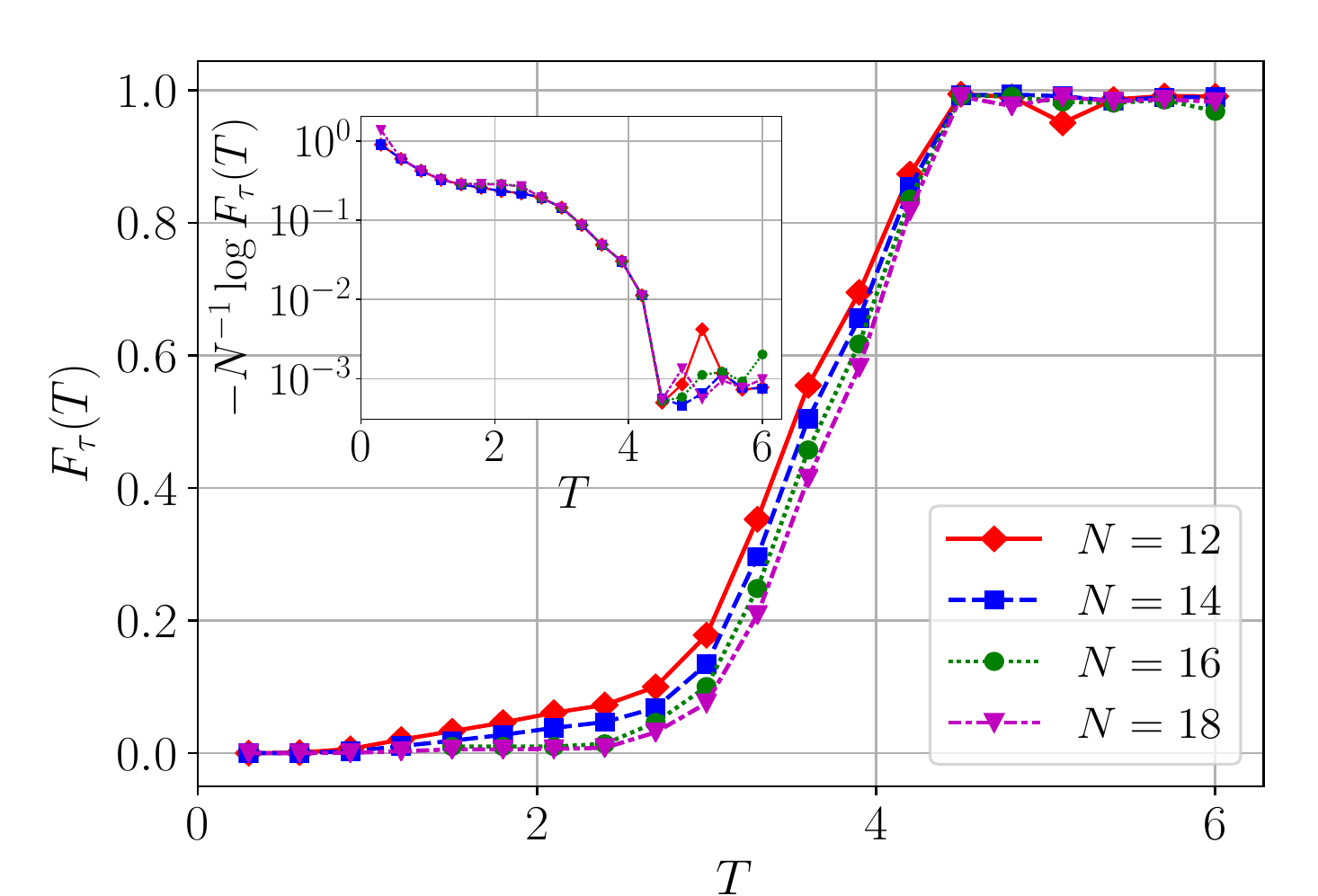}
    \caption{\label{fig:Ising_fidelity_scaling}
        Spin-$1/2$ Ising model: 
        many-body fidelity maximization and corresponding quantity [inset, log scale] against protocol duration $T$ for different system sizes $N$.
        The QAOA parameters are $q\!=\!3$ and $\mathcal{A}=\{Z|Z\!+\!Z, X;Y, X|Y,Y|Z \}$. The model parameters are the same as in Fig.~\ref{fig:Ising_energy}.
    }
\end{figure}

Although the fidelity $F_\tau(T)$ is anticipated to vanish for $T<T_\mathrm{QSL}$ in the thermodynamic limit, the negative log-fidelity density, $-N^{-1}\log F_\tau(T)$, is more likely to. Figure~\ref{fig:Ising_fidelity_scaling} [inset] shows the finite size scaling of the fidelity curves. Similar to the energy density [Fig.~\ref{fig:Ising_energy_scaling}], we obtain an almost perfect scale collapse. We verified that maximizing the fidelity produces similar results as minimizing the negative log-fidelity density for the spin-$1/2$ chain: at first sight, this is nontrivial because $F_\tau(T)$ is exponentially suppressed with the system size $N$ for $T<T_\mathrm{QSL}$; however, this behavior is likely explained by the generalization capabilities of the RL agent from small to large system sizes [cf.~Sec.~\ref{sec:generalization}].

\begin{figure*}[t!]
    \centering
    \includegraphics[width=\textwidth]{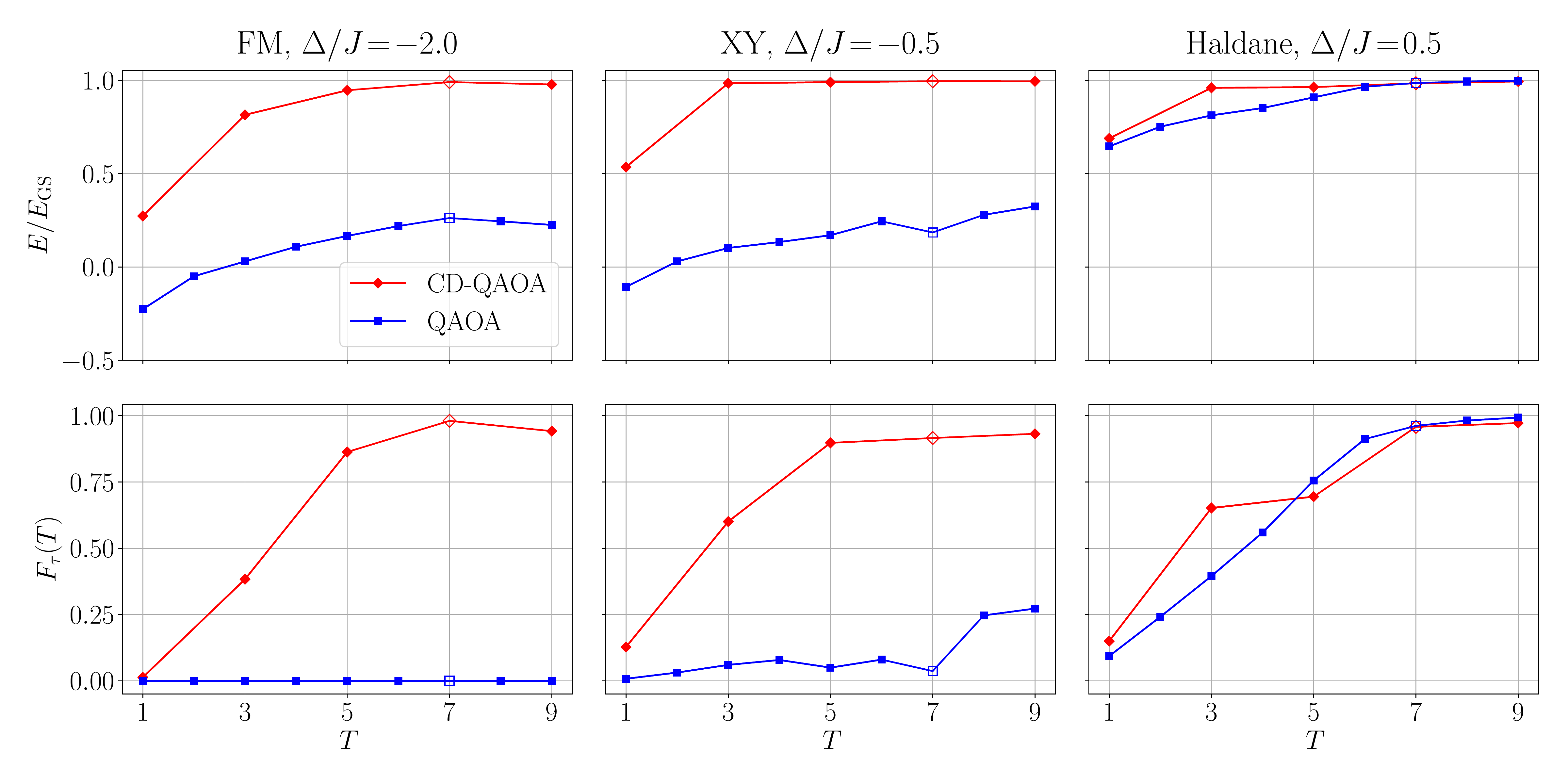}
    \caption{\label{fig:comparison-eng-fid-heisenberg-app} Anisotropic Heisenberg spin-$1$ chain:
    energy minimization against protocol duration $T$ ---  
    the corresponding energy (top row) and many-body fidelity (bottom row) for three ordered target states, corresponding to the ground state of the ferromagnetic (left, $\Delta/J\!=\!-2.0$), XY (middle, $\Delta/J\!=\!-0.5$), and Haldane (right, $\Delta/J\!=\!0.5$) target states, respectively.
    The empty symbols mark the duration at which we show the evolution of the system in Fig.~\ref{fig:comparison-heisenberg-stat}.
    The model parameters are the same as in Fig.~\ref{fig:comparison-eng-heisenberg}.
    }
\end{figure*}

\subsection{\label{app:Heisenberg}Anisotropic Spin-\texorpdfstring{$1$}{1} Heisenberg Chain}

Next, we discuss in detail the ground state preparation process in the anisotropic Heisenberg spin-$1$ chain:
\begin{eqnarray}
    H &\!=\!&H_1\!+\!H_2,\\
    H_1\!&\!=\!&\! J\sum_{j=1}^N (S_{j+1}^xS_j^x \!+\! S_{j+1}^yS_j^y),\quad
    H_2= \Delta \sum_{j=1}^N S_{j+1}^zS_j^z, \nonumber
\end{eqnarray}
where the model parameters are defined in the main text.

An important detail worth mentioning is that the ferromagnetic ground state at $\Delta/J=-2.0$ is two-fold degenerated (one state, corresponding to one of the two $z$-polarizations). While being a trivial observation, this requires certain care when analyzing the physics of the protocols the agent found. In particular, notice that energy minimization is insensitive to this degeneracy, and hence the final state can appear as an arbitrary superposition of the two ferromagnetic states, and still have the correct ground-state energy.
This leads to ambiguity when computing the fidelity of being in the target state: related to this, the cost function landscape likely develops a continuous one-dimensional structure for the (degenerate) global minima. Because we are interested in energy minimization, here we define the fidelity using the projector to the ground state manifold $P=|\psi_*^{(1)}\rangle \langle\psi_*^{(1)}|+|\psi_*^{(2)}\rangle \langle\psi_*^{(2)}|$:
\begin{eqnarray*}
    F_\tau(T) = F(\{\alpha_j\}_{j=1}^q, \tau ) \!&=&\! |\langle\psi_\ast^{(1)}|U(\{\alpha_j\}_{j=1}^q ,\tau) |\psi_i\rangle|^2 \nonumber\\
    && +|\langle\psi_\ast^{(2)}|U(\{\alpha_j\}_{j=1}^q ,\tau) |\psi_i\rangle|^2
\end{eqnarray*}
where $|\psi_\ast^{(1)}\rangle, |\psi_\ast^{(2)}\rangle$ are any two orthonormal states which span the doubly degenerate ground state manifold (e.g., the two FM ground states).

Figure~\ref{fig:comparison-eng-fid-heisenberg-app} shows a comparison between CD-QAOA and conventional QAOA for FM, XY, and Haldane target states: the top row shows the result of energy density minimization [cf.~Fig.~\ref{fig:comparison-eng-heisenberg}]. The bottom row, on the other hand, displays the many-body fidelity associated with the same protocols. 
For $\Delta/J=0.5$, CD-QAOA allows reaching the target topological Haldane state already faster, as compared to conventional QAOA.
Notice also that the gauge potential ansatz appears essential for reaching the target for both the XY ($\Delta/J=-0.5$) and FM states ($\Delta/J=-2.0$); this becomes particularly obvious from the many-body fidelity curves. The latter also reveals an interesting detail: at $\Delta/J=0.5$, a regime emerges around $T\approx 5$, where the QAOA fidelity is better than the CD-QAOA fidelity. However, this peculiarity below the quantum speed limit can be explained, recalling that the RL agent is given the (negative) energy density as the reward signal, and not the fidelity (note that CD-QAOA does outperform QAOA in energy).

\begin{figure*}[!htbp]
    \centering
    \includegraphics[width=\textwidth]{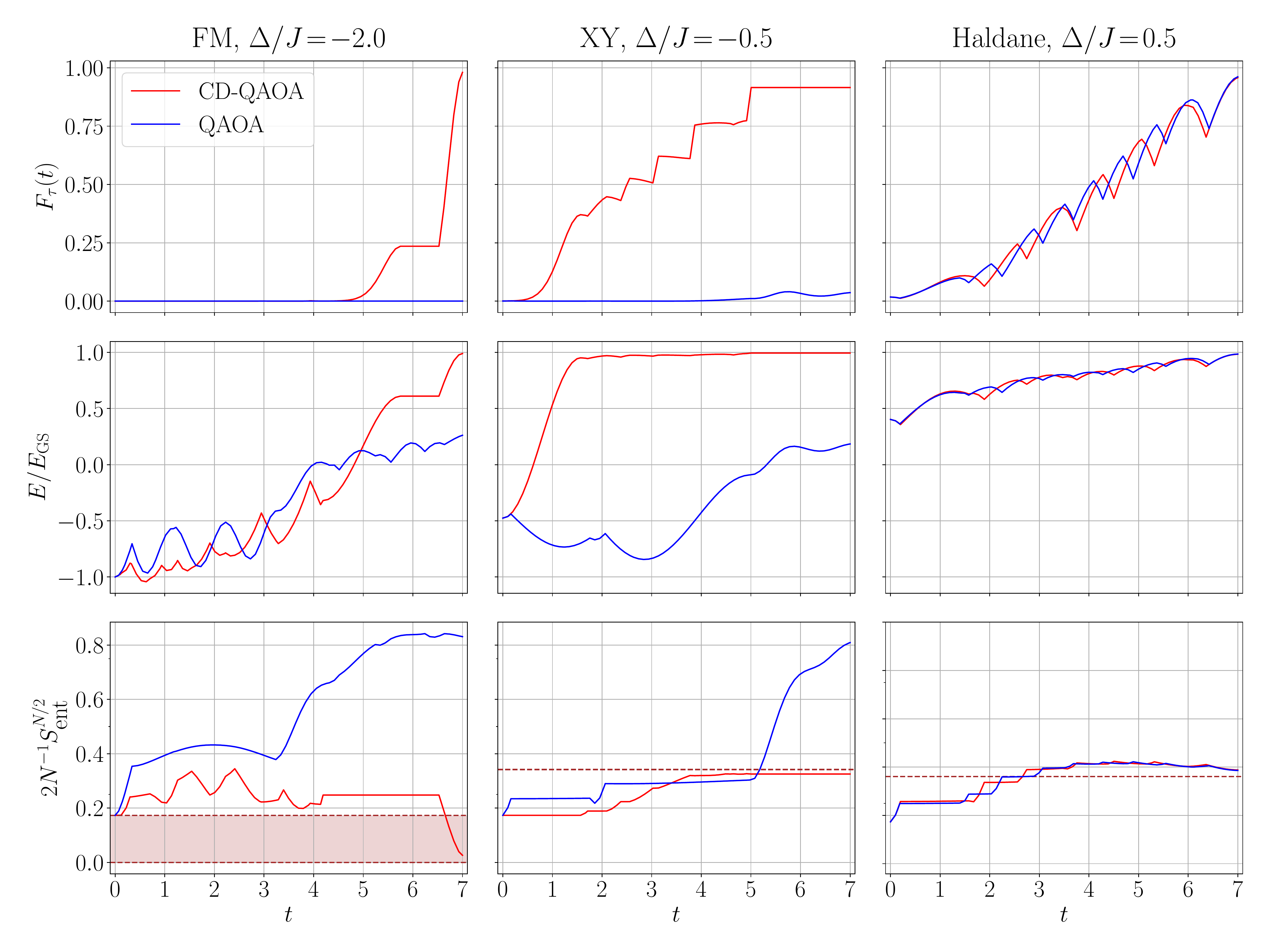}
    \caption{Anisotropic Heisenberg spin-$1$ chain: 
        time evolution generated by the protocol given by CD-QAOA (blue line), and conventional QAOA (red line) for the three target states, corresponding to the ferromagnetic ($\Delta/J\!=\!-2.0$), XY ($\Delta/J\!=\!-0.5$), and Haldane ($\Delta/J\!=\!0.5$) target state, respectively. Three quantities are shown: many-body fidelity (first row), energy ratio (second row), and the entanglement entropy density of the half chain (third row). 
        The horizontal dashed line in the entanglement entropy curve shows the value in the target state, while the shaded area for the FM state denotes that in the span of the doubly degenerate ground state manifold. 
        The protocols correspond to the duration $T=7$ in Fig.~\ref{fig:comparison-eng-heisenberg}. The related CD-QAOA protocol sequences are given in Table~\ref{tab:proto-delta=-2.0} [ferromagnetic ($\Delta/J\!=\!-2.0$)], Table~\ref{tab:proto-delta=-0.5} [XY ($\Delta/J\!=\!-0.5$)] and Table~\ref{tab:proto-delta=0.5} [Haldane ($\Delta/J\!=\!0.5$)]. The simulation parameters are the same as in Fig.~\ref{fig:comparison-eng-heisenberg}.
        }
    \label{fig:comparison-heisenberg-stat}
\end{figure*}

In order to investigate in detail in the protocols found by CD-QAOA, we fix a duration $T$, and consider the time evolution of the state, $|\psi(t)\rangle = U(\{\alpha_j\}_{j=1}^q ,\tau)|\psi_i\rangle$, for three physical quantities:
\begin{itemize}

    \item[(i)]  the energy
          \begin{equation*}
              E(t) =  \langle\psi(t)|H_\ast|\psi(t)\rangle
          \end{equation*}
          provides a measure of how far away in the cost function landscape the state is, at any given time $t\in[0,T]$.

    \item[(ii)] the instantaneous fidelity
          \begin{equation*}
              F_\tau(t) = |\langle\psi_\ast|\psi(t)\rangle|^2
          \end{equation*}
          (and its generalization to the doubly-degenerate ground state manifold), measures how far the current state is, from the target state $|\psi_\ast\rangle$ in the Hilbert space; typically, we choose the ground state as the target state $|\psi_\ast\rangle = |\psi_{\mathrm{GS}}(H)\rangle$.

    \item[(iii)] the entanglement entropy of the half chain
          \begin{equation*}
              S_\mathrm{ent}^{N/2}(t) = -\mathrm{tr}_{A} \left[\rho_A(t)\log\rho_A(t)\right],\ \rho_A(t) =\mathrm{tr}_{\bar A} |\psi(t)\rangle\langle\psi(t)|,
          \end{equation*}
          where $A$ denotes a contiguous spacial region with a complement $\bar{A}$ comprising half the periodic chain, and $\rho_A(t)$ is the reduced density matrix on $A$ at time $t$.  
          For many-body systems, it is common to look at the entanglement entropy per site, which for spin-$1$ systems lies within the interval $2N^{-1}S_\mathrm{ent}^{N/2}\in[0,\log 3]$.

\end{itemize}

Figure~\ref{fig:comparison-heisenberg-stat} shows the time evolution of the energy, fidelity and entropy density, for all three target states of interest. For $\Delta/J=0.5$, transferring the population from the AFM initial state to the Haldane state can be obtained equally well using either QAOA or CD-QAOA. Table~\ref{tab:proto-delta=0.5} shows the optimal protocol found by the RL agent: notice the three vanishing durations $\alpha_2=\alpha_{17}=\alpha_{18}=0$; factoring them out, we recover precisely the conventional QAOA sequence (albeit with $q$ odd). Thus, we see that the CD-QAOA may converge to conventional QAOA whenever the latter provides a high-reward sequence. This result exemplifies our claim that CD-QAOA generalizes QAOA successfully. Of course, it is not clear whether this is the true global minimum of the cost function landscape (the RL agent does make use of the additional gauge potential terms for $T<7$).  
Nevertheless, all physical quantities are expected to be prepared with similar accuracy under both protocols: to see this, notice that the entanglement entropy density depends only on the quantum state (unlike expectation values of observables), and that its value at $t=T$ is close to the value for the target state (dashed horizontal line).
Importantly, the entanglement remains area-law (as seen by the values being much smaller than the maximum entropy per site, $\log(3)$, suggesting the existence of a local effective Hamiltonian which generates the population transfer process dynamically.

The best sequence for targeting the XY state at $\Delta/J\!=\!-0.5$ is shown in Table~\ref{tab:proto-delta=-0.5}. Although its structure is more complicated, factoring out the vanishing $\alpha_j$, we can discern two clear patterns: (i) the sequence starts and ends with two different single-particle basis rotations, and (ii) there is an alternating subsequence based on the subset $\{X|X+Y|Y,Y\}\subsetneq\mathcal{A}_\mathrm{CD-QAOA}$. Interestingly, the only gauge potential term used by the RL agent is the experimentally accessible single-particle $Y$ rotation, and it is sufficient to reach the target with a very high many-body fidelity. For comparison, conventional QAOA appears insufficient to prepare the target state for the circuit depth of $q=18$ ($p=9$). The advantage of CD-QAOA is also visible in the entanglement entropy density curve: QAOA can easily lead to volume-law entanglement, while CD-QAOA manages to generate as little entanglement as needed for the target state.

The discrepancy between conventional QAOA and CD-QAOA is best visible in the FM state preparation at $\Delta/J=-2.0$. In this case, a na\"ive application of QAOA with the set $\mathcal{A}_\mathrm{QAOA}=\{X|X+Y|Y, Z|Z\}$ is a priori doomed to fail: starting from the initial AFM state, which is orthogonal to the target FM manifold, the resulting QAOA unitaries leave the target AFM manifold invariant; in other words, transitions between the initial and the target states are forbidden by selection rules within the QAOA dynamics. Therefore, the many-body fidelity remains zero at all times during the QAOA evolution. The energy and entanglement entropy curves certify that the state does undergo nontrivial dynamics: similar to the XY state, QAOA creates volume-law entanglement and cannot reach the FM ground state manifold in energy, while CD-QAOA is well-behaved and sufficient to prepare the target. The CD-QAOA protocol sequence is shown in Table~\ref{tab:proto-delta=-2.0}: while we do not discern an obvious pattern, we emphasize that this time the RL agent makes use of both single-particle and two-body gauge potential terms.

\begin{figure}[t!]
    \centerline{
        \includegraphics[width=1.1\columnwidth]{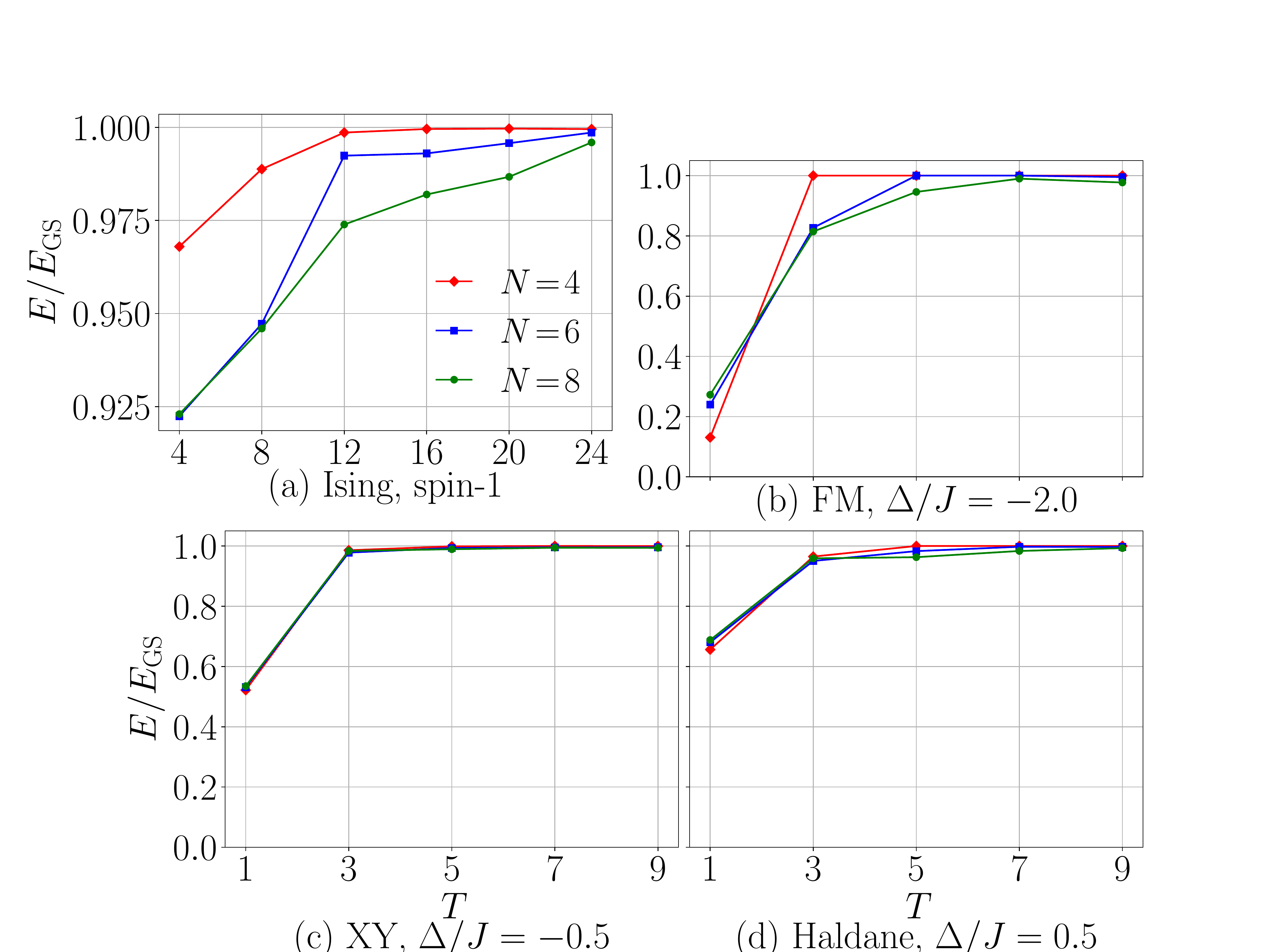}}
    \caption{system-size scaling of the energy minimization against protocol duration $T$ for different system sizes $N$:
        (a) spin-$1$ Ising chain,
        (b-d) anisotropic Heisenberg spin-$1$ chain for $\Delta/J=-2.0$, $\Delta/J=-0.5$, $\Delta/J=0.5$, respectively.
        Note that the $y$-axis scale is different for the spin-$1$ Ising model in panel (a). The model parameters are the same as in (a) Fig.~\ref{fig:comparison-eng} and (b-d)  Fig.~\ref{fig:comparison-eng-heisenberg} correspondingly.
    }
    \label{fig:E-Nscaling}
\end{figure}

Last, we show the system-size scaling of the energy curves for the three target states in Fig.~\ref{fig:E-Nscaling}(b-d). Similar to the spin$-1/2$ Ising chain, we find very little system-size dependence for the Haldane (b) and XY states (c). However, we cannot extrapolate the results to the thermodynamic limit due to the relatively small system sizes we were able to investigate. system-size effects are more pronounced for the ferromagnetic state (d), which is the one furthest away in Hilbert space from the initial perfect antiferromagnet.

Last, we mention in passing that we do not show results on preparing the AFM ground state at $\Delta/J\!=\!2.0$ since this problem is somewhat trivial: indeed, starting from a perfect AFM in the $z$-direction, the AFM ground state of the spin-$1$ Heisenberg model can be easily reached even using adiabatic evolution because it lies within the AFM phase.

\subsection{\label{app:LMG_physics}Lipkin-Meshkov-Glick model}

In the main text, we also introduced the ferromagnetic Lipkin-Meshkov-Glick (LMG) model, described by the total spin Hamiltonian
\begin{equation*}
    H = -\frac{J}{N} (S^x)^2 + h\left(S^z + \frac{N}{2}\right).
\end{equation*}

\begin{figure}[t!]
    \begin{minipage}[b]{\columnwidth}
	    \includegraphics[width=1.0\columnwidth]{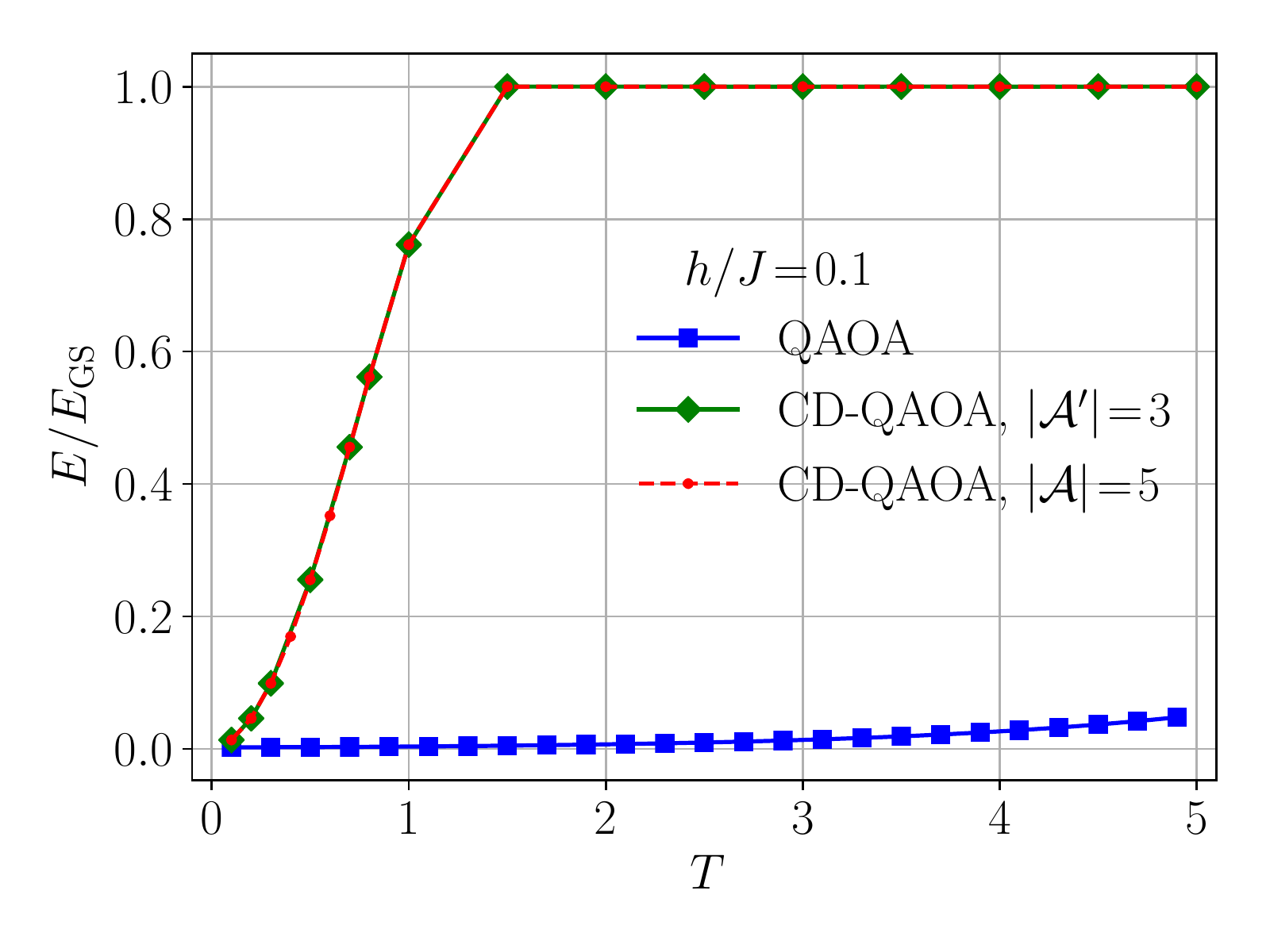}
    \end{minipage}~\\
    \begin{minipage}[b]{\columnwidth}
	    \includegraphics[width=1.0\textwidth]{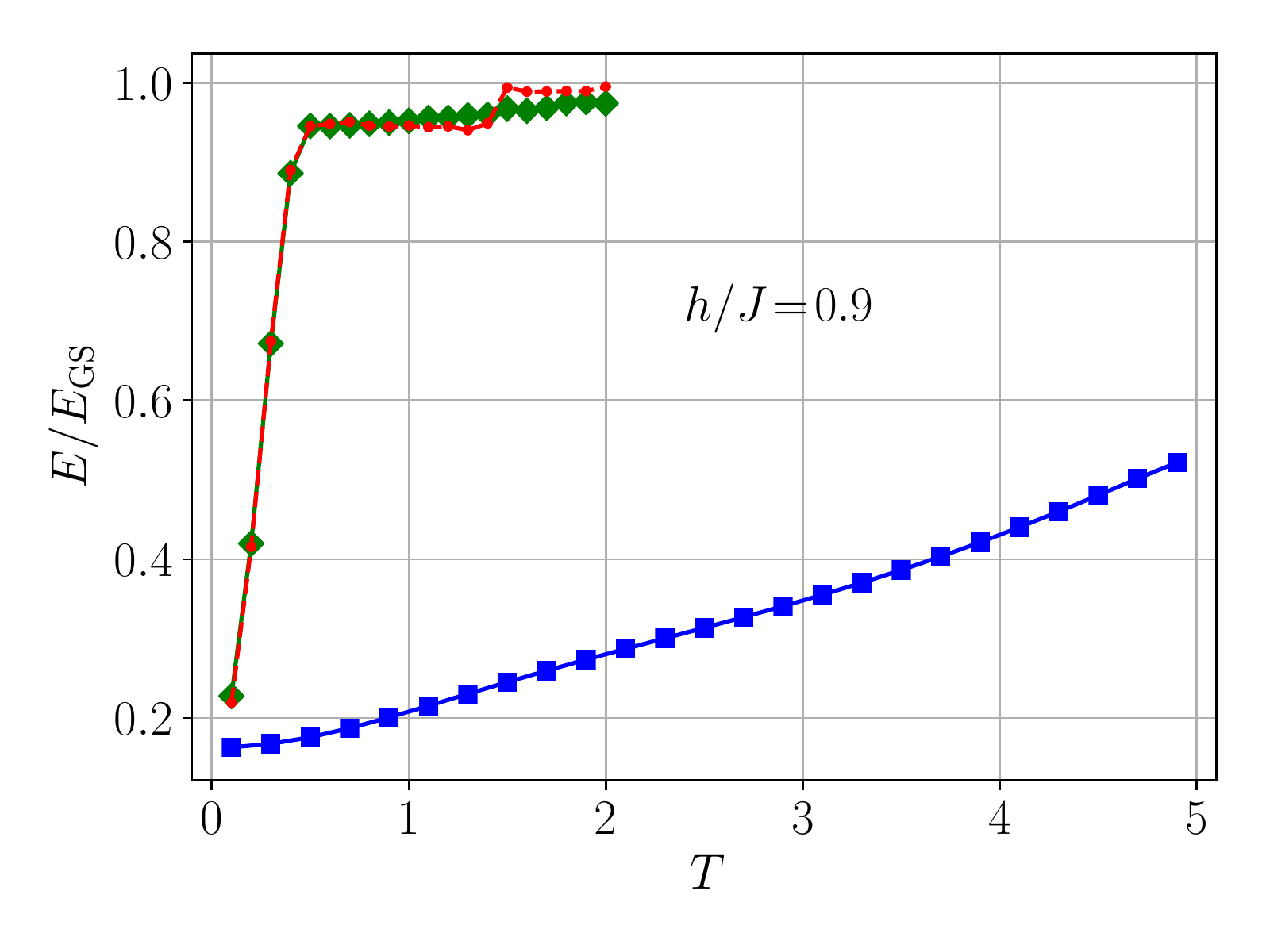}
    \end{minipage}%
    \caption{LMG model: energy minimization against protocol duration $T$ using conventional QAOA (blue square) and CD-QAOA (red dashed line, green solid line).
    The model parameters are the same from the settings in Fig.~\ref{fig:LMG_energy} but for $h/J\eq 0.1$ (top panel), and $h/J\eq 0.9$ (bottom panel). 
    \label{fig:LMG_physics}
    }
\end{figure}

Figure~\ref{fig:LMG_physics} shows the comparison between CD-QAOA and QAOA for two more values of $h/J=0.1$ (deep in the ferromagnetic regime), and $h/J=0.9$ (close to the critical point at $h/J=1.0$). While the behavior for $h/J=0.1$ is qualitatively similar to $h/J=0.5$ (discussed in the main text), we do see that close to the critical point the two-body gauge potential terms $\hat{XY}$ and $\hat{ZY}$ may offer some degree of improvement below the quantum speed limit, as compared to using only using the single-body $\hat{Y}$ term. We mention in passing that we observed a stronger system-size dependence in the optimal protocol found by the RL agent in the immediate vicinity of the critical point $h_c/J=1$.

\subsection{\label{app:spin-1}Spin-\texorpdfstring{$1$}{1} Ising Chain}

Finally, let us turn to the spin-$1$ Ising chain:
\begin{eqnarray}
    H(\lambda) \!&=&\! \lambda(t)H_1 \! + \! H_2, \\
    H_1 \!&=&\! \sum_{j=1}^N J S^z_{j+1}S^z_j + h_x S^x_j,\qquad H_2\!=\! \sum_{j=1}^N h_zS^z_j, \nonumber
\end{eqnarray}
see main text for discussion of the model parameters. Using this model, we compare four state preparation techniques: CD-QAOA, conventional QAOA, CD-driving using a variational gauge potential, and adiabatic evolution.

\begin{figure}[t!]
    \includegraphics[width=0.5\textwidth]{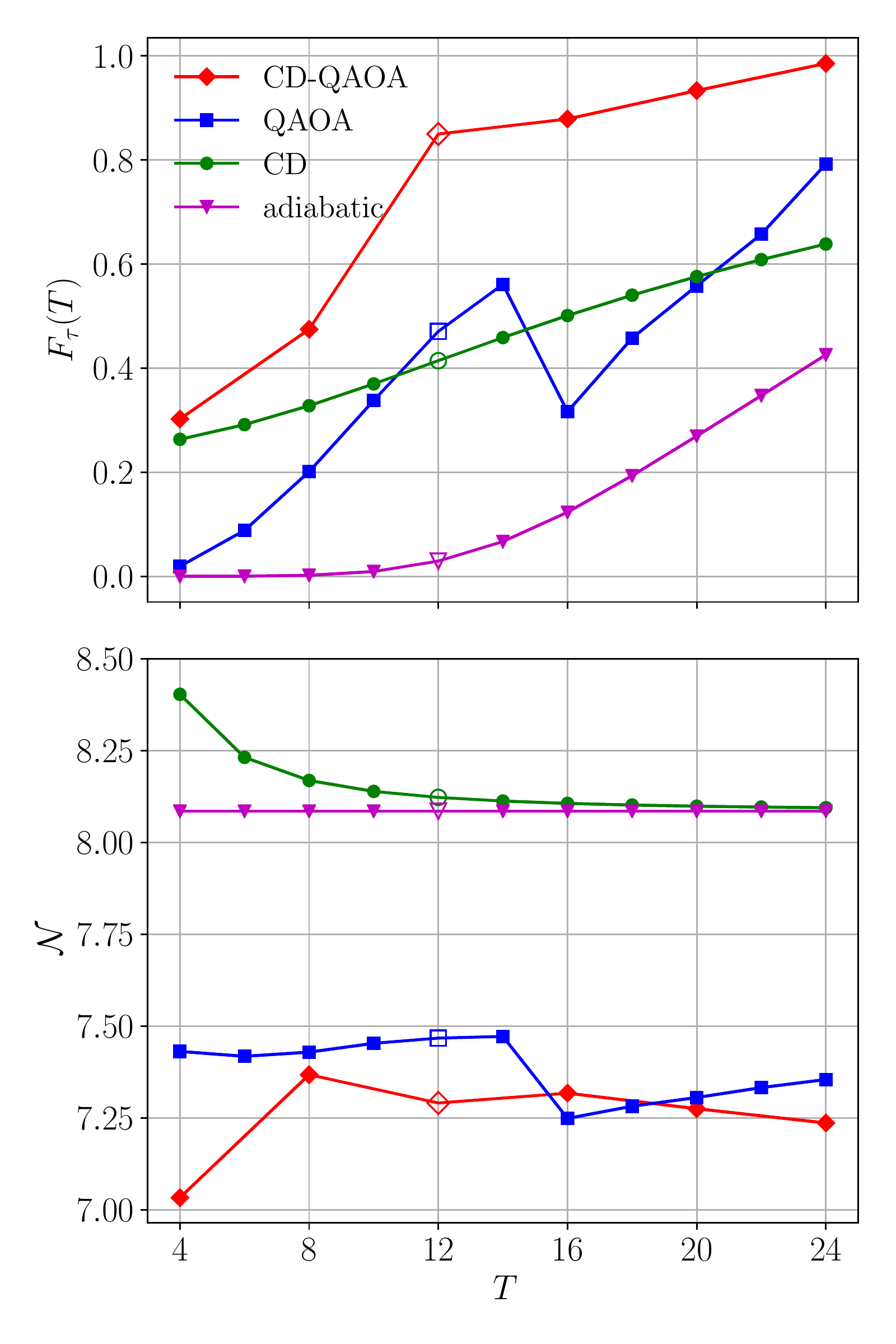}
    \caption{\label{fig:comparison-ising1-fid} Spin-$1$ Ising model:
    energy minimization against different protocol duration $T$ for four different optimization methods: CD-QAOA (red line), conventional QAOA (blue line), variational gauge potential (green) and adiabatic evolution (magenta). 
    Two associated quantities are shown: 
    many-body fidelity $F_\tau$ (top)
    and normalized time-averaged energy density $\mathcal{N}$ over the protocol (bottom).
    The empty symbols mark the duration for which the evolution of physical quantities is shown in Fig.~\ref{fig:comparison-ising1-stat}.
    The parameters are the same as in Fig.~\ref{fig:comparison-eng}.
    }
\end{figure}

\begin{figure*}[t!]
    \includegraphics[width=1.0\textwidth]{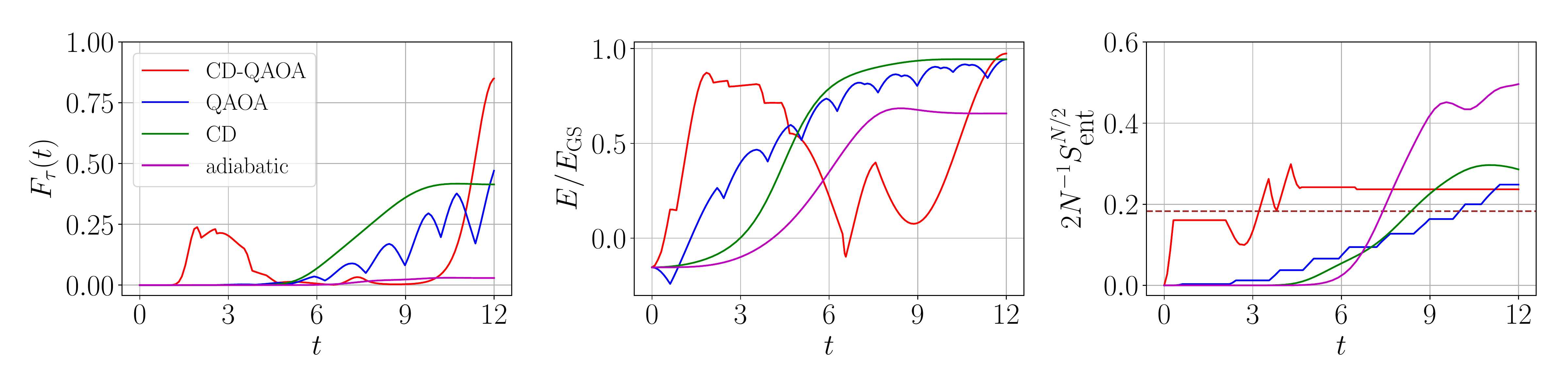}

    \caption{\label{fig:comparison-ising1-stat} Spin-$1$ Ising model:
        time evolution generated by the four different methods: CD-QAOA (red), conventional QAOA (blue), CD driving using the variational gauge potential (green) and adiabatic evolution (magenta). The three quantities are shown: the many-body fidelity (left), energy (middle), and entanglement entropy of the half chain (right). The protocols correspond to the empty symbols during $T\!=\!12$ in Fig.~\ref{fig:comparison-eng}.
        We compare 
        The horizontal dashed line in the entanglement entropy curve shows the value in the target state.
        The CD-QAOA protocol sequence is given in Table~\ref{tab:proto-ising1}.
        The model parameters are the same as in Fig.~\ref{fig:comparison-eng}. 
    }
\end{figure*}

In order to compare these four methods, we first investigate their energy budget, i.e.~the amount of energy required by the corresponding protocols. This is necessary, since variational CD-driving does not put any constraints on the magnitude of the expansion parameters $\beta_j(\lambda)$ [cf.~App.~\ref{app:varl_gauge}], and we know that larger energies (i.e.~generators of unitaries $H_j$ with large norms) in general allow for a faster population transfer. To measure quantitatively the energy budget of a protocol, we use the average energy density along the protocol trajectory
\begin{equation}
    \label{eq:norm_density}
    \mathcal{N}=\frac{1}{T}\int_0^T\mathrm{d}t\frac{\Vert H(t) \Vert}{N},
\end{equation}
where $H(t)$ is a unified notation for the continuous protocols in the case of adiabatic or CD driving, and the piecewise-constant (in time) sequences in CD-QAOA and conventional QAOA; $\Vert H \Vert$ denotes the Hilbert-Schmidt norm of the operator $H$. Since we are interested in many-body systems, it is also natural to look at the energy density, i.e.~$\Vert H(t) \Vert/N$.
Figure~\ref{fig:comparison-ising1-fid} [bottom] shows that $\mathcal{N}$ is on a similar scale for all four methods within the range of durations of interest, which allows for a meaningful comparison between them. As expected, CD-driving approaches adiabatic driving at large $T$, since the gauge potential term comes with a pre-factor $\dot\lambda$ which vanishes for $T\to\infty$; in the opposite limit of $T\to 0$, the energy budget of CD-driving blows up, as a result of $\beta_j(\lambda)$ being unconstrained.

In Fig.~\ref{fig:comparison-ising1-fid} [top], we see that the many-body fidelity, associated with the protocols obtained using energy density minimization, increases the performance contrast between the performance of the different methods [cf.~Fig.~\ref{fig:comparison-eng}, main text]. Since the fidelity is defined as the overlap square of the final with the target states [Eq.~\eqref{eq:fidelity}], like the entanglement entropy, it is insensitive to any specific observable; this implies that CD-QAOA outperforms the other three methods on \emph{all} observables, not just energy. This is anticipated, because CD-QAOA combines the variational power of QAOA with physical insights from CD driving. Despite its better performance, notice how CD-QAOA also has a smaller energy budget than either of CD- and adiabatic driving.

To demonstrate the nonequilibrium character of the optimal protocols found by the RL agent in this setup, we fix $T=12$, and look at the time evolution of the energy, the fidelity, and the entanglement entropy within the learned protocol, cf.~Fig.~\ref{fig:comparison-ising1-stat}.
While the protocol sequence [Table~\ref{tab:proto-ising1}] appears impenetrable, we remark that (i) the RL agent makes use of both single-particle and two-body gauge potential terms, and (ii) some step durations $\alpha_j$ are found to vanish identically, suggesting that the value of $q$ may be reduced.
As anticipated, the behavior of the dynamics generated by the CD and adiabatic driving is smooth, in contrast to the circuit-like piece-wise continuous curves of QAOA and CD-QAOA. The highly non-monotonic behavior of the energy curve shows that the CD-QAOA dynamics can be highly nonequilibrium: this likely stems from the RL objective [cf.~App.~\ref{app:algo}] -- the total expected return: the agent only cares about maximizing the reward at $t=T$ and is insensitive to any intermediate values. This allows the agent to drive the system through various states which are very far away from the target (e.g.~w.r.t.~the fidelity) [Curiously, these bad-energy states are all distinct, since they have different entanglement entropy, and the system does not visit the same quantum state twice during the evolution]. The non-smooth and non-monotonic behavior of the CD-QAOA solution raises the question about how robust the protocol is, to small external perturbations -- a topic of future studies.

\begin{table*}%
    \centering
    \captionsetup[subtable]{position=top}
    \begin{subtable}{0.3\linewidth}
        \centering
        \caption{{\bf Ising spin-$1$}}
        \begin{tabular}{c|c} 
\hline
\textbf{Hamiltonian} & \textbf{Duration} \\
\hline\hline
\rowcolor{GrayBG}$X|Y$ & 0.312\\
\rowcolor{GrayBG}$Y$ & 0.299\\
$Z$ & 0.216 \\
\rowcolor{GrayBG}$Y$ & 0.717\\
\textcolor{Gray}{$Z$} & \textcolor{Gray}{0.000} \\
\rowcolor{GrayBG}$Y$ & 0.537\\
$Z|Z\!+\!X$ & 0.477 \\
\rowcolor{GrayBG}$Y$ & 0.054\\
$Z|Z\!+\!X$ & 0.657 \\
\textcolor{Gray}{$Z$} & \textcolor{Gray}{0.000} \\
$Z|Z\!+\!X$ & 0.269 \\
\rowcolor{GrayBG}$Y|Z$ & 0.274\\
$Z|Z\!+\!X$ & 0.478 \\
\rowcolor{GrayBG}$Y|Z$ & 0.372\\
\textcolor{Gray}{$Z|Z\!+\!X$} & \textcolor{Gray}{0.000} \\
$Z$ & 1.794 \\
\rowcolor{GrayBG}$X|Y$ & 0.072\\
$Z$ & 0.039 \\
\rowcolor{GrayBG}$Y$ & 1.007\\
$Z$ & 4.426 \\
\hline
\end{tabular}
        \label{tab:proto-ising1}
    \end{subtable}
    \hspace*{5em}
    \begin{subtable}{0.3\linewidth}
        \centering
        \caption{{ \bf Ferromagnetic ($\Delta/J\!=\!-2.0$)}}
        \begin{tabular}{c|c} 
\hline
\textbf{short-hand notation} & \textbf{Duration} \\
\hline\hline
\rowcolor{GrayBG}$Y|Z\!-\!YZ$ & 0.122\\
$X|X\!+\!Y|Y$ & 0.178 \\
\rowcolor{GrayBG}$YZ$ & 0.027\\
$Z|Z$ & 0.376 \\
\rowcolor{GrayBG}$Y|Z\!-\!YZ$ & 0.234\\
\textcolor{Gray}{$Z|Z$} & \textcolor{Gray}{0.000} \\
$X|X\!+\!Y|Y$ & 0.323 \\
$Z|Z$ & 0.284 \\
\rowcolor{GrayBG}$Y|Z\!-\!YZ$ & 0.366\\
\textcolor{Gray}{$Z|Z$} & \textcolor{Gray}{0.000} \\
$X|X\!+\!Y|Y$ & 0.314 \\
$Z|Z$ & 0.188 \\
\rowcolor{GrayBG}$Y|Z\!-\!YZ$ & 0.535\\
\rowcolor{GrayBG}$Y$ & 0.001\\
$X|X$ & 0.342 \\
$Z|Z$ & 0.105 \\
\rowcolor{GrayBG}$Y|Z\!-\!YZ$ & 0.538\\
$X|X$ & 0.208 \\
\textcolor{Gray}{$Y$} & \textcolor{Gray}{0.000} \\
$Z|Z$ & 0.051 \\
\rowcolor{GrayBG}$Y$ & 0.658\\
\rowcolor{GrayBG}$Y|Z\!-\!YZ$ & 0.002\\
\rowcolor{GrayBG}$Y$ & 0.900\\
$Z$ & 0.771 \\
\rowcolor{GrayBG}$Y$ & 0.005\\
\rowcolor{GrayBG}$X|Y\!-\!XY$ & 0.474\\
\textcolor{Gray}{$Y|Z\!-\!YZ$} & \textcolor{Gray}{0.000} \\
\textcolor{Gray}{$X|X\!+\!Y|Y$} & \textcolor{Gray}{0.000} \\
\hline
\end{tabular}
        \label{tab:proto-delta=-2.0}

    \end{subtable} \\
    \vspace*{2em}
    \begin{subtable}{0.3\linewidth}
        \centering
        \caption{{ \bf XY ($\Delta/J\!=\!-0.5$)}}
        \begin{tabular}{c|c} 
\hline
\textbf{short-hand notation} & \textbf{Duration} \\
\hline\hline
\rowcolor{GrayBG}$Y$ & 0.795\\
\textcolor{Gray}{$X|X$} & \textcolor{Gray}{0.000} \\
\rowcolor{GrayBG}$Y$ & 0.772\\
$X|X\!+\!Y|Y$ & 0.143 \\
$X|X$ & 0.383 \\
\rowcolor{GrayBG}$Y$ & 0.001\\
$X|X\!+\!Y|Y$ & 0.284 \\
$X|X$ & 0.180 \\
$X|X\!+\!Y|Y$ & 0.467 \\
$X|X$ & 0.113 \\
$X|X\!+\!Y|Y$ & 0.635 \\
$X|X$ & 0.097 \\
$X|X\!+\!Y|Y$ & 0.617 \\
\textcolor{Gray}{$Y$} & \textcolor{Gray}{0.000} \\
$Z|Z$ & 0.162 \\
$X|X\!+\!Y|Y$ & 0.265 \\
$X|X$ & 0.092 \\
$Z$ & 1.995 \\
\hline
\end{tabular}
        \label{tab:proto-delta=-0.5}
    \end{subtable}
    \hspace*{5em}
    \begin{subtable}{0.3\linewidth}
        \centering
        \caption{{ \bf Haldane ($\Delta/J\!=\!0.5$)}}
         \begin{tabular}{c|c} 
\hline
\textbf{short-hand notation} & \textbf{Duration} \\
\hline\hline
$X|X\!+\!Y|Y$ & 0.149 \\
\textcolor{Gray}{$X|X$} & \textcolor{Gray}{0.000} \\
$X|X\!+\!Y|Y$ & 0.052 \\
$Z|Z$ & 1.376 \\
$X|X\!+\!Y|Y$ & 0.313 \\
$Z|Z$ & 0.668 \\
$X|X\!+\!Y|Y$ & 0.187 \\
$Z|Z$ & 0.723 \\
$X|X\!+\!Y|Y$ & 0.289 \\
$Z|Z$ & 0.528 \\
$X|X\!+\!Y|Y$ & 0.218 \\
$Z|Z$ & 0.561 \\
$X|X\!+\!Y|Y$ & 0.254 \\
$Z|Z$ & 0.684 \\
$X|X\!+\!Y|Y$ & 0.360 \\
$Z|Z$ & 0.639 \\
\textcolor{Gray}{$X|X$} & \textcolor{Gray}{0.000} \\
\textcolor{Gray}{$Z$} & \textcolor{Gray}{0.000} \\
\hline
\end{tabular}
        \label{tab:proto-delta=0.5}
    \end{subtable}%
    \vspace{1.5em}
    \caption{Ising spin-$1$ chain and Anisotropic Heisenberg spin-$1$ chain: the protocol sequences and corresponding durations given by CD-QAOA. The protocol \subref{tab:proto-ising1} correspond to Ising spin-$1$ in Fig.~\ref{fig:comparison-ising1-stat}; the \subref{tab:proto-delta=-2.0}, \subref{tab:proto-delta=-0.5}, \subref{tab:proto-delta=0.5} three sequences correspond to the three phases in the same setting as Fig.~\ref{fig:comparison-heisenberg-stat}. The short-hand notation is the same in Table~\ref{table:gauge_pot}. 
    We use a shaded cell background whenever terms from the CD gauge potential are used in the protocol sequence. Terms of zero durations are marked in light grey. 
    }
\end{table*}

\end{document}